\documentclass[aps,prb,twocolumn,showpacs,floatfix]{revtex4-1}
\usepackage{graphicx,amsfonts,amssymb,amsmath,hyperref}
\usepackage{textcomp}

\newif\ifhyper
\hypertrue
\ifhyper
\hypersetup{
  citecolor = {green},
  colorlinks = {true}, 
  urlcolor = {blue} 
}

\begin{document}

\graphicspath{{./figures_submit/}}


\def\Sloc{S_{\rm loc}} 
\newcommand{\sgn}{{\rm sgn}} 
\def\intw{\int_{-\infty}^\infty \frac{d\w}{2\pi}}
\newcommand{\ket}[1]{|#1\rangle}
\newcommand{\bra}[1]{\langle#1|}
\newcommand{\const}{{\rm const}} 
\def\beq{\begin{equation}}
\def\eeq{\end{equation}}
\def\llbrace{\left\lbrace}
\def\rrbrace{\right\rbrace}
\def\lbraket{\left[}
\def\rbraket{\right]}

\newcommand{\Tr}{{\rm Tr}} 
\newcommand{\mean}[1]{\langle #1 \rangle}

\newcommand{\ie}{i.e. }
\newcommand{\eg}{e.g. }
\newcommand{\cc}{{\rm c.c.}} 
\newcommand{\hc}{{\rm h.c.}} 

\def\eps{\epsilon}
\def\gam{\gamma} 
\def\phibf{\boldsymbol{\phi}}
\def\varphibf{\boldsymbol{\varphi}}
\def\psibf{\boldsymbol{\psi}}
\def\lamb{\lambda}
\def\sig{\sigma}

\def\half{\frac{1}{2}}

\def\a{{\bf a}}
\def\b{{\bf b}}
\def\e{{\bf e}}
\def\f{{\bf f}}
\def\g{{\bf g}}
\def\h{{\bf h}}
\def\k{{\bf k}}
\def\l{{\bf l}}
\def\m{{\bf m}}
\def\n{{\bf n}} 
\def\p{{\bf p}} 
\def\q{{\bf q}}
\def\r{{\bf r}}
\def\t{{\bf t}}
\def\u{{\bf u}}
\def\v{{\bf v}}
\def\x{{\bf x}}
\def\y{{\bf y}} 
\def\z{{\bf z}} 
\def\A{{\bf A}}
\def\B{{\bf B}}
\def\D{{\bf D}} 
\def\E{{\bf E}} 
\def\F{{\bf F}} 
\def\H{{\bf H}}  
\def\J{{\bf J}}
\def\K{{\bf K}} 

\def\G{{\bf G}}
\def\L{{\bf L}}
\def\M{{\bf M}}  
\def\O{{\bf O}} 
\def\P{{\bf P}} 
\def\Q{{\bf Q}} 
\def\R{{\bf R}}
\def\S{{\bf S}}
\def\nablabf{\boldsymbol{\nabla}}

\def\para{\parallel}

\def\w{\omega}
\def\wn{\omega_n}
\def\wnu{\omega_\nu}
\def\wp{\omega_p} 
\def\dmu{{\partial_\mu}}
\def\dl{{\partial_l}}  
\def\dt{\partial_t} 
\def\tdt{\tilde\partial_t}
\def\dk{\partial_k}
\def\tdk{\tilde\partial_k}
\def\dx{\partial_x}
\def\dy{\partial_y} 
\def\dtau{{\partial_\tau}}  
\def\det{{\rm det}} 
\def\Pf{{\rm Pf}}

\def\intr{\int d^dr}  
\def\dintr{\displaystyle \int d^dr} 
\def\dinttau{\displaystyle \int_0^\beta d\tau}
\def\inttau{\int_0^\beta d\tau}
\def\intx{\int d^{d+1}x} 
\def\inttaur{\int_0^\beta d\tau \int d^dr}
\def\intinf{\int_{-\infty}^\infty}

\def\calA{{\cal A}} 
\def\calC{{\cal C}} 
\def\dt{\partial_t}
\def\calD{{\cal D}}
\def\calF{{\cal F}} 
\def\calG{{\cal G}}
\def\calH{{\cal H}}
\def\calJ{{\cal J}}
\def\calL{{\cal L}} 
\def\calN{{\cal N}}
\def\calO{{\cal O}}
\def\calP{{\cal P}}  
\def\calR{{\cal R}} 
\def\calS{{\cal S}}
\def\calT{{\cal T}}
\def\calU{{\cal U}}
\def\calY{{\cal Y}} 
\def\calZ{{\cal Z}} 

\def\dw{\partial_\w}
\def\Gamloc{\Gamma_{\rm loc}}
\def\Vloc{V_{\rm loc}}
\def\Zloc{Z_{\rm loc}}
\def\Gn{G_{\rm n}}
\def\Gan{G_{\rm an}}
\def\nbarloc{\bar n_{\rm loc}} 
\def\Ibarll{\bar I_{\rm ll}}
\def\Ibartt{\bar I_{\rm tt}}
\def\Jbarllll{\bar J_{\rm ll,ll}}
\def\Jbartttt{\bar J_{\rm tt,tt}}
\def\Jbarltlt{\bar J_{\rm lt,lt}}
\def\Jbarlltt{\bar J_{\rm ll,tt}}
\def\Jbarttll{\bar J_{\rm tt,ll}}
\def\Jbarlllt{\bar J_{\rm ll,lt}}
\def\Jbarltll{\bar J_{\rm lt,ll}}
\def\Jbarttlt{\bar J_{\rm tt,lt}}
\def\Jbarlttt{\bar J_{\rm lt,tt}}
\def\gllb{\bar{G}_{k,{\rm ll}}}
\def\gttb{\bar{G}_{k,{\rm tt}}}
\def\gltb{\bar{G}_{k,{\rm lt}}}

\title{Nonperturbative renormalization-group approach to strongly-correlated lattice bosons} 

\author{A. Ran\c{c}on and  N. Dupuis}
\affiliation{
 Laboratoire de Physique Th\'eorique de la Mati\`ere Condens\'ee, 
CNRS UMR 7600, \\ Universit\'e Pierre et Marie Curie, 4 Place Jussieu, 
75252 Paris Cedex 05,  France}

\date{October 10, 2011}

\begin{abstract} 
We present a nonperturbative renormalization-group approach to the Bose-Hubbard model. By taking as initial condition of the renormalization-group flow the (local) limit of decoupled sites, we take into account both local and long-distance fluctuations in a nontrivial way. This approach yields a phase diagram in very good quantitative agreement with quantum Monte Carlo simulations, and reproduces the two universality classes of the superfluid--Mott-insulator transition. The critical behavior near the multicritical points, where the transition takes place at constant density, agrees with the original predictions of Fisher {\it et al.} [Phys. Rev. B {\bf 40}, 546 (1989)] based on simple scaling arguments. At a generic transition point, the critical behavior is mean-field like with logarithmic corrections in two dimensions. In the weakly-correlated superfluid phase (far away from the Mott insulating phase), the renormalization-group flow is controlled by the Bogoliubov fixed point down to a characteristic (Ginzburg) momentum scale $k_G$ which is much smaller than the inverse healing length $k_h$. In the vicinity of the multicritical points, when the density is commensurate, we identify a sharp crossover from a weakly- to a strongly-correlated superfluid phase where the condensate density and the superfluid stiffness are strongly suppressed and both $k_G$ and $k_h$ are of the order of the inverse lattice spacing.
\end{abstract}
\maketitle

\section{Introduction}

In the last two decades, the nonperturbative renormalization group (NPRG) approach has been successfully applied to many areas of physics,\cite{Berges02,Delamotte07} from high-energy physics to statistical and condensed-matter physics. It has proven to be a powerful tool to study not only the low-energy long-distance properties in the vicinity of second-order phase transitions but also non-universal quantities. In particular, the NPRG approach has been implemented in lattice models and used to  compute the transition temperature and the magnetization in classical spin models (Ising, XY and Heisenberg models).\cite{Machado10} This implementation of the NPRG is referred to as the lattice NPRG.

The strategy of the NPRG is to build a family of models indexed by a momentum scale $k$, such that fluctuations are smoothly taken into account as $k$ is lowered from a microscopic scale $\Lambda$ down to 0. In practice this is achieved by adding to the action $S$ of the system an infrared regulator term $\Delta S_k$ which vanishes for $k=0$. For a scalar field theory, the regulator term is a mass-like term $\Delta S_k[\varphi]=\half \sum_\q \varphi_{-\q} R_k(\q) \varphi_\q$, where the cutoff function $R_k(\q)$ is chosen such that $R_k(\q) \sim k^2$ for $|\q|\lesssim k$ and $R_k(\q)\sim 0$ for $|\q|\gtrsim k$, which effectively suppresses the low-energy modes $|\q|\lesssim k$. One can then define a scale-dependent partition function $Z_k[J]$ and a scale-dependent effective action $\Gamma_k[\phi]$ defined as a slightly modified Legendre transform (see Sec.~\ref{subsec_ea} for the precise definition) of $-\ln Z_k[J]$. Here $J$ is an external source which couples linearly to the $\varphi$ field and $\phi(\r)=\delta \ln Z_k[J]/\delta J(\r)$. In the standard implementation of the NPRG, at the microscopic scale $k=\Lambda$, all fluctuations are frozen by the $\Delta S_\Lambda$ term so that $\Gamma_\Lambda[\phi]=S[\phi]$ as in Landau's (mean-field) theory of phase transitions. The effective action of the original model is obtained for $k=0$ ($\Delta S_{k=0}=0$) and can be determined by (approximately) solving the RG equation satisfied by $\Gamma_k$.\cite{Berges02,Delamotte07}

The lattice NPRG differs from the standard implementation in the initial condition.\cite{Machado10}
The cutoff function $R_k(\q)$ is chosen such that at the microscopic scale $k=\Lambda$ the action $S+\Delta S_\Lambda$ corresponds to the local limit of decoupled sites. Local fluctuations are therefore included from the very beginning of the RG procedure. The intersite coupling is then gradually restored as $k$ decreases from $\Lambda$ down to 0. In the low-energy limit $k\ll\Lambda$, $R_k(\q)$ acts as an infrared regulator suppressing fluctuations with momenta $|\q|\lesssim k$. The lattice NPRG is then equivalent to the standard NPRG and yields identical results for the critical properties. The hallmark of the lattice NPRG is thus to take into account both local and critical fluctuations in a nontrivial way. 

In this paper, we present a NPRG study of the Bose-Hubbard model\cite{Fisher89} at zero temperature and in dimension $d=2$ or $d=3$. This model has been intensively studied in the last years following the experimental observation of the superfluid--Mott-insulator transition of an ultracold bosonic gas in an optical lattice.\cite{Jaksch98,Greiner02,Stoferle04,Spielman07} Phase diagram and thermodynamic quantities are known from the numerically exact lattice quantum Monte Carlo (QMC) simulations.\cite{Capogrosso07,Capogrosso08} On the other hand few studies have addressed the critical behavior at the superfluid--Mott-insulator transition,\cite{Krauth91} and most of our understanding goes back to the seminal work of Fisher {\it et al.}\cite{Fisher89} 

The standard NPRG scheme does not capture the superfluid--Mott-insulator transition in the Bose-Hubbard model. The reason is that near the transition the mean-field solution is too far away from the actual state of the system to provide a reliable initial condition for the NPRG procedure. The two-pole structure of the local (on-site) single-particle propagator is crucial for the very existence of the transition (see, e.g., Ref.~\onlinecite{Fisher89}).  It is however impossible to reproduce this structure from a RG approach starting from the mean-field (Bogoliubov) theory within standard approximations of the RG equation satisfied by the effective action $\Gamma_k$. This prevents a straightforward generalization of recent NPRG studies\cite{Dupuis07,Dupuis09a,Dupuis09b,Wetterich08,Floerchinger08,Sinner09,Sinner10}  of interacting bosons to the Bose-Hubbard model.

By contrast the lattice NPRG, which takes into account local fluctuations, is able to describe the superfluid--Mott-insulator transition.\cite{Rancon11} Since the starting action $S+\Delta S_\Lambda$ is purely local, this approach is to some extent reminiscent of various $t/U$ expansions of the Bose-Hubbard model.\cite{Freericks94,*Freericks96,*Freericks09,Buonsante05,Santos09,Teichmann09a,*Teichmann09b,Koller06,Knap10,Knap11,Arrigoni11} Moreover, the lattice NPRG is not restricted to the computation of thermodynamic quantities and allows us to study the critical behavior at the superfluid--Mott-insulator transition and compare with the predictions of Fisher {\it et al.}\cite{Fisher89} based on scaling arguments. 

In addition to the phase diagram and the critical behavior at the superfluid--Mott-insulator transition, the NPRG approach can also address the superfluid phase. Deep in the superfluid phase, localization effects are negligible and we expect the Bogoliubov theory to provide a good description of the system. However, even in this weak correlation limit, it is known that the Bogoliubov approximation breaks down below a characteristic (Ginzburg) momentum scale $k_G$. In perturbation theory about the Bogoliubov approximation, the Ginzburg scale manifests itself by the appearance of infrared divergences below three dimensions ($d\leq 3$). Although these divergences cancel out in local gauge invariant quantities (condensate density, sound mode velocity, etc.),\cite{Beliaev58a,Beliaev58b,Hugenholtz59,Gavoret64} they do have a physical origin: they result from the coupling between longitudinal and transverse (phase) fluctuations and reflect the divergence of the longitudinal susceptibility\cite{Nepomnyashchii78,Nepomnyashchii83} -- a general phenomenon in systems with a continuous broken symmetry.\cite{Patasinskij73,Fisher73,Anishetty99,Sachdev99,Zwerger04,Dupuis11} The normal and anomalous self-energies, $\Sigma_{\rm n}(\q,\w)$ and $\Sigma_{\rm an}(\q,\w)$, are non-analytic functions of $\q$ and $\w$ when $|\q|,|\w|/c\ll k_G$ and $d\leq 3$ ($c$ denotes the velocity of the sound mode), while $\Sigma_{\rm an}(0,0)$ vanishes,\cite{Nepomnyashchii75} in marked contrast with the Bogoliubov approximation where the linear spectrum and the superfluidity rely on a finite value of the anomalous self-energy. A weakly-correlated superfluid is defined by the condition $k_G\ll k_h$ where the healing scale $k_h$ is the inverse of the healing length $\xi_h=k_h^{-1}$.\cite{Pistolesi04,Dupuis09b,Dupuis11,Capogrosso10} In this case, the Bogoliubov theory applies to a large part of the spectrum where the dispersion is linear ($|\q|\ll k_h$) and breaks down only at very low momenta $|\q|\ll k_G$. The Goldstone regime $|\q|\ll k_G$, dominated by phase fluctuations, is conveniently described by Popov's hydrodynamic theory (free of infrared divergences) based on a density-phase representation of the boson field $\psi=\sqrt{n}e^{i\theta}$.\cite{Popov72a,Popov79,Popov_book_1} The NPRG approach yields a unified description of superfluidity which includes both Bogoliubov theory (valid for $|\q|\gg k_G$) and Popov's hydrodynamic approach (valid for $|\q|\ll k_h$).\cite{Dupuis09a,Dupuis09b,Dupuis11} The Bose-Hubbard model gives us the opportunity to understand the fate of the weakly-correlated superfluid phase as we increase the strength of the interactions and  move closer to the Mott insulating phase in the phase diagram. 

The paper is organized as follows. In Sec.~\ref{sec_nprg}, we derive the lattice NPRG formalism for the Bose-Hubbard model. We introduce the scale-dependent effective action $\Gamma_k$ and compute its initial value $\Gamma_\Lambda$ by solving the single-site Bose-Hubbard model. We show that $\Gamma_\Lambda$ reproduces the result of the strong-coupling random-phase approximation (RPA).\cite{Sheshadri93,Oosten01,Sengupta05,Ohashi06,Menotti08} We also discuss the approximations used to solve the flow equation satisfied by $\Gamma_k$. The phase diagram obtained from the NPRG equations is in very good quantitative agreement with the QMC results (Sec.~\ref{sec_phase_dia}). Furthermore, the critical behavior derived from the NPRG analysis is in complete agreement with the predictions of Fisher {\it et al.} based on scaling arguments (Sec.~\ref{sec_crit}).\cite{Fisher89} We find multicritical points in the universality class of the $(d+1)$-dimensional XY model where the transition takes place at constant density. The XY critical behavior is observed in the Mott gap, the condensate density, the compressibility and the superfluid stiffness when a multicritical point is approached at constant chemical potential by varying the ratio $t/U$ between the hopping amplitude and the local repulsion between particles. At a generic transition point, we observe mean-field behavior, with logarithmic corrections in dimension $d=2$ (corresponding to the upper critical dimension). The superfluid phase is discussed in Sec.~\ref{sec_sf}. In the dilute limit, the renormalization-group flow is controlled by the Bogoliubov fixed point down to a characteristic (Ginzburg) momentum scale $k_G$ which is much smaller than the inverse healing length $k_h$. The Goldstone regime $k\ll k_G$, dominated by phase fluctuations, is characterized by a (relativistic) Lorentz invariance of the effective action $\Gamma_k$.\cite{Wetterich08,Dupuis07} In the vicinity of the multicritical points, when the density is commensurate, we identify a sharp crossover from a weakly- to a strongly-correlated superfluid phase where the condensate density and the superfluid stiffness are strongly suppressed and both $k_G$ and $k_h$ are of the order of the inverse lattice spacing. The main results are summarized in Sec.~\ref{sec_conclu}.

\section{Lattice NPRG}
\label{sec_nprg}

The Bose-Hubbard model on a $d$-dimensional hypercubic lattice is defined by the (Euclidean) action
\begin{align}
S = \inttau \biggl\lbrace & \sum_\r \Bigl[ \psi_\r^* (\dtau-\mu)\psi_\r + \frac{U}{2} (\psi_\r^*\psi_\r)^2 \Bigr] \nonumber \\ & - t \sum_{\mean{\r,\r'}} \left(\psi_\r^* \psi_{\r'}+\mbox{c.c.}\right) \biggr\rbrace ,
\label{action}
\end{align}
where $\psi_\r(\tau)$ is a complex field and $\tau\in [0,\beta]$ an imaginary time with $\beta\to\infty$ the inverse temperature. $\lbrace\r\rbrace$ denotes the $N$ sites of the lattice. $U$ is the on-site repulsion, $t$ the hopping amplitude between nearest-neighbor sites $\mean{\r,\r'}$ and $\mu$ the chemical potential. In the following we will sometimes write the boson field 
\begin{equation}
\psi_{\r} = \frac{1}{\sqrt{2}} \left( \psi_{1\r} + i \psi_{2\r} \right) 
\label{psi12} 
\end{equation}
in terms of two real fields $\psi_{1\r}$ and $\psi_{2\r}$. 

We set $\hbar=k_B=1$ and take the lattice spacing as the unit length throughout the paper.

\subsection{Scale-dependent effective action} 
\label{subsec_ea}

\begin{figure}
\centerline{
\includegraphics[width=2.65cm,clip]{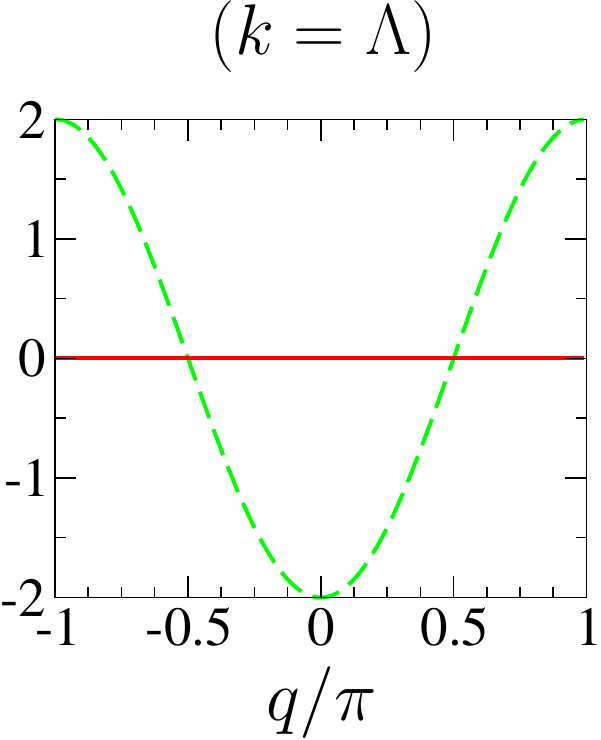}
\includegraphics[width=2.5cm,clip]{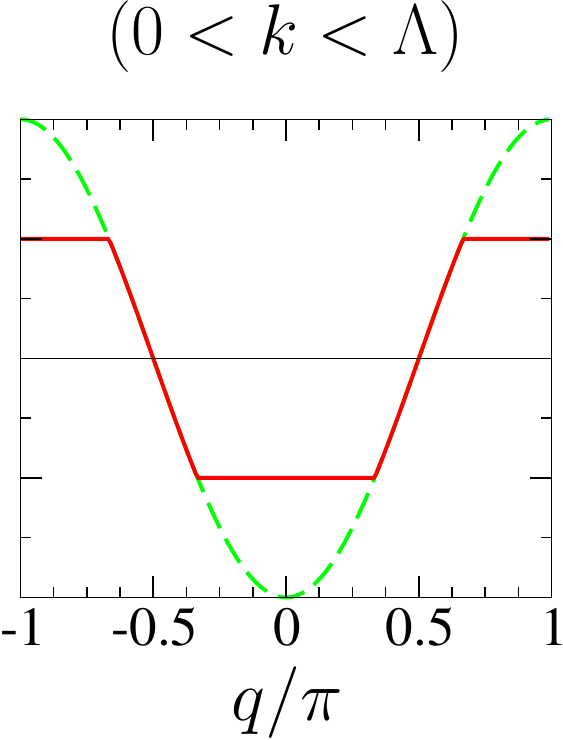}
\includegraphics[width=2.5cm,clip]{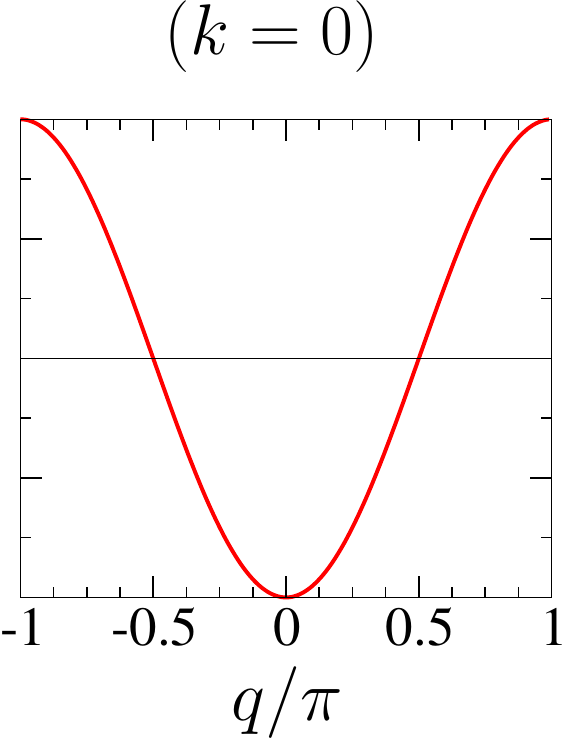}}
\caption{(Color online) Effective (bare) dispersion $t_q+R_k(q)$ for $k=\Lambda$, $0<k<\Lambda$ and $k=0$ with cutoff function (\ref{cutoff}). The (green) dashed line shows the bare dispersion $t_q=-2t\cos q$. $d=1$ and $t$ is taken as the energy unit.}
\label{fig_disp}  
\end{figure}

Following the general strategy of the NPRG, we consider a family of models with action $S_k=S+\Delta S_k$ indexed by a momentum scale $k$ varying from a microscopic scale $\Lambda$ down to 0. The regulator term is defined by  
\begin{equation}
\Delta S_k=\inttau \sum_\q \psi^*_\q R_k(\q) \psi_\q,
\end{equation}
where $\psi_\q$ is the Fourier transform of $\psi_\r$ and the sum over $\q$ runs over the first Brillouin zone $]-\pi,\pi]^d$ of the reciprocal lattice. The cutoff function $R_k(\q)$ modifies the bare dispersion $t_\q=-2t\sum_{i=1}^d \cos q_i$ of the bosons. $R_\Lambda(\q)$ is chosen such that the effective (bare) dispersion $t_\q+R_\Lambda(\q)$ vanishes.\cite{Machado10} The action $S_\Lambda=S+\Delta S_\Lambda$ then corresponds to the local limit of decoupled sites (vanishing hopping amplitude). By choosing $R_\Lambda(\q)+t_\q=0$, rather than $R_\Lambda(\q)+t_\q=2dt$ as in Ref.~\onlinecite{Machado10}, we ensure that $R_\Lambda(\q)$ does not modify the chemical potential but only the kinetic energy. 

In practice, we choose the cutoff function
\begin{equation} 
R_k(\q) = - Z_{A,k} \eps_k \mbox{sgn}(t_\q) (1-y_\q)\Theta(1-y_\q) ,
\label{cutoff} 
\end{equation}
with $\Lambda=\sqrt{2d}$, $\eps_k=tk^2$, $y_\q=(2dt-|t_\q|)/\eps_k$ and $\Theta(x)$ the step function (see Fig.~\ref{fig_disp}). The $k$-dependent constant $Z_{A,k}$ is defined below ($Z_{A,\Lambda}=1$). 
Since $R_{k=0}(\q)=0$, the action $S_{k=0}$ coincides with the action~(\ref{action}) of the original model. For small $k$, the function $R_k(\q)$ gives a mass $\sim k^2$ to the low-energy modes $|\q|\lesssim k$ and acts as an infrared regulator as in the standard NPRG scheme.\cite{Berges02,Delamotte07} 

The scale-dependent effective action 
\begin{align}
\Gamma_k[\phi^*,\phi] ={}& - \ln Z_k[J^*,J] + \inttau \sum_\r (J^*_\r\phi_\r+\mbox{c.c.}) \nonumber \\ & - \Delta S_k[\phi^*,\phi] 
\end{align}
is defined as a (slightly modified) Legendre transform which includes the explicit subtraction of $\Delta S_k[\phi^*,\phi]$. Here $Z_k[J^*,J]$ is the partition function obtained from the action $S+\Delta S_k$, $J_\r$ a complex external source which couples linearly to the bosonic field $\psi_\r$, and 
\begin{equation}
\phi_\r(\tau)=\frac{\delta\ln Z_k[J^*,J]}{\delta J^*_\r(\tau)} , \quad 
\phi^*_\r(\tau)=\frac{\delta\ln Z_k[J^*,J]}{\delta J_\r(\tau)} 
\end{equation}
the superfluid order parameter. The variation of the effective action with $k$ is governed by Wetterich's equation,\cite{Wetterich93} 
\begin{equation}
\partial_k \Gamma_k[\phi^*,\phi] = \half \Tr\biggl\lbrace \partial_k R_k\left(\Gamma^{(2)}_k[\phi^*,\phi] + R_k\right)^{-1} \biggr\rbrace ,
\label{wetteq}
\end{equation}
where $\Gamma^{(2)}_k$ is the second-order functional derivative of $\Gamma_k$. In Fourier space, the trace in (\ref{wetteq}) involves a sum over momenta and frequencies as well as the two components of the complex field $\phi$. 

We are primarily interested in two quantities. The first one is the effective potential defined by
\begin{equation}
V_k(n)=\frac{1}{\beta N}\Gamma_k[\phi^*,\phi] \biggl|_{\phi\;\const} 
\end{equation}
where $\phi$ is a constant (uniform and time-independent) field. The U(1) symmetry of the action implies that $V_k(n)$ is a function of $n=|\phi|^2$. Its minimum determines the condensate density $n_{0,k}$ and the thermodynamic potential (per site) $V_{0,k}=V_k(n_{0,k})$ in the equilibrium state. 

The second quantity of interest is the two-point vertex 
\begin{equation}
\Gamma^{(2)}_{k,ij}(\r-\r',\tau-\tau';\phi) = \frac{\delta^{(2)} \Gamma[\phi]}{\delta\phi_{i\r}(\tau) \delta\phi_{j\r'}(\tau')} \biggl|_{\phi\;\const}  
\end{equation}
which determines the one-particle propagator $G_k=-\Gamma^{(2)-1}_k$. Here the indices $i,j$ refer to the real and imaginary parts of $\phi$ [see Eq.~(\ref{psi12})]. Because of the U(1) symmetry of the action~(\ref{action}), the two-point vertex in a constant field takes the form\cite{Dupuis09b} 
\begin{equation}
\Gamma_{k,ij}^{(2)}(q;\phi) = \delta_{i,j}\Gamma_{A,k}(q;n) + \phi_i\phi_j \Gamma_{B,k}(q;n) + \eps_{ij} \Gamma_{C,k}(q;n) 
\label{gam2}
\end{equation}
in Fourier space, where $q=(\q,i\w)$, $\w$ is a Matsubara frequency and $\eps_{ij}$ the antisymmetric tensor. For $q=0$, we can relate $\Gamma^{(2)}_k$ to the derivative of the effective potential, 
\begin{equation}
\Gamma^{(2)}_{k,ij}(q=0;\phi) = \frac{\partial^2 V_k(n)}{\partial\phi_i\partial\phi_j} = \delta_{i,j}V_k'(n) + \phi_i\phi_j V_k''(n), 
\end{equation}
so that
\begin{equation}
\begin{split}
\Gamma_{A,k}(q=0;n) &= V_k'(n) , \\
\Gamma_{B,k}(q=0;n) &= V_k''(n) , \\
\Gamma_{C,k}(q=0;n) &= 0 . \\
\end{split}
\end{equation}
Parity and time-reversal invariance imply\cite{Dupuis09b}
\begin{equation}
\begin{split}
\Gamma_{A,k}(q;n) &= \Gamma_{A,k}(-q;n) = \Gamma_{A,k}(\q,-i\w;n) , \\ 
\Gamma_{B,k}(q;n) &= \Gamma_{B,k}(-q;n) = \Gamma_{B,k}(\q,-i\w;n) , \\ 
\Gamma_{C,k}(q;n) &= -\Gamma_{C,k}(-q;n) = - \Gamma_{C,k}(\q,-i\w;n) .
\end{split}
\label{sym}
\end{equation}

The one-particle propagator $G_k=-\Gamma^{(2)-1}_k$ can be written in a form analogous to~(\ref{gam2}) or in terms of its longitudinal and transverse components, 
\begin{align}
G_{k,ij}(q;\phi) ={}& \frac{\phi_i\phi_j}{2n} G_{k,\rm ll}(q;n) + \left( \delta_{i,j} -  \frac{\phi_i\phi_j}{2n} \right) G_{k,\rm tt}(q;n) \nonumber \\ & + \eps_{ij}  G_{k,\rm lt}(q;n) ,
\end{align}
where
\begin{equation}
\begin{split}
G_{k,\rm ll}(q;n) &= - \frac{\Gamma_{A,k}(q;n)}{D_k(q;n)} , \\ 
G_{k,\rm tt}(q;n) &= - \frac{\Gamma_{A,k}(q;n)+2n\Gamma_{B,k}(q;n)}{D_k(q;n)} , \\ 
G_{k,\rm lt}(q;n) &= \frac{\Gamma_{C,k}(q;n)}{D_k(q;n)} ,
\end{split}
\label{green}
\end{equation}
with
$D_k=\Gamma_{A,k}^2+2n\Gamma_{A,k}\Gamma_{B,k}+\Gamma_{C,k}^2$. Note that the single-particle propagator entering the flow equation~(\ref{wetteq}) is defined by $-(\Gamma_k^{(2)}+R_k)^{-1}$, which is the propagator associated with the true Legendre transform, rather than $-\Gamma_k^{(2)-1}$. 

\subsection{Initial conditions}
\label{subsec_init}

Since the action $S+\Delta S_\Lambda\equiv S_{\rm loc}$ corresponds to the local limit, the initial value of the effective action reads 
\begin{equation}
\Gamma_\Lambda[\phi^*,\phi] = \Gamloc[\phi^*,\phi] + \inttau \sum_\q \phi^*(\q) t_\q \phi(\q) ,
\label{GamLambda}
\end{equation}
where 
\begin{equation}
\Gamloc[\phi^*,\phi] = -\ln \Zloc[J^*,J] + \inttau \sum_\r (J^*_\r\phi_\r+\mbox{c.c.}) 
\label{Gamloc}
\end{equation}
is the Legendre transform of the thermodynamic potential $-\ln \Zloc[J^*,J]$ in the local limit. In Eq.~(\ref{Gamloc}), $J$ is related to $\phi$ by the relation $\phi_r(\tau)=\delta \ln \Zloc[J^*,J]/\delta J^*_r(\tau)$ and $\Zloc$ is the partition function obtained from $S_{\rm loc}$. 

It is not possible to compute the functional $\Gamloc[\phi^*,\phi]$ for arbitrary time-dependent fields. One can however easily obtain the effective potential $\Vloc(n)$ and the two-point vertex $\Gamloc^{(2)}$ in a time-independent field $\phi$. These quantities are sufficient to specify the initial conditions of the flow within the approximations that we consider in Sec.~\ref{subsec_rgeq}. 

\begin{figure}
\centerline{\includegraphics[width=6.5cm,clip]{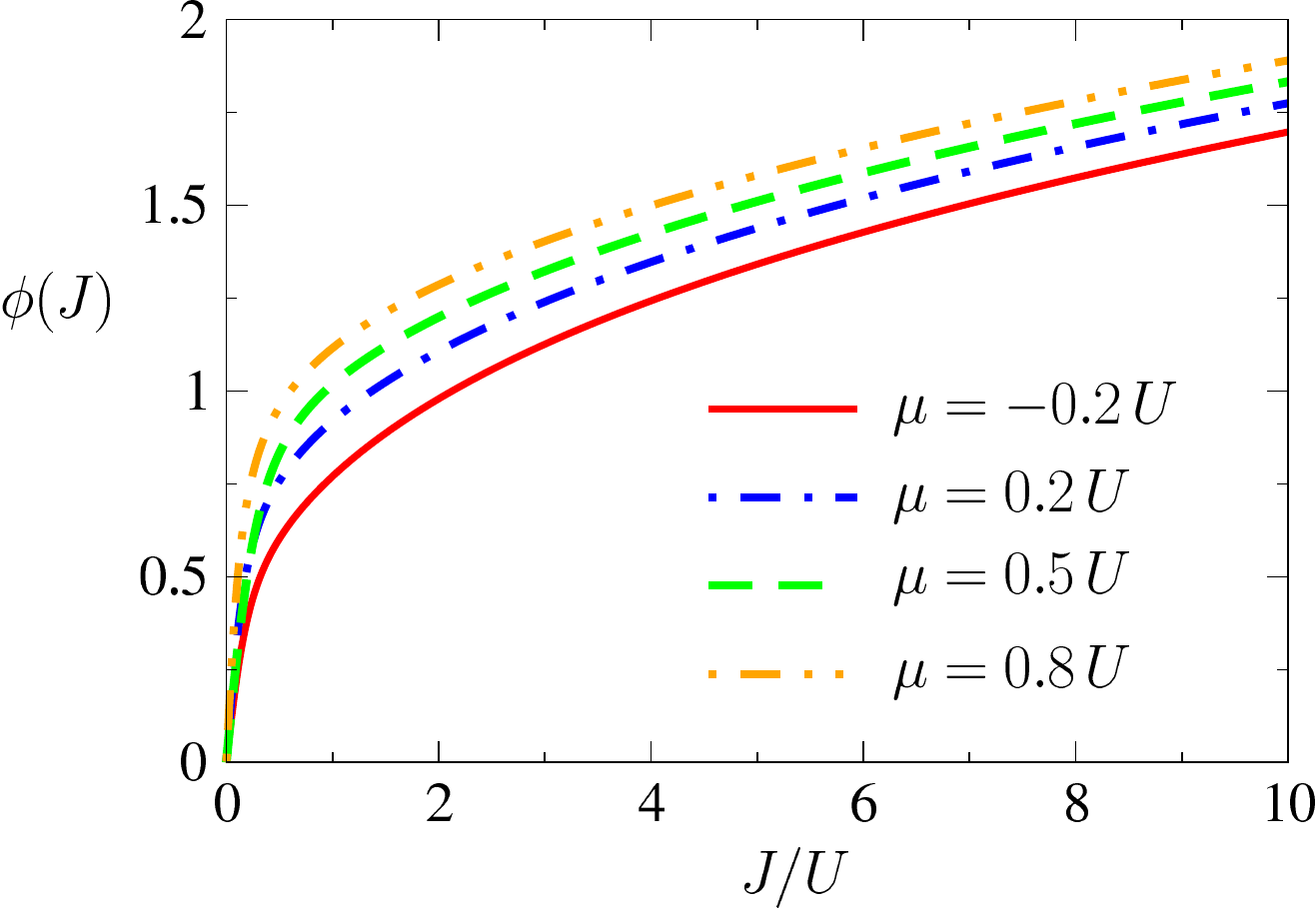}}
\vspace{0.25cm}
\centerline{\includegraphics[width=6.5cm,clip]{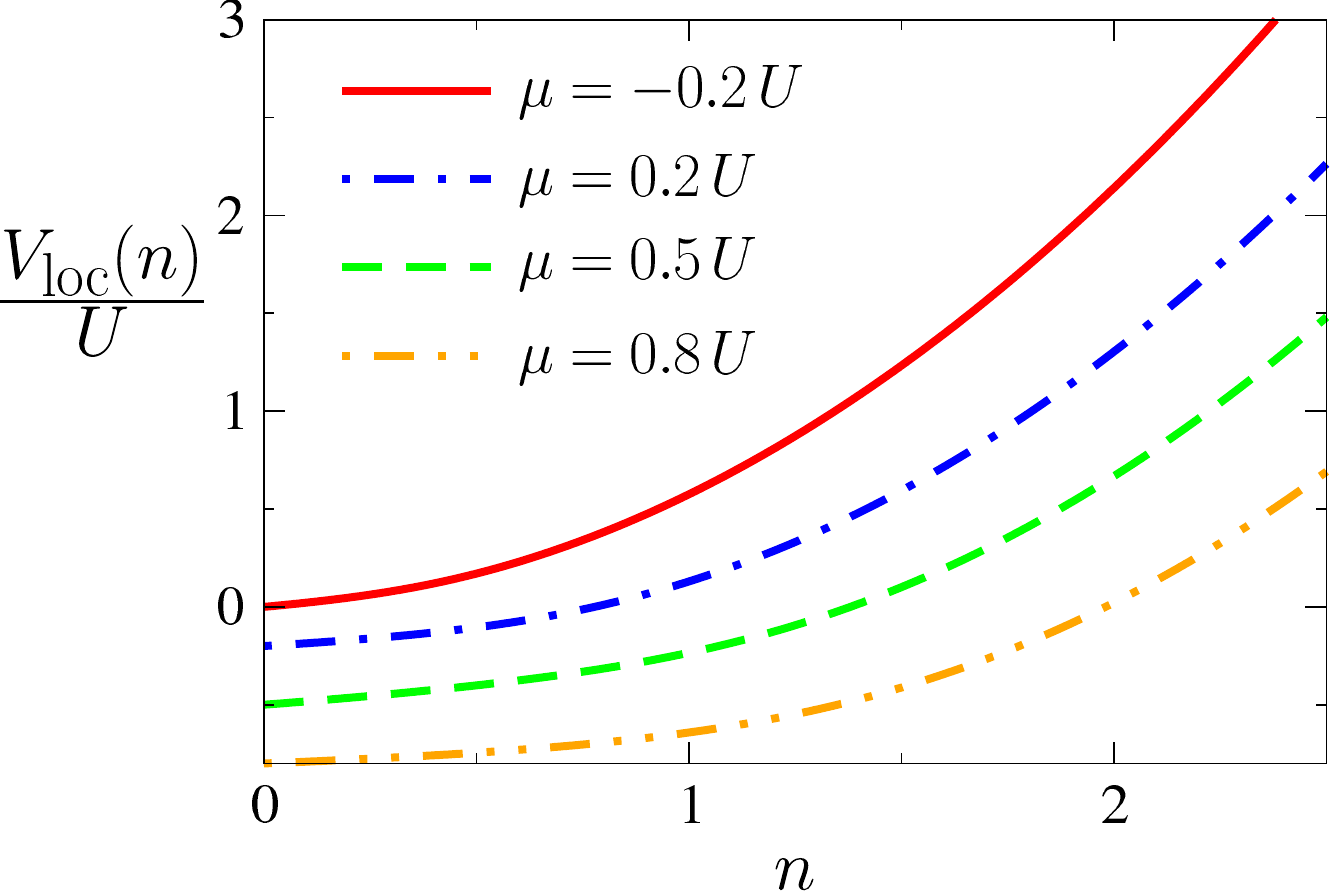}}
\caption{(Color online) (Top) Superfluid order parameter $\phi$ vs external source $J$ (here assumed real) in the local limit for various values of the chemical potential $\mu$. (Bottom) Effective potential $V_{\rm loc}(n)$.}
\label{fig_Vloc}
\end{figure}

To obtain $\Vloc$ and $\Gamloc^{(2)}$ in a time-independent field, it is sufficient to consider a single site with time-independent complex external source $J$. The corresponding Hamiltonian reads
\begin{equation}
\hat H = -\mu \hat n + \frac{U}{2}\hat n(\hat n -1) - J^* \hat b - J \hat b^\dagger ,
\label{Hloc}
\end{equation}
where $\hat b^\dagger$ ($\hat b$) is a creation (annihilation) operator and $\hat n=\hat b^\dagger \hat b$. In the basis $\lbrace \ket{m}\rbrace$ [$\hat n\ket{m}=m\ket{m}$ with $m$ integer], the Hamiltonian is represented by a tridiagonal matrix,
\begin{align}
\bra{m} \hat H \ket{m'} ={}& \delta_{m,m'} \left[-\mu m + \frac{U}{2}m(m-1) \right] \nonumber \\ & - \delta_{m+1,m'} J^* \sqrt{m+1} - \delta_{m-1,m'} J \sqrt{m} ,
\end{align}
which can be numerically diagonalized in the truncated Hilbert space $m\leq m_{\rm max}$. The low-energy eigenstates are independent of $m_{\rm max}$ if the latter is large enough. If we denote by $\lbrace \ket{\alpha},E_\alpha\rbrace$ the source-dependent eigenstates and eigenvalues -- with $\lbrace\ket{0},E_0\rbrace$ the ground state -- we obtain the  superfluid order parameter
\begin{equation}
\phi = - \frac{\partial E_0}{\partial J^*} , \quad  \phi^* = - \frac{\partial E_0}{\partial J} ,
\label{phiJ}
\end{equation}
and the effective potential 
\begin{equation}
\Vloc(n) = E_0 + J^*\phi + J \phi 
\end{equation} 
($n=|\phi|^2$) in the zero-temperature limit $\beta\to\infty$. Figure~\ref{fig_Vloc} shows the superfluid order parameter $\phi$ as a function of the external source $J$, and the local effective potential $V_{\rm loc}(n)$ obtained by numerically inverting~(\ref{phiJ}). The special case where the ground state in the local limit is degenerate for $J=0$ ($\mu/U$ integer) is discussed in Appendix~\ref{app_local}. 

\begin{figure}
\centerline{\includegraphics[width=4cm,clip]{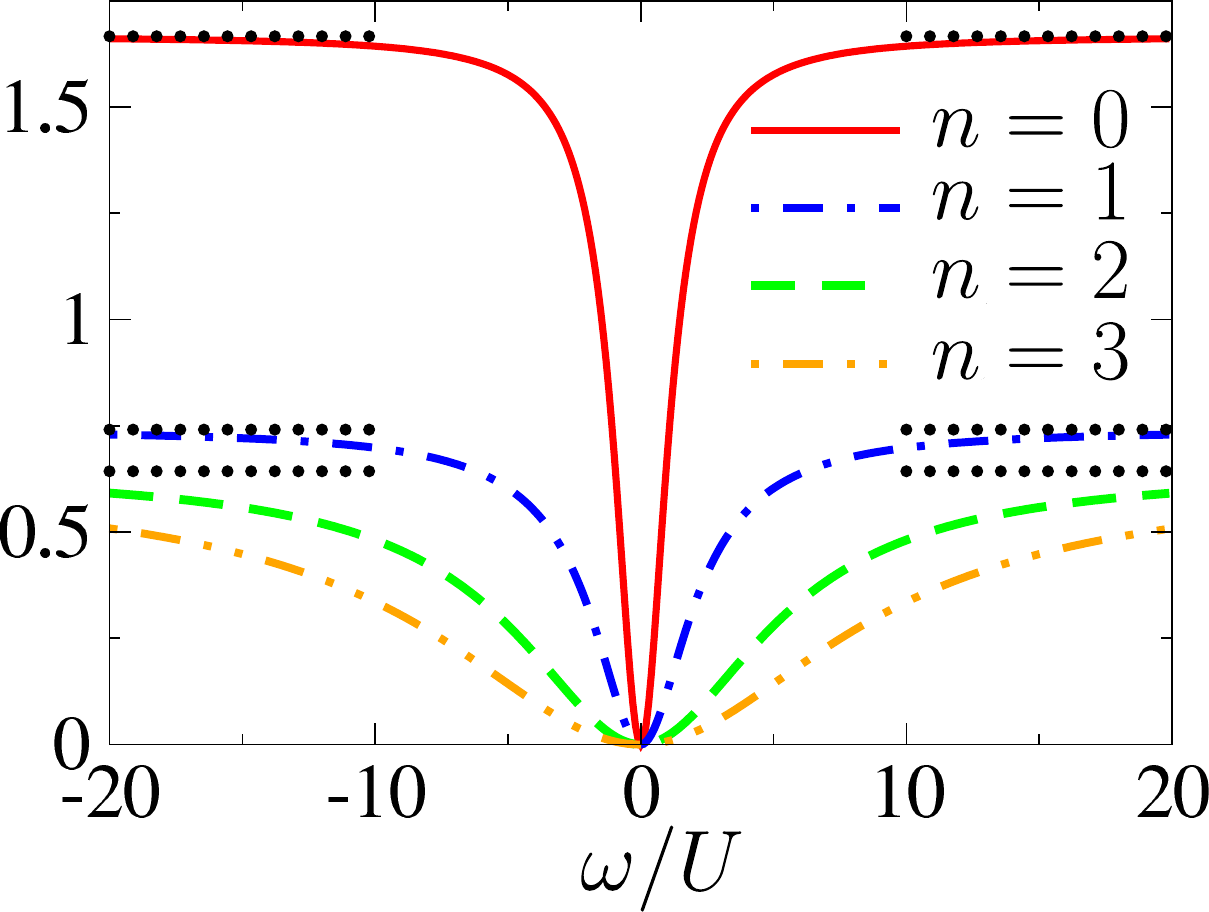}
\includegraphics[width=4cm,clip]{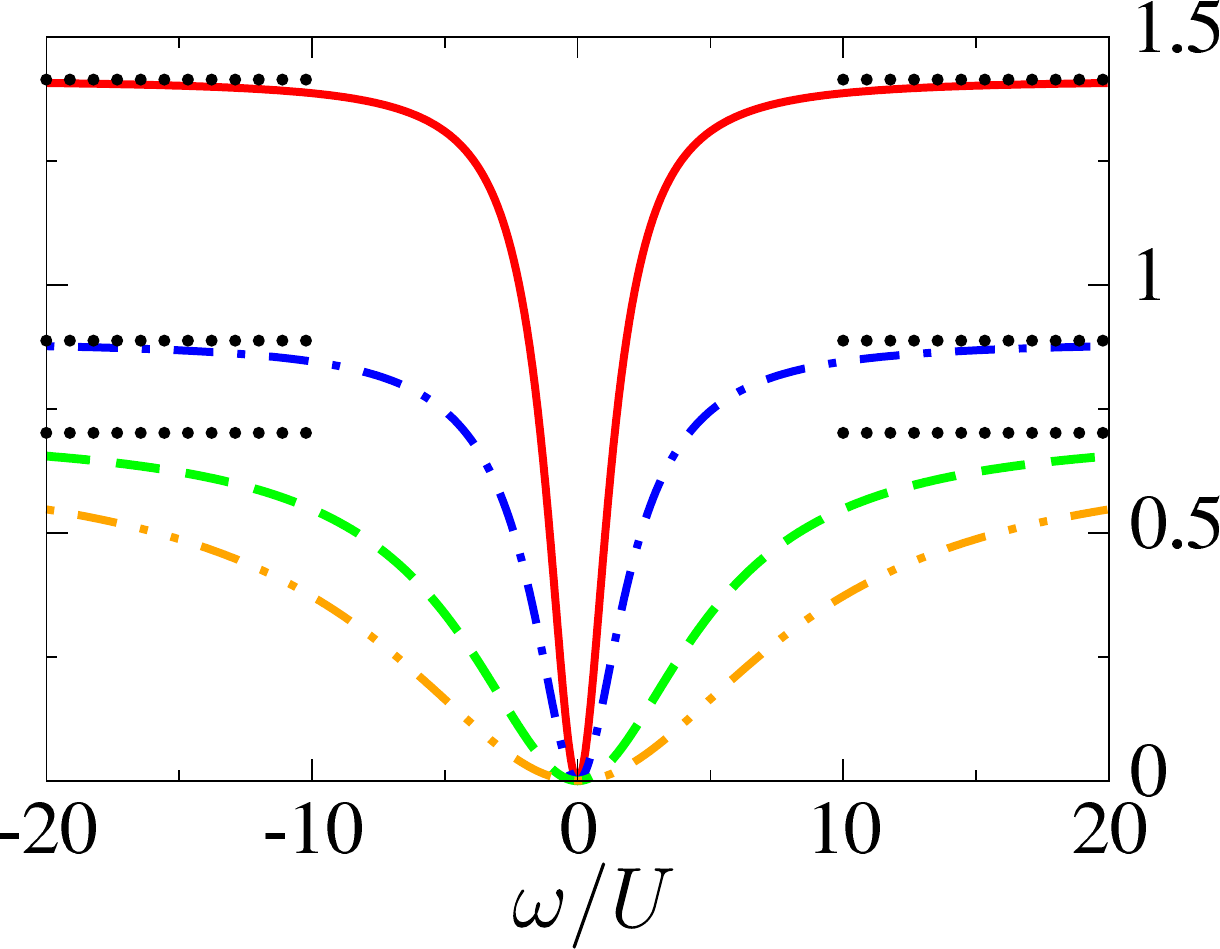}}
\caption{(Color online) $[\Gamma_{{\rm loc},A}(i\w;n)-V'_{\rm loc}(n)]/U$ vs $\w/U$ for various values of $n$. $\mu=0.2U$ (left) and $\mu=(\sqrt{2}-1)U$ (right). The dotted lines show the large frequency limit~(\ref{Dinf}).}
\label{fig_deltaA}
\centerline{\includegraphics[width=4.cm,clip]{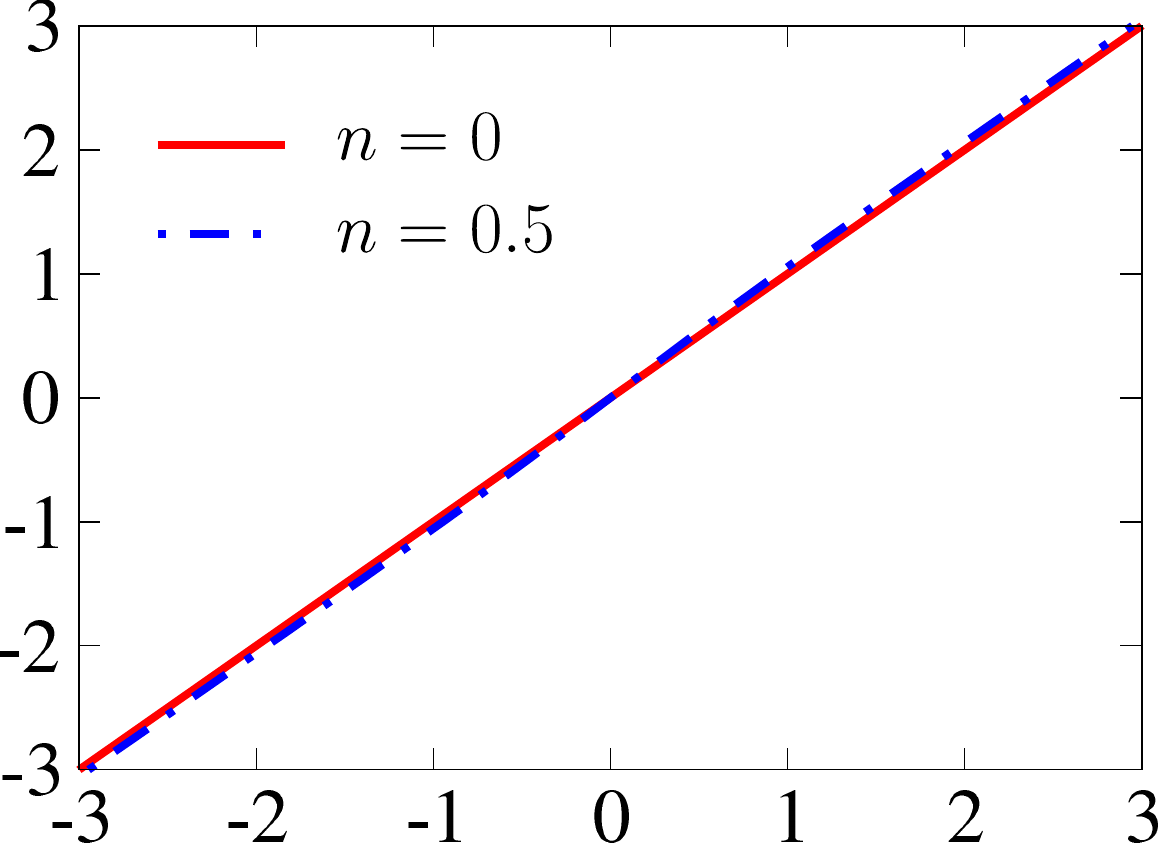}
\includegraphics[width=4.cm,clip]{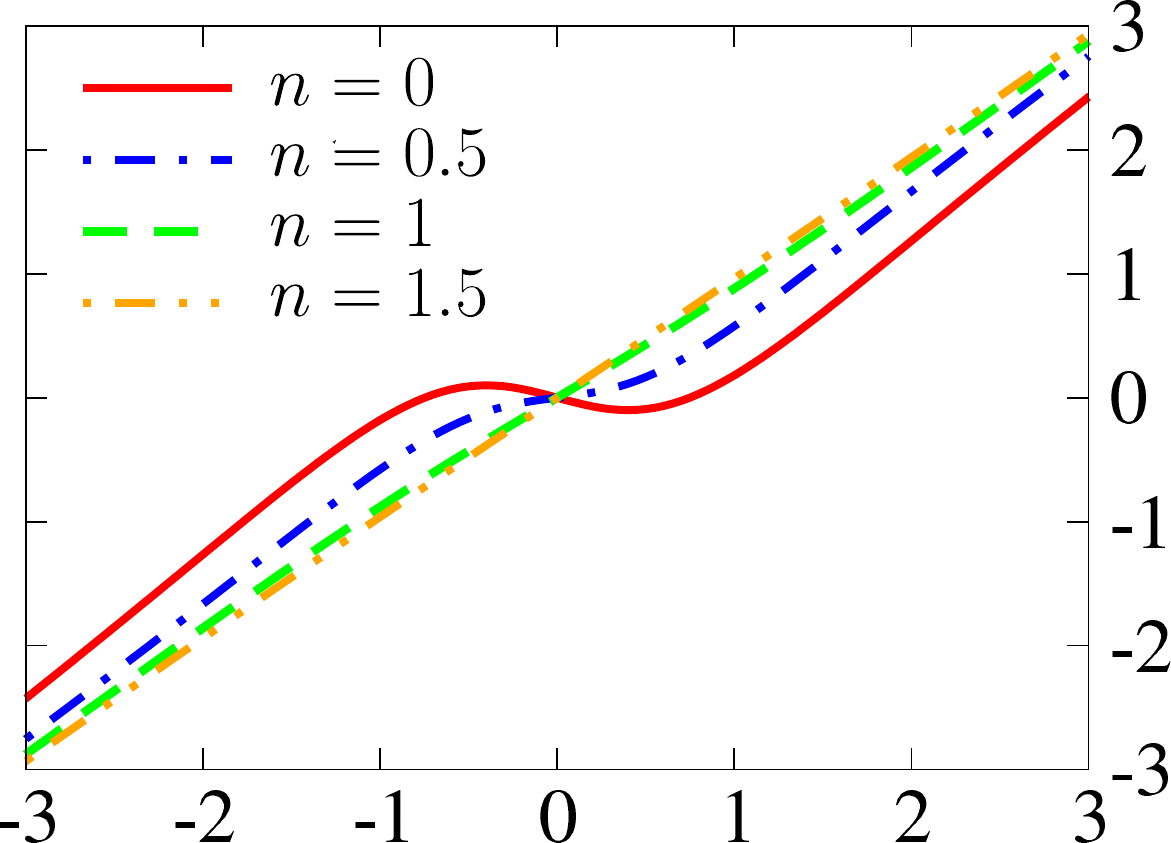}}
\centerline{\includegraphics[width=4.cm,clip]{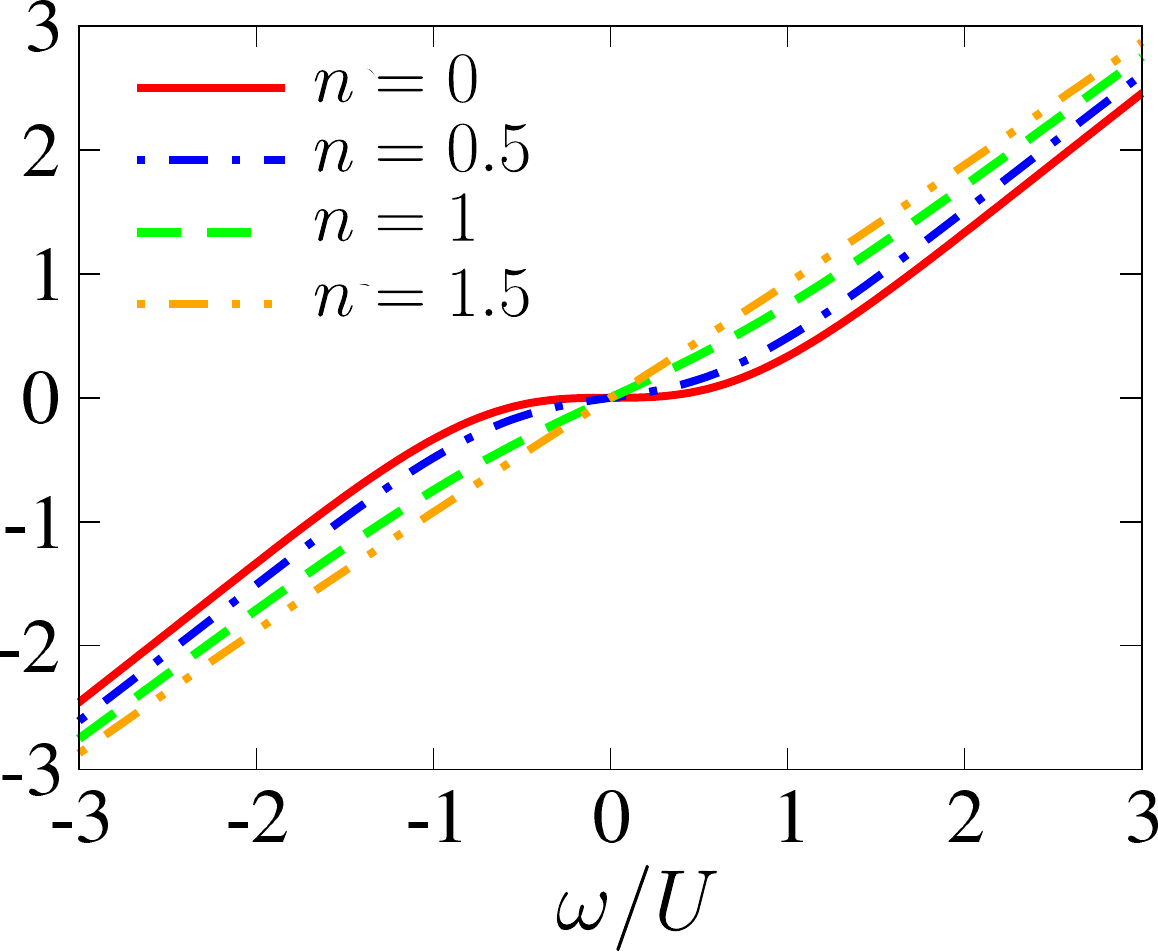}
\includegraphics[width=4.cm,clip]{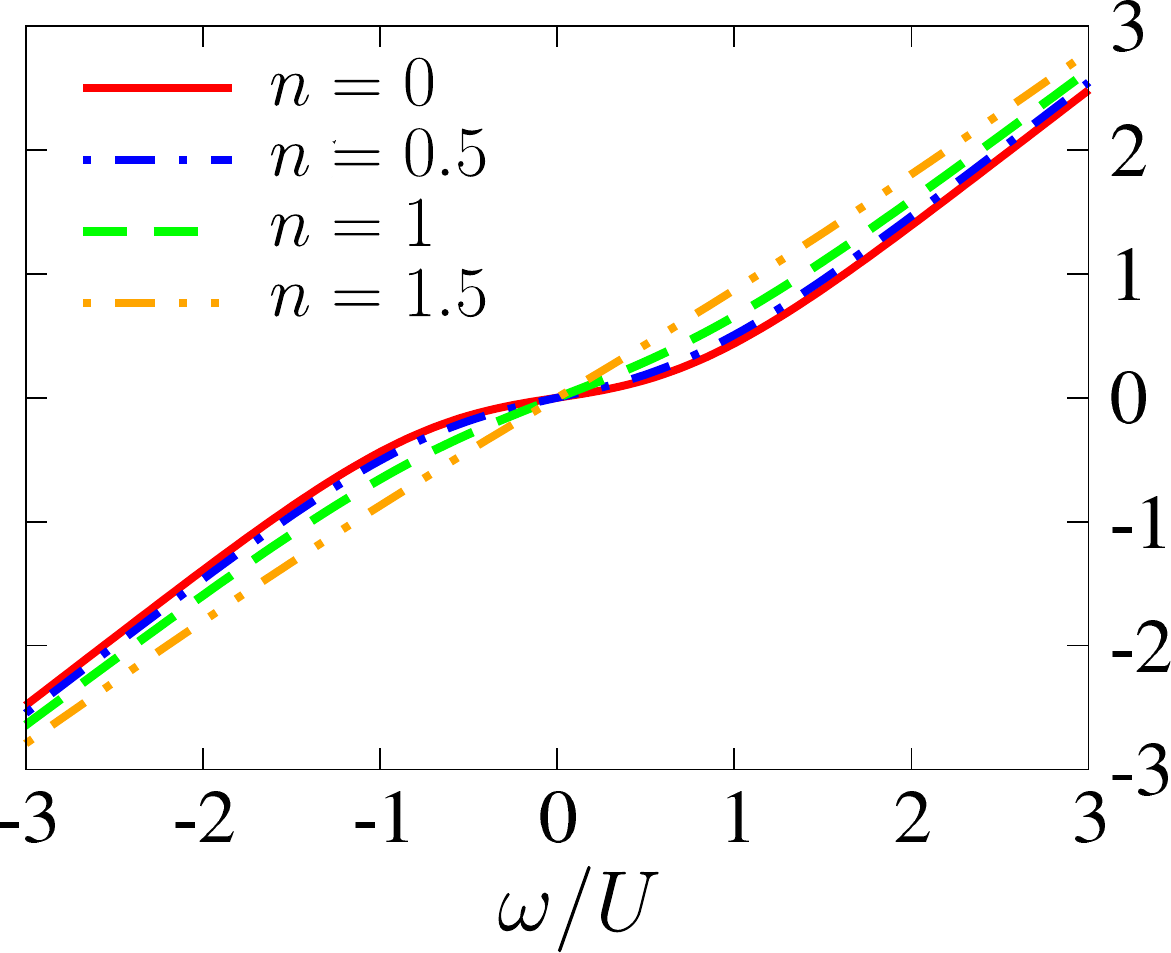}}
\caption{(Color online) $\Gamma_{{\rm loc},C}(i\w;n)/U$ vs $\w/U$ for various values of $n$. $\mu=-0.2U$, $0.2U$, $(\sqrt{2}-1)U$ and $0.6U$ (from top left to bottom right). In the large frequency limit, $\Gamma_{{\rm loc},C}(i\wn;n)=\w$ [Eqs.~(\ref{Dinf})].}
\label{fig_deltaC}
\end{figure}

To determine the two-point vertex $\Gamma^{(2)}_{\rm loc}$, we start from the (source-dependent) normal and anomalous local Green functions
\begin{equation}
\begin{split}
\Gn(\tau) &= - \mean{T_\tau \hat b(\tau) \hat b^\dagger(0)} + |\mean{\hat b}|^2 , \\ 
\Gan(\tau) &= - \mean{T_\tau \hat b(\tau) \hat b(0)} + \mean{\hat b}^2 ,
\end{split}
\end{equation}
where $\hat b^{(\dagger)}(\tau) = e^{\tau\hat H} \hat b^{(\dagger)} e^{-\tau\hat H}$ and $T_\tau$ is a time-ordering operator. The Fourier transforms $\Gn(i\w)$ and $\Gan(i\w)$ are easily expressed in terms of the eigenstates $\ket{\alpha}$ of the Hamiltonian,
\begin{equation}
\begin{split}
\Gn(i\w) &= - \sum_{\alpha\neq 0} \left[ \frac{|\bra{\alpha} \hat b \ket{0}|^2}{i\w+E_\alpha-E_0} - \frac{|\bra{0} \hat b \ket{\alpha}|^2}{i\w+E_0-E_\alpha} \right] , \\ 
\Gan(i\w) &= - \sum_{\alpha\neq 0} \bra{\alpha} \hat b \ket{0}\bra{0} \hat b \ket{\alpha}  \frac{2(E_\alpha-E_0)}{\w^2+(E_\alpha-E_0)^2}  .
\end{split}
\end{equation}
From the relation $\Gamma^{(2)}=-G^{-1}$, we obtain
\begin{equation}
\begin{split}
\Gamma_{{\rm loc},A}(i\w;n) &= -\frac{1}{2D} [\Gn(i\w)+\Gn(-i\w)+2\Gan(i\w)] , \\ 
\Gamma_{{\rm loc},B}(i\w;n) &= \frac{\Gan(i\w)}{nD} , \\ 
\Gamma_{{\rm loc},C}(i\w;n) &= \frac{i}{2D} [\Gn(i\w)-\Gn(-i\w)] , 
\end{split}
\end{equation}
where $D=\Gn(i\w)\Gn(-i\w)-\Gan(i\w)^2$. $\Gamma^{(2)}$ is expressed in terms of the condensate density $n$ (rather than the external source $J$) by inverting (\ref{phiJ}). 

The large frequency limit of the two-point vertex is given by\cite{note1}
\begin{equation}
\begin{split}
\lim_{|\w|\to\infty} \Gamma_{{\rm loc},A}(i\w;n) &= - \mu - U\mean{\psi_\r^2} + 2U \mean{\psi_\r^*\psi_\r}  , \\
\lim_{|\w|\to\infty} \Gamma_{{\rm loc},B}(i\w;n) &=  U \frac{\mean{\psi_\r^2}}{n} , \\
\lim_{|\w|\to\infty} \Gamma_{{\rm loc},C}(i\w;n) &= \w , 
\end{split}
\label{Dinf} 
\end{equation}
where $\mean{\psi_\r}=\phi$ and $\mean{\psi_\r^2}$ are assumed real (which corresponds to a real external source $J$), 
\begin{align}
\mean{\psi_\r^*\psi_\r} &= - G_{\rm n}(\tau=0^-) + |\phi|^2 \nonumber \\ 
&= - \intw G_{\rm n}(i\w) e^{i\w 0^+} + |\phi|^2 ,
\end{align}
and
\begin{align} 
\mean{\psi_\r^2} &= - G_{\rm an}(\tau=0^-) + \phi^2 \nonumber \\ 
&= - \intw G_{\rm an}(i\w) e^{i\w 0^+} + \phi^2 .
\end{align} 
The asymptotic forms~(\ref{Dinf}) are reached for $|\w|\gg U$. $\Gamma_{{\rm loc},A}(i\w;n)$ and  $\Gamma_{{\rm loc},C}(i\w;n)$ are shown in Figs.~\ref{fig_deltaA} and \ref{fig_deltaC}.

\subsubsection{Large-field limit} 
\label{subsubsec_largefield}

To obtain the large-field limit of the effective potential $\Vloc(n)$, we must compute the local partition function $Z_{\rm loc}[J^*,J]$ for $|J|\to\infty$. Using a loop expansion about the saddle-point approximation (see Appendix~\ref{app_largefield}), we find
\begin{equation}
\Vloc(n) = -\bar\mu n + \frac{U}{2} n^2 + \calO(n^0) ,
\label{Vlocinf}
\end{equation}
where $\bar\mu=\mu+U(1-\sqrt{3}/2)$.

\subsubsection{Strong-coupling RPA} 
\label{subsubsec_rpa}

The initial effective action $\Gamma_\Lambda$ [Eq.~(\ref{GamLambda})] treats the local fluctuations exactly but includes the intersite hopping term at the mean-field level, thus reproducing the strong-coupling RPA.\cite{Sheshadri93,Oosten01,Sengupta05,Ohashi06,Menotti08}  The effective potential reads
\begin{equation}
V_\Lambda(n) = \Vloc(n) - 2dtn , 
\end{equation}
while the two-point vertex takes the RPA-like form 
\begin{equation}
\Gamma^{(2)}_{\Lambda,ij}(q;n) = \Gamma_{{\rm loc},ij}^{(2)}(i\w;n) + \delta_{i,j}t_\q . 
\end{equation}
Expanding $V_\Lambda(n)$ about $n=0$, we find 
\begin{equation}
V_\Lambda(n) = \Vloc(0) +  \left[\Gamma_{{\rm loc},11}^{(2)}(i\w=0;n=0)-2dt \right] n + \calO(n^2) , 
\label{Vexpand} 
\end{equation}
where 
\begin{equation}
\Gamma_{{\rm loc},ii}^{(2)}(i\w=0;n=0) = - \Gn(i\w=0;n=0)^{-1} 
\end{equation}
is determined by the local Green function 
\begin{equation}
\Gn(i\w;n=0) = \frac{\nbarloc+1}{i\w+\mu-U\nbarloc} - \frac{\nbarloc}{i\w+\mu-U(\nbarloc-1)} 
\end{equation}
for vanishing source ($J^*=J=0$). Here $\nbarloc$ is the number of bosons per site in the local limit: $\nbarloc-1\leq \mu/U\leq \nbarloc$ if $\mu\geq 0$ and $\nbarloc=0$ if $\mu\leq 0$. The ground state is a Mott insulator as long as $V_\Lambda'(0)\geq 0$. Thus the transition to the superfluid state is determined by the criterion $V'_\Lambda(0)=0$, \ie
\begin{equation}
\Gn(i\w=0;n=0)^{-1} + 2dt = 0, 
\label{mfcrit}
\end{equation}
which reproduces the mean-field (or strong-coupling RPA) phase diagram.\cite{Fisher89} Equation~(\ref{mfcrit}) can also be obtained from the condition $\det\,\Gamma^{(2)}_\Lambda(\q=i\w=0;n=0)$, which signals the appearance of a pole at zero momentum and frequency in the one-particle propagator $G_\Lambda=-\Gamma^{(2)-1}_\Lambda$. 

In the strong-coupling RPA, the condensate density $n_0$ in the superfluid phase is determined by
\begin{equation}
V_\Lambda'(n_0) = \Vloc'(n_0) - 2dt = 0. 
\end{equation}
The hopping amplitude $t$ acts as a source term for the local potential $\Vloc(n)$. For $t/U\gg 1$, \ie deep in the superfluid phase, the source term is large and we are effectively in the large field limit discussed in Sec.~\ref{subsubsec_largefield}. From Eqs.~\eqref{GamLambda}, and the fact that $\bar\mu+2dt\simeq\mu+2dt$ when $t/U\gg 1$, we then obtain $V_\Lambda(n)\simeq \frac{1}{\beta N} S[\phi^*,\phi]$ (with $|\phi|^2=n$), which is nothing but the result of the Bogoliubov approximation.\cite{Dupuis09b,note13} The strong-coupling RPA reduces to the Bogoliubov theory in the limit $t/U\gg 1$.\cite{Menotti08} Table~\ref{table_init} compares the initial conditions given by the Bogoliubov theory and the strong-coupling RPA. 

It should be noted that the true Legendre transform is $\Gamma_{\rm loc}$ for $k=\Lambda$ since the action $S_\Lambda=S+\Delta S_\Lambda=S_{\rm loc}$ is local. The lattice NPRG is an expansion about the local limit. The scale-dependent effective action $\Gamma_k$ is however the right quantity to consider to analyze the physical properties of the system with action $S$ (without the regulator term). In $\Gamma_k$, the regulator term $\Delta S_k$ is compensated, in a mean-field manner, by subtracting $\Delta S_k[\phi^*,\phi]$ from the true Legendre transform. It follows that the physical quantities at scale $k$, such as the condensate density $n_{0,k}$, are obtained from $\Gamma_k$ rather than from the true Legendre transform. 

\begin{table}
\renewcommand{\arraystretch}{1.5}
\begin{center}
\begin{tabular}{ccc}
\hline \hline
& Bogoliubov & RPA \\
\hline
$V_\Lambda(n)$ & $-(\mu+2dt)n+\frac{U}{2}n^2$ & $\Vloc(n)-2dtn$  \\
$\Gamma_{A,\Lambda}(q;n)$ & $-\mu+Un+t_\q$ & $\Gamma_{{\rm loc},A}(i\w;n)+t_\q$ \\
$\Gamma_{B,\Lambda}(q;n)$ & $U$ & $\Gamma_{{\rm loc},B}(i\w;n)$ \\
$\Gamma_{C,\Lambda}(q;n)$ & $\w$ & $\Gamma_{{\rm loc},C}(i\w;n)$ \\
\hline \hline
\end{tabular}
\end{center}
\caption{Initial conditions given by the Bogoliubov approximation and the strong-coupling RPA. In the weak-coupling limit the two approximations become equivalent.}
\label{table_init}
\end{table}

\subsection{Gauge invariance and Ward identities}
\label{subsec_gauge_inv}

The invariance of the action $S+\Delta S_k$ in the local (time-dependent) gauge transformation $\psi_\r\to e^{i\alpha}\psi_\r$, $\psi^*_\r\to e^{-i\alpha}\psi^*_\r$ and $\mu\to \mu+i\dtau\alpha$ imposes important constraints on the effective action $\Gamma_k$. In the superfluid phase, this implies that the two-point vertex satisfies the Ward identities\cite{Dupuis09b}
\begin{equation}
\begin{split} 
\frac{\partial}{\partial\w} \Gamma_{C,k}(q;n_{0,k}) \biggl|_{q=0} &=  - \frac{\partial^2 V_k}{\partial n \partial\mu} \biggl|_{n_{0,k}} , \\ 
\frac{\partial^2}{\partial\w^2} \Gamma_{A,k}(q;n_{0,k}) \biggl|_{q=0} &=  - \frac{1}{2n_{0,k}} \frac{\partial^2 V_k}{\partial\mu^2} \biggl|_{n_{0,k}} ,
\end{split}
\label{wardid}
\end{equation}
where the effective potential $V_k(n,\mu)$ is considered as a function of both $n$ and $\mu$, the condensate density $n_{0,k}\equiv n_{0,k}(\mu)$ being then defined by
\begin{equation}
\frac{\partial V_k(n,\mu)}{\partial n}\biggl|_{n_{0,k}} = 0. 
\label{n0def}
\end{equation}
Since Eq.~\eqref{n0def} is valid for any $\mu$, we deduce the relation
\begin{equation}
0 = \frac{d}{d\mu} \frac{\partial V_k}{\partial n}\biggl|_{n_{0,k}} = \frac{\partial^2 V_k}{\partial n \partial\mu} \biggl|_{n_{0,k}} +
\frac{\partial^2 V_k}{\partial n^2} \biggl|_{n_{0,k}} \frac{dn_{0,k}}{d\mu} ,
\label{wardid1}
\end{equation}
which will be used below together with~\eqref{wardid}. 

In the Mott insulator ($n_{0,k}=0$), the Ward identities~(\ref{wardid}) become
\begin{equation}
\begin{gathered}
\frac{\partial}{\partial\w} \Gamma_{C,k}(q;n=0) \biggl|_{q=0} = - \frac{\partial^2 V_k}{\partial n \partial\mu} \biggl|_{n=0} , \\
\frac{\partial^2 V_{0,k}}{\partial \mu^2} = \frac{d^2 V_{0,k}}{d\mu^2} = 0 ,
\end{gathered}
\label{wardid2}
\end{equation}
which implies that the compressibility\cite{note4}
\begin{equation}
\kappa_k = \frac{d\bar n_k}{d\mu} = - \frac{d^2 V_{0,k}}{d\mu^2}
\label{kappa}
\end{equation}
vanishes. $\bar n_k=-dV_{0,k}/d\mu$ denotes the boson density (mean boson number per site). 

\subsection{Derivative expansion and infrared behavior}
\label{subsec_de} 

The low-energy behavior of the system is best understood from a derivative expansion of the two-point vertex. Since the cutoff function~(\ref{cutoff}) acts as an infrared regulator, $\Gamma^{(2)}_k(q;n)$ is a regular function of $q$ for $q\to 0$. In the infrared limit, we can therefore use the derivative expansion
\begin{equation}
\begin{split}
\Gamma_{A,k}(q;n) &= Z_{A,k}(n) t\q^2 + V_{A,k}(n) \w^2 + V_k'(n) , \\ 
\Gamma_{B,k}(q;n) &= V''_k(n), \\ 
\Gamma_{C,k}(q;n) &= Z_{C,k}(n)\w , 
\end{split}
\label{gamde}
\end{equation}
($V_k'(n)=\partial V_k/\partial n$, etc.) in agreement with the symmetry properties~(\ref{sym}). For the following discussion, it is convenient to introduce 
\begin{equation}
\delta_k = \frac{\partial V_k}{\partial n}\biggl|_{n_{0,k}} , \quad
\lamb_k = \frac{\partial^2 V_k}{\partial n^2}\biggl|_{n_{0,k}} ,
\label{deltalamb} 
\end{equation}
with $\delta_k$ vanishing in the superfluid phase. If the spectrum is gapped, Eqs.~(\ref{gamde}) will always be valid for energy scales below the gap. Otherwise their validity requires $|\q|\lesssim k$ and $|\w|\lesssim \w^-_k$ where $\w^-_k$ is the lowest excitation energy for $|\q|\sim k$ (see Sec.~\ref{subsec_rgeq}).

\subsubsection{Superfluid phase}

Using~(\ref{n0def}) and (\ref{wardid1}), we can rewrite the Ward identities~(\ref{wardid}) as 
\begin{equation}
\begin{split} 
Z_{C,k}(n_{0,k}) &= \lamb_k \frac{dn_{0,k}}{d\mu} , \\ 
V_{A,k}(n_{0,k}) &= - \frac{1}{2n_{0,k}} \frac{\partial^2 V_k}{\partial\mu^2} \biggl|_{n_{0,k}} ,
\end{split}
\end{equation}
while the compressibility~\eqref{kappa} is expressed as 
\begin{align}
\kappa_k &= - \frac{\partial^2 V_{0,k}}{\partial \mu^2}\biggl|_{n_{0,k}}  - \frac{\partial^2 V_{0,k}}{\partial n \partial\mu}\biggl|_{n_{0,k}}  \frac{dn_{0,k}}{d\mu} \nonumber \\ 
&= 2 n_{0,k} V_{A,k} + \frac{Z_{C,k}^2}{\lamb_k} .
\label{dndmu} 
\end{align} 

The superfluid stiffness $\rho_{s,k}$, defined as the rigidity with respect to a twist of the phase of the order parameter, can be obtained from the transverse part of the two-point vertex,\cite{Dupuis09b}
\begin{equation}
\Gamma_{A,k}(\q,\w=0;n_{0,k}) = \frac{\rho_{s,k}}{2n_{0,k}} \q^2 \quad (\q\to 0), 
\end{equation}
which leads to 
\begin{equation}
\rho_{s,k} = 2t Z_{A,k}(n_{0,k})n_{0,k} .
\label{stiffness}
\end{equation}

The excitation spectrum is given by the zeros of the determinant of the $2\times 2$ matrix $\Gamma^{(2)}_k(q;n_{0,k})$ (after analytical continuation $i\w\to \w+i0^+$),
\begin{align}
\det\,\Gamma^{(2)}_k(q) &= \Gamma_{A,k}(q) \left[\Gamma_{A,k}(q) + 2n_{0,k} \Gamma_{B,k}(q) \right] +  \Gamma_{C,k}(q)^2 \nonumber \\ 
&\simeq 2 \lamb_k n_{0,k} (Z_{A,k}t\q^2 + V_{A,k} \w^2) + (Z_{C,k}\w)^2 
\end{align} 
(all quantities are evaluated for $n=n_{0,k}$) for $|\q|,|\w|\to 0$. This equation yields a gapless (Goldstone) mode $\w=c_k|\q|$ with a velocity 
\begin{align}
c_k &= \left(\frac{Z_{A,k}(n_{0,k})t}{V_{A,k}(n_{0,k})+Z_{C,k}(n_{0,k})^2/(2\lambda_k n_{0,k})}\right)^{1/2} \nonumber \\ 
&= \left( \frac{\rho_{s,k}}{\kappa_k}\right)^{1/2} 
\label{velocity}
\end{align}
which can be expressed in terms of the compressibility and superfluid stiffness.\cite{Fisher89} The existence of a gapless mode is a consequence of the Hugenholtz-Pines theorem\cite{Hugenholtz59} which, in our formalism, reads\cite{Dupuis09b}
\begin{equation}
\Gamma_{A,k}(q=0;n_{0,k}) = V'_k(n_{0,k}) = 0. 
\end{equation}
Equations~(\ref{stiffness},\ref{velocity}) are identical to those obtained in continuum models if we identify $1/2t$ with the (effective) mass $m$ of the bosons in the lattice potential. From $\det\,\Gamma^{(2)}_k(\q,\w+i0^+)=0$, we also obtain a gapped mode, with a gap which is however larger than $\w_k^-\sim c_kk$, and therefore beyond the domain of validity of the derivative expansion ($|\q|,|\w|/c_k\ll k$). The existence of two modes in the superfluid phase follows from $\det\,\Gamma^{(2)}_k(q)$ being of order $\w^4$. Pushing the derivative expansion to higher order in $\w^2$ would yield additional modes. These modes are not in the domain of validity of the derivative expansion and do not show up in the spectral function.\cite{Dupuis09b,Sinner10}

\subsubsection{Mott insulator} 
\label{subsubsec_mi}

In the Mott insulator ($n_{0,k}=0$), the Ward identities~(\ref{wardid2}) yield
\begin{equation}
Z_{C,k} = - \frac{d\delta_k}{d\mu} .
\label{wardid3}
\end{equation}
Since $G_{{\rm an},k}(q)=0$, the excitation spectrum is obtained from $G_{{\rm n},k}^{-1}(q)=-\Gamma_{A,k}(q)+i\Gamma_{C,k}(q)=0$ after analytical continuation $i\w\to \w+i0^+$. This gives two gapped modes 
\begin{align}
\w_\pm(\q) &= - \frac{Z_{C,k}}{2V_{A,k}} \pm \frac{1}{2V_{A,k}} \left[ Z_{C,k}^2+4V_{A,k}(Z_{A,k}t\q^2 + \delta_k) \right]^{1/2} \nonumber \\ 
&= \Delta_{k\pm} \pm \frac{Z_{A,k}t\q^2}{( Z_{C,k}^2+4V_{A,k}\delta_k)^{1/2}} + \calO(|\q|^4), 
\label{wq1}
\end{align}
where 
\begin{equation}
\Delta_{k\pm} = - \frac{Z_{C,k}}{2V_{A,k}} \pm \frac{1}{2V_{A,k}} {\left( Z_{C,k}^2+4V_{A,k}\delta_k\right)^{1/2}} .
\label{gap}
\end{equation}

When $Z_{C,k}\neq 0$, both modes have a quadratic dispersion for small $\q$. The modes $\w_+(\q)$ and $\w_-(\q)$ have positive and negative effective mass, respectively. Thus $\w_+(\q)$ ($\w_-(\q)$) corresponds to a particle-like (hole-like) excitation. At the transition to the superfluid phase ($\delta_{k=0}\to 0$), $\Delta_{k=0,+}$ ($\Delta_{k=0,-}$) vanishes if $Z_{C,k=0}>0$ ($Z_{C,k=0}<0$), but the particle-hole excitation gap $\Delta_{k=0}=\Delta_{k=0,+}-\Delta_{k=0,-}$ remains finite. The critical mode energy $\w\sim \q^2$ being quadratic in $\q$, the dynamical critical exponent takes the value $z=2$.

When $Z_{C,k}=0$, the excitation spectrum takes the particle-hole symmetric form 
\begin{align}
\w_\pm(\q) &= \pm \left(\frac{Z_{A,k}t\q^2+\delta_k}{V_{A,k}}\right)^{1/2} \nonumber \\
&= \pm \left(c_k^2 \q^2 + \Delta^2_{k}\right)^{1/2} ,
\label{wq2}
\end{align}
where 
\begin{equation}
\begin{split}
\Delta_k &= \left(\frac{\delta_k}{V_{A,k}}\right)^{1/2} , \\ 
c_k &= \left(\frac{Z_{A,k}t}{V_{A,k}}\right)^{1/2} .
\end{split}
\end{equation}
At the transition ($\delta_{k=0}=0$), the particle-hole excitation gap $2\Delta_{k=0}$ vanishes and the dispersion $\w_\pm(\q)=\pm c_{k=0}|\q|$ becomes linear, which implies that the critical dynamical exponent takes the value $z=1$.\cite{note15} We can also understand this result as a direct consequence of the (relativistic) Lorentz invariance of the vertex $\Gamma^{(2)}_{k=0,ij}$ [Eq.~(\ref{gamde})] when $Z_{C,k=0}$ vanishes. The quantum critical point $\delta_{k=0}=0$ then coincides with the critical point of the $(d+1)$-dimensional XY model.  

We conclude that the universality class of the superfluid--Mott-insulator transition depends on whether $Z_{C,k=0}(n=0)$ vanishes or not.\cite{Fisher89,Sachdev_book} The same conclusion can be reached from the superfluid phase by considering the spectrum in the limit $n_{0,k=0}\to 0$.

\subsection{Flow equations} 
\label{subsec_rgeq} 

Since we do not have an explicit (approximate) form of the effective action $\Gamma_k[\phi^*,\phi]$, we cannot directly use Eq.~(\ref{wetteq}) to obtain the RG equations satisfied by $V_k(n)$ and $\Gamma_{\alpha,k}(n)$ ($\alpha=A,B,C$). We can nevertheless obtain  RG equations for the effective potential $V_k(n)$ and the two-point vertex $\Gamma^{(2)}_k(q;n)$ in a constant field within a simplified Blaizot--M\'endez-Galain--Wschebor (BMW) scheme.\cite{Blaizot06,Benitez09} 

Equation~(\ref{wetteq}) leads to the RG equations 
\begin{equation}
\dl V_k(n) = - \half \int_q \dl R_k(\q) [G_{k,\rm ll}(q;n)+G_{k,\rm tt}(q;n)] 
\label{flowpot} 
\end{equation}
and 
\begin{multline}
\dl \Gamma^{(2)}_{k,ij}(p;\phi) = \\ 
-\half \sum_{q,i_1,i_2} \tilde\dl G_{k,i_1i_2}(q;\phi) \Gamma^{(4)}_{k,iji_2i_1}(p,-p,q,-q;\phi)  \\ 
- \half \sum_{q,i_1\cdots i_4} \Bigl\lbrace \Gamma^{(3)}_{k,ii_2i_3}(p,q,-p-q;\phi) \Gamma^{(3)}_{k,ji_4i_1}(-p,p+q,-q;\phi) \\
\times [\tilde\dl G_{k,i_1i_2}(q;\phi)]G_{k,i_3i_4}(p+q;\phi) + (p\leftrightarrow -p, i\leftrightarrow j) \Bigr\rbrace 
\label{flow1} 
\end{multline}
($l=\ln(k/\Lambda)$) for the effective potential and the two-point vertex in a constant field $\phi$. $G_k=-(\Gamma^{(2)}_k+R_k)^{-1}$ is the single-particle propagator. We use the notation 
\begin{equation}
\frac{1}{\beta N} \sum_q \equiv \int_q = \int_\q \int_\w = \int \frac{d^dq}{(2\pi)^d} \intinf \frac{d\w}{2\pi} , 
\end{equation}
where the momentum integral is restricted to the Brillouin zone $]-\pi,\pi]^d$. The operator $\tilde\dl=(\dl R_k)\partial_{R_k}$ acts only on the $l$ dependence of the cutoff function $R_k$. The BMW approximation is based on the following two observations. i) For a given momentum $\q$, the frequency integral in (\ref{flow1}) is dominated by the region $|\w|\lesssim \w^-_{k}(\q)$ where $\w^-_{k}(\q)$ is the lowest excitation energy defined by the propagator $G_k$. Since the function $\tilde\dl G_{ij}(q;\phi)$ is proportional to $\dl R_k(q)$, the integral over the loop momentum $\q$ in (\ref{flow1}) is dominated by values of $|\q|$ of the order or smaller than $k$. It follows that the important frequency range for the loop integral is $|\w|\lesssim \w^-_{k}$ where $\w^-_{k}$ is the typical value of $\w^-_{k}(\q)$ for $|\q|\sim k$. In the superfluid phase $\w_k^-\sim c_kk$ ($c_k$ is the velocity of the Goldstone mode), while in the Mott insulating phase $\w_k^-$ can be deduced from~(\ref{wq1},\ref{wq2}). ii) Because of the cutoff function $R_k(\q)$, the vertices $\Gamma_k^{(n)}(q_1\cdots q_n)$ are smooth functions of momenta and frequencies in the range $|\q_i|/k,|\w_i|/\w^-_{k}\ll 1$. These two properties allow us to expand the vertices in the rhs of~(\ref{flow1}) in powers of $\q^2/k^2$ and $\w^2/\w^-_k{}^2$. To leading order, one simply sets $q=0$ in the three- and four-point vertices in Eq.~(\ref{flow1}). We can then obtain a closed equation for $\Gamma^{(2)}_k$ by noting that\cite{Blaizot06}
\begin{equation}
\begin{split}
\Gamma^{(3)}_{k,ijl}(p,-p,0;\phi) &= \frac{1}{\sqrt{\beta N}} \frac{\partial}{\partial\phi_l} \Gamma^{(2)}_{k,ij}(p;\phi) , \\ 
\Gamma^{(4)}_{k,ijlm}(p,-p,0,0;\phi) &= \frac{1}{\beta N} \frac{\partial^2}{\partial\phi_l \partial\phi_m} \Gamma^{(2)}_{k,ij}(p;\phi) .
\end{split}
\label{gam_bmw} 
\end{equation}

Furthermore, properties (i) and (ii) allow us to use the derivative expansion of the two-point vertex $\Gamma_k^{(2)}$ [Eq.~\eqref{gamde}] to obtain the propagator $G_k$ to be used in the RG equations~(\ref{flowpot},\ref{flow1}). Since it is however crucial to retain the full lattice structure at the beginning of the RG flow ($k\simeq\Lambda$), we take
\begin{equation}
\Gamma_{A,k}(q;n) = Z_{A,k}(n)\eps_\q + V_{A,k}(n)\w^2 + V'_k(n) , 
\label{gamde1}
\end{equation}
which coincides with \eqref{gamde} for $|\q|\ll\Lambda$. We have introduced the shifted dispersion $\eps_\q=t_\q+2dt$ ($\eps_\q\simeq t\q^2$ for $|\q|\ll\Lambda$). Following Ref.~\onlinecite{Machado10}, we define $Z_{A,k}(n)$ as 
\begin{equation}
Z_{A,k}(n) = \frac{1}{t} \lim_{q\to 0} \frac{\partial}{\partial \q^2} \Gamma_{A,k}(q;n) , 
\label{ZAdef}
\end{equation}
so that $Z_{A,k}(n_{0,k})$ has the meaning of a field renormalization factor (and should not be confused with a renormalization of the hopping amplitude between nearest-neighbor sites\cite{Machado10}). For $k\ll\Lambda$, Eqs.~(\ref{gamde1}) and (\ref{ZAdef}) are equivalent to the so-called LPA' approximation (LPA stands for local potential approximation). For $k\simeq\Lambda$, these equations can be justified by noting that in this limit $Z_{A,k}(n)\simeq Z_{A,\Lambda}(n)=1$ so that approximating the renormalized dispersion by $Z_{A,k}(n)\eps_\q$, which is valid for small $\q$ when $Z_{A,k}$ is defined by~(\ref{ZAdef}), is expected to remain approximately valid in the whole Brillouin zone.\cite{Machado10} The derivative expansion of the local vertex $\Gamma^{(2)}_{\rm loc}$ is further discussed in Appendix~\ref{app_de}. 

Although the final flow equations rely on a derivative expansion of the vertices, they cannot be derived directly from a simple Ansatz of the effective action $\Gamma_k$. The reason is that it is not possible to propose an Ansatz for the initial effective action $\Gamma_\Lambda[\phi^*,\phi]$ [Eq.~(\ref{GamLambda})] based on a derivative expansion, since we do not know its expression for arbitrary time-dependent fields. In the BMW approach, we deal only with quantities computed in a constant field which, for $k=\Lambda$, can  easily be obtained from the Hamiltonian in the local limit as explained in Sec.~\ref{subsec_init}.

As far as the momentum dependence of the vertices is concerned, our BMW approximation (supplemented with a derivative expansion of $\Gamma_k^{(2)}$ to obtain $G_k$) is as legitimate as the original one.\cite{Blaizot06,Benitez09,Benitez08} It is however more questionable regarding the $\w$ dependence. Contrary to the momentum integral, the frequency integral in Eq.~(\ref{flow1}) is not exponentially cut off by the regulator $R_k(\q)$. The integrand typically decays as a power of $1/|\w|$ for $|\w|\gg \w_-(\q)$, so that the contribution of large frequencies is small but not negligible. The reason why the BMW approximation nevertheless leads to accurate results (see Sec.~\ref{sec_phase_dia}) can be understood as follows. In the weak-coupling limit, the frequency dependence of the vertices $\Gamma^{(3)}_k$ and $\Gamma_k^{(4)}$ is weak,\cite{note5} so that setting  the loop frequency $\w$ to zero in $\Gamma^{(3)}_k$ and $\Gamma_k^{(4)}$, as well as using a derivative expansion of $\Gamma_k^{(2)}$, should be justified. In the strong-coupling limit, $\Gamma^{(3)}_k$ and $\Gamma_k^{(4)}$ do depend on frequency but this dependence is controlled by $U$ as in the local limit. Since $U\gg \w_-(\q)$, except deep in the Mott phase where the strong-coupling RPA is already a good approximation to the $k=0$ results, it appears again justified to set the loop frequency to zero in three- and four-point vertices and use a derivative expansion to obtain the propagator $G_k$. The use of a cutoff function $R_k(q)$ acting both on momentum and frequency would put the BMW approximation on a firmer basis,\cite{Dupuis09b} but such a cutoff function would be incompatible with the initial condition $S+\Delta S_\Lambda=S_{\rm loc}$ of the lattice NPRG.\cite{note6}

The numerical solution of the flow equations can be further simplified by approximating $V_{A,k}(n)$ and $Z_{A,k}(n)$ by $V_{A,k}\equiv V_{A,k}(n_{0,k})$ and $Z_{A,k}\equiv Z_{A,k}(n_{0,k})$.\cite{[{For similar approximations in the classical O($N$) model, see }] Guerra07} To obtain an accurate description of the phase diagram, it is nevertheless necessary to keep the full $n$-dependence of $Z_{C,k}(n)$ and $V_k(n)$.\cite{Rancon11} The $n$-dependence of $Z_{C,k}(n)$ is also necessary for a good description of the critical behavior at the multicritical points (Sec.~\ref{sec_crit}).\cite{note7} Away from the multicritical points, and when accuracy is not the primary goal, it is possible to approximate $Z_{C,k}(n)$ by $Z_{C,k}(n_{0,k})$, and expand the effective potential to quadratic order about its minimum, 
\begin{equation}
V_k(n) = \left\lbrace 
\begin{array}{lcc}
V_{0,k} + \frac{\lambda_k}{2}(n-n_{0,k})^2 & \mbox{if} & n_{0,k}>0 , \\ 
V_{0,k} + \delta_k n + \frac{\lambda_k}{2}n^2 & \mbox{if} & n_{0,k}=0 ,
\end{array}
\right. 
\label{trunc} 
\end{equation}
where $\delta_k$ and $\lamb_k$ are defined in~\eqref{deltalamb}. The BMW equations and their various approximations are detailed in Appendix~\ref{app_floweq}.

\section{Phase diagram}
\label{sec_phase_dia}

To alleviate the notations, we drop the subscript $k$ whenever we refer to a $k=0$ quantity (e.g. $n_0\equiv n_{0,k=0}$).  

\begin{figure}
\centerline{\includegraphics[width=6.5cm]{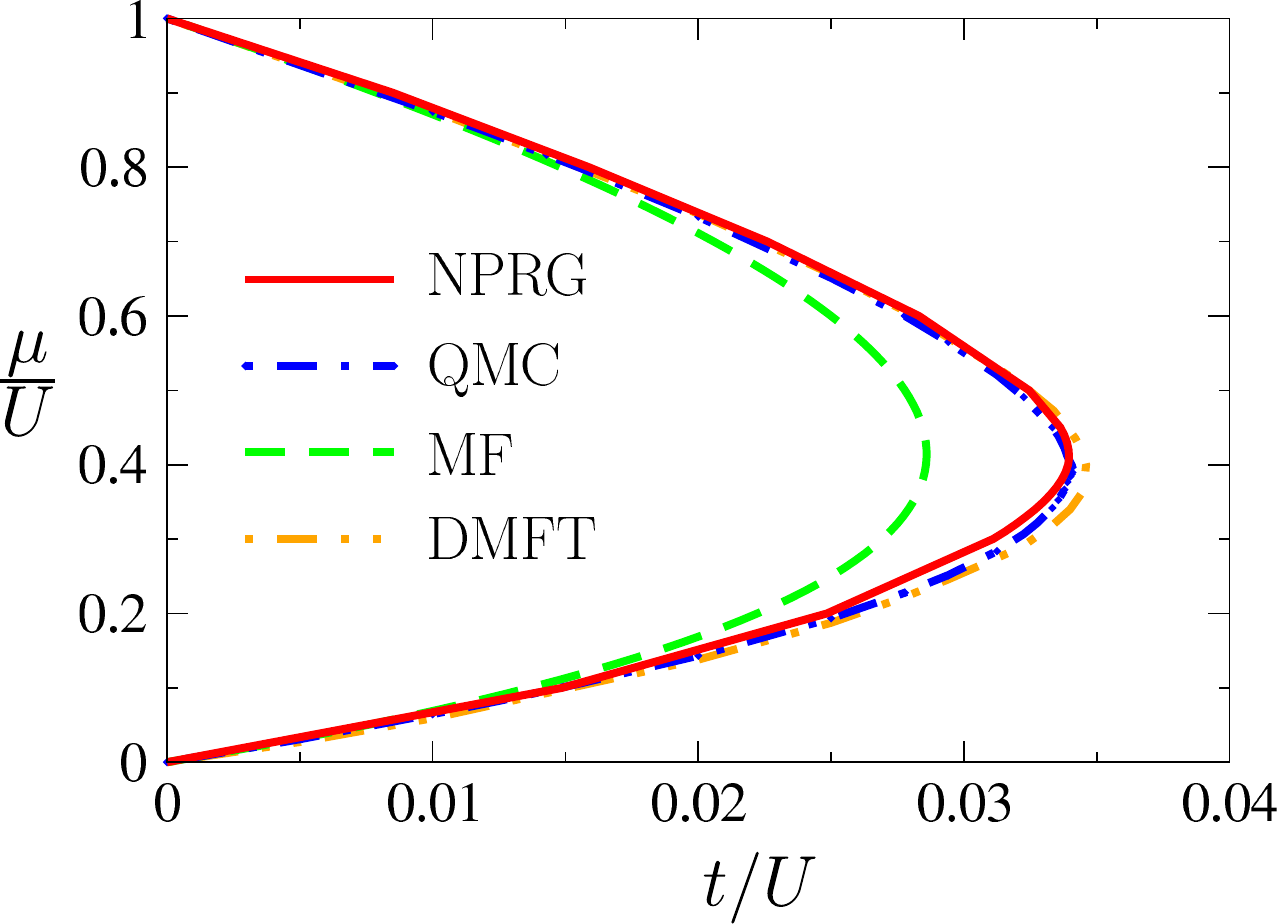}}
\caption{(Color online) Phase diagram of the 3D Bose-Hubbard model. Only the first Mott lobe ($\bar n=1$) is shown. The (green) dashed line shows the mean-field (or strong-coupling RPA) phase diagram. The QMC data are obtained from Ref.~\onlinecite{Capogrosso07} and the DMFT data from Ref.~\onlinecite{Anders11}.} 
\label{fig_dia3D}
\vspace{0.25cm}	
\centerline{\includegraphics[width=6.5cm]{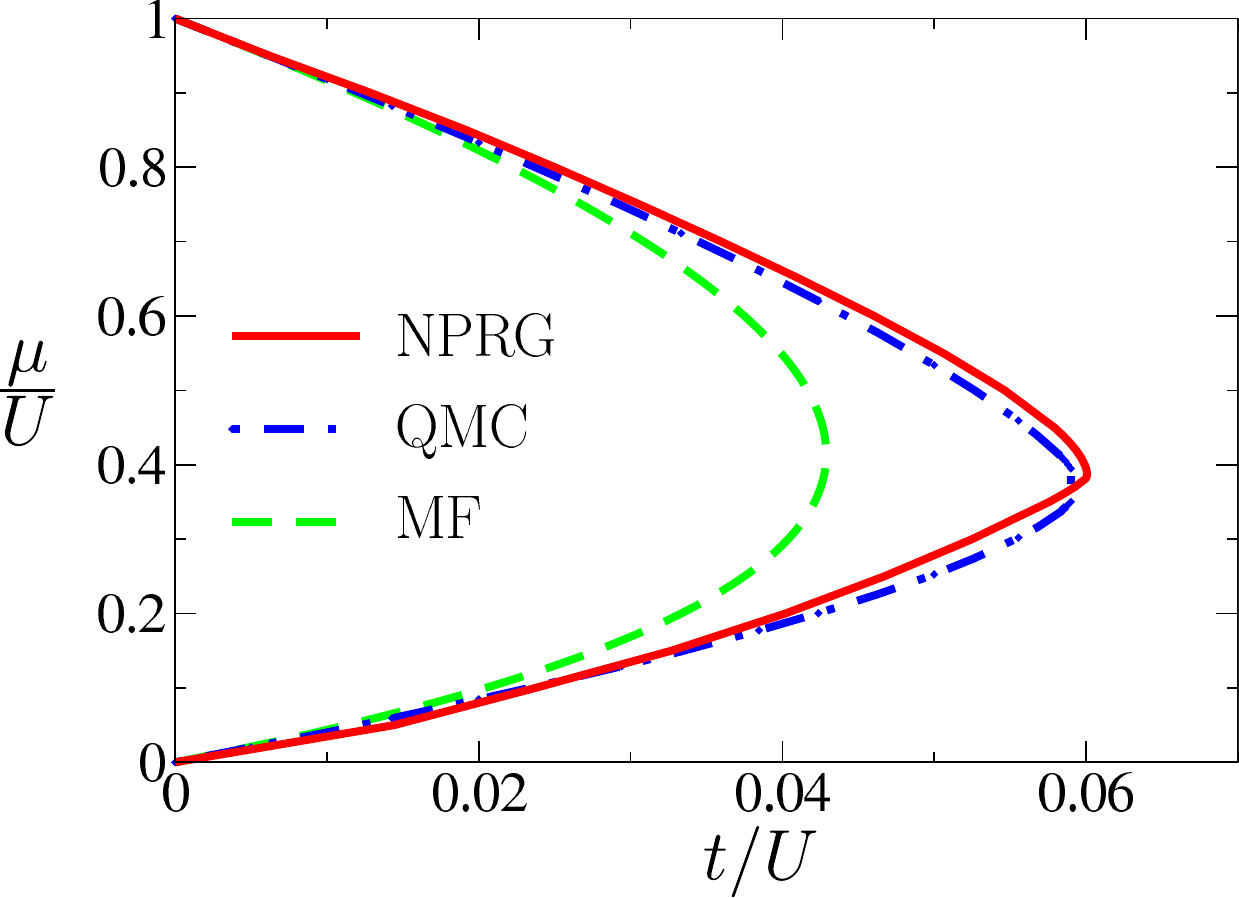}}
\caption{(Color online) Phase diagram of the 2D Bose-Hubbard model. The QMC data are obtained from Ref.~\onlinecite{Capogrosso08}.}
\label{fig_dia2D}
\end{figure}

For given values of $t$, $U$ and $\mu$, the ground state can be deduced from the values of the condensate density $n_{0}$ ($n_{0}>0$ in the superfluid phase). To obtain thermodynamic quantities, it is sufficient to integrate the RG flow down to $k\sim 10^{-5}$. In Ref.~\onlinecite{Rancon11}, we have shown that by increasing the functional character of the NPRG equations (e.g. by retaining the full $n$-dependence of $V_k(n)$ rather than using the truncation~(\ref{trunc})), we observe a nice convergence of our results, which we therefore expect to be close to the exact ones. The most accurate results, obtained by keeping the full $n$ dependence of $V_k(n)$ and $Z_{C,k}(n)$ are shown in Figs.~\ref{fig_dia3D} and \ref{fig_dia2D}. Both in three and two dimensions, the transition line between the superfluid phase and the Mott insulator is very close to the QMC result:\cite{Capogrosso07,Capogrosso08} the tip of the Mott lobe ($t/U=0.0339$, $\mu/U=0.3992$) differs from the QMC data only by ($0.001\%$, $3\%$) in three dimensions, while in two dimensions the tip is located at $(t/U=0.060,\mu/U=0.387)$, which corresponds to a relative error of order ($1.5\%$, $4\%$). For comparison, in Figs.~\ref{fig_dia3D} and \ref{fig_dia2D} we also show the mean-field (or strong-coupling RPA) phase diagram as well as the one obtained from Dynamical Mean-Field Theory (DMFT).\cite{Anders10,Anders11}  

\begin{figure}
\centerline{\includegraphics[width=6.5cm,clip,angle=0]{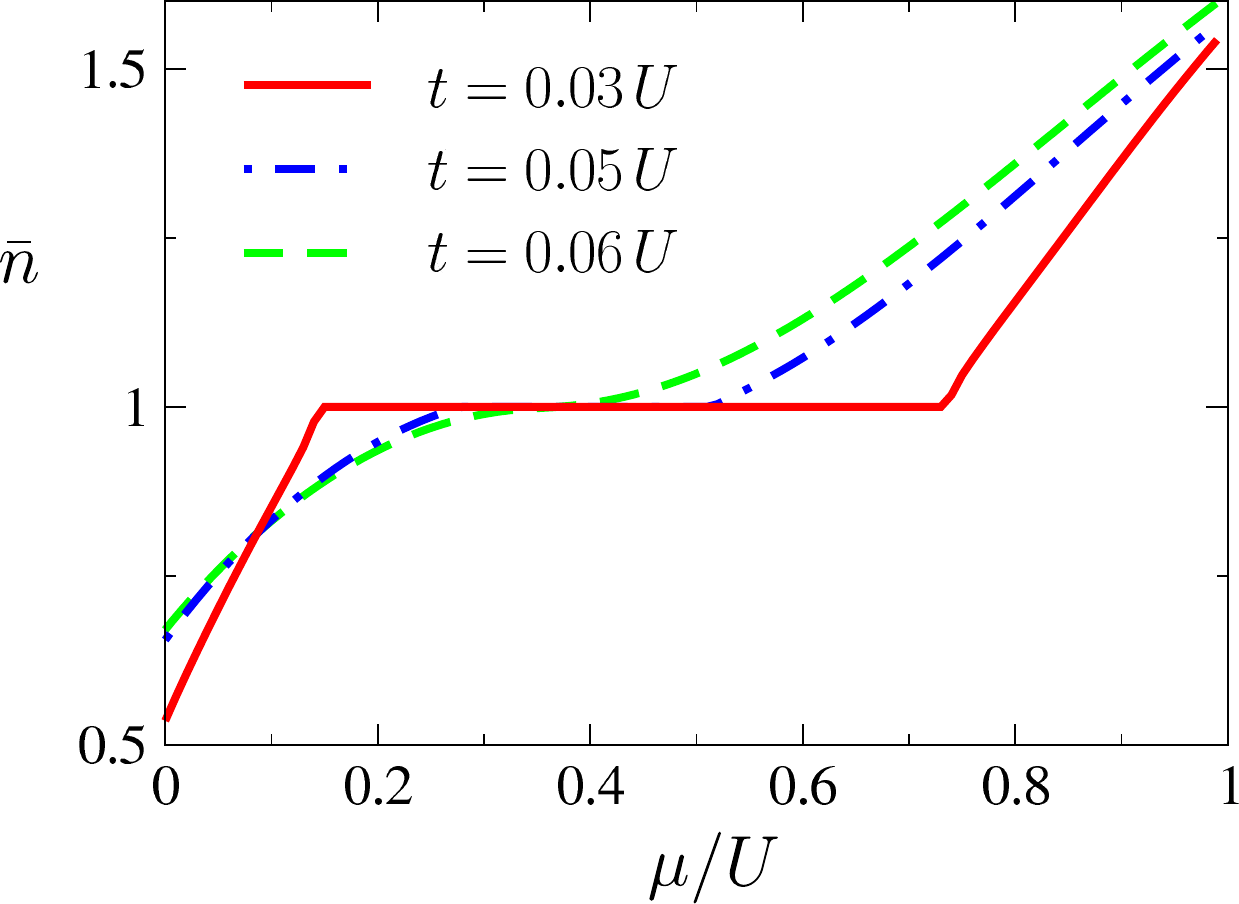}}
\caption{(Color online) Density $\bar n\equiv \bar n_{k=0}$ vs $\mu/U$ for various values of $t/U$ and $d=2$.} 
\label{fig_den}
\end{figure}

An important characteristic of the Mott insulating phases is the vanishing compressibility $\kappa=d\bar n/d\mu=0$. The expression~\eqref{dndmu} enables us to determine the boson density $\bar n_k$ directly from $n_{0,k}$, $V_{A,k}$, $Z_{C,k}$ and $\lamb_k$ by integrating $d\bar n_k/d\mu$. The unknown integration constant is easily fixed since we know that the density vanishes for $\mu=-2dt$. Alternatively, one can use the fact that the density is integer in the Mott insulator. This method not only avoids to numerically compute $dV_{0,k}/d\mu$ (which requires to solve the RG equations for nearby values of $\mu$) but also turns out to be much less sensitive to numerical noise. Figure~\ref{fig_den} shows the density $\bar n$ as a function of the chemical potential $\mu$ for various values of $t/U$ and $d=2$. The vanishing compressibility $\kappa=0$ in the Mott insulating phase $\bar n=1$ is clearly visible. In the figure, the density is obtained from $\kappa$ and the condition $\bar n=1$ in the Mott phase. If we use the condition $\bar n(\mu=-2dt)=0$, we obtain $\bar n=1\pm 0.03$ ($\bar n=1\pm 0.045$) in the three-dimensional (two-dimensional) Mott phase $\bar n=1$. The error is more pronounced near the tip of the Mott lobe. 

\section{Critical behavior}
\label{sec_crit}

In Sec.~\ref{subsec_de}, we have seen that the universality class of the superfluid--Mott-insulator transition depends on whether $Z_{C}(n=0)$ vanishes or not. We must therefore determine the value of $Z_{C}(n=0)$ at the transition.

Let us consider the parameter $\delta(t,\mu)=V'(0)$ as a function of $t$ and $\mu$ (with $U$ fixed). In the Mott insulator, the Ward identity~(\ref{wardid3}) becomes 
\begin{equation}
Z_C \equiv Z_C(0) = -\frac{\partial\delta}{\partial\mu}\biggl|_t .  
\label{wardid4}
\end{equation}
Since the transition line is defined by $\delta(t,\mu)=0$, $Z_C$ vanishes at the tip of the Mott lobe ($\mu=\mu_c$) where the tangent to the transition line is vertical. Moreover, since $\delta(t,\mu)\geq 0$ in the Mott phase, on the transition line $Z_C$ is positive (negative) for $\mu>\mu_c$ ($<\mu_c$). We deduce that at the tip of the Mott lobe, the quantum critical point coincides with the critical point of the $(d+1)$-dimensional XY model. The dynamical critical exponent takes the value $z=1$ in agreement with the Lorentz invariance of the effective action when $Z_C=0$ (Sec.~\ref{subsubsec_mi}). The upper and lower critical dimensions are therefore $d_c^+=3$ and $d_c^-=1$, respectively. The transition from the Mott insulator to the superfluid phase is driven by the vanishing of the particle-hole excitation gap, while the density is conserved.\cite{Fisher89,note14} The critical point is a multicritical point as two parameters ($t/U$ and $\mu/U$) have to be fine tuned. Away from the Mott lobe tip, $Z_C$ is nonzero and the dynamical critical exponent takes the value $z=2$. This transition, which is driven by a density change, is mean-field like for $d\geq 2$ (with logarithmic corrections at the upper critical dimension $d_c^+=2$). 

\subsection{Multicritical point} 
\label{subsec_multicrit}

The critical behavior at the tip of a Mott lobe can be understood from the linearized flow equations. If we set $Z_{C,k}(n)=0$, we recover the flow equations of the $(d+1)$-dimensional XY model with one relevant direction in the space of parameters of the effective action. The flow of the corresponding scaling field (which we denote by $r$) determines the exponent $\nu$. Since $Z_{C,k}(n)$ enters the propagators quadratically, it does not enter the linearized flow equations (except of course its own RG equation). Thus $Z_{C,k}(n_{0,k})$ corresponds to the second relevant direction of the flow and is orthogonal (in the parameter space of the action) to the critical surface. 

The behavior of the system near the multicritical point $(t_c,\mu_c)$ is best understood by considering the singular part $V_s(r,Z_C)$ of the effective potential ($Z_C\equiv Z_C(n_0)$).\cite{Fisher89} For small $r$ and $Z_C$, and $d<3$:
\begin{align}
V_s(r,Z_C) &= s^{-d-z} V_s(s^{1/\nu}r,sZ_C) \nonumber \\ 
&= |r|^{\nu(d+z)} \tilde V_s(|r|^{-\nu}Z_C) .
\label{Vsing}
\end{align}
Here we anticipate that the eigenvalue related to the scaling field $Z_C$ is equal to one (see below).
$V_s$ being finite and nonzero in the limits $r\to 0$ and $Z_C\to 0$, $\tilde V_s(x)$ must behave like a constant when $x\to 0$ and like $x^{d+z}$ when $x\to\infty$. Moreover, $r$ and $Z_C$ are presumably analytic functions of $t-t_c$ and $\mu-\mu_c$, and must vanish linearly with $t-t_c$ as we approach the multicritical point on a typical path (\ie a path which is not vertical in the $(t/U,\mu/U)$ plane). Since the critical exponent $\nu$ of the XY model satisfies $1-\nu>0$ for all $d+1\geq 3$, the argument of $\tilde V_s$ in (\ref{Vsing}) vanishes as $t-t_c\to 0$. Given that $\tilde V_s(x)\to\const$ as $x\to 0$, we conclude that $Z_C$ drops out of the scaling relation~(\ref{Vsing}) and the multicritical point looks like an ordinary XY critical point as shown explicitly below by the NPRG results. 

\begin{table}
\renewcommand{\arraystretch}{1.5}
\begin{center}
\begin{tabular}{ccc}
\hline \hline
& multicritical point  & generic transition \\
\hline
$\tilde \q$ & $\q/k$ & $\q/k$  \\
$\tilde \w$ & $\left(\frac{V_{A,k}}{Z_{A,k}\eps_k}\right)^{1/2}\w$ & $\left(\frac{Z_{C,k}}{Z_{A,k}\eps_k}\right)\w$ \\ 
$\tilde n$ & $k^{-d} (V_{A,k}Z_{A,k}\eps_k)^{1/2}n$ & $k^{-d} Z_{C,k} n$\\ 
$\tilde V_k(\tilde n)$ & $k^{-d} \left(\frac{V_{A,k}}{Z_{A,k}\eps_k}\right)^{1/2}V_k(n)$ & $k^{-d}\left(\frac{Z_{C,k}}{Z_{A,k}\eps_k}\right) V_k(n)$ \\ 
$\tilde \delta_k$ & $(Z_{A,k}\eps_k)^{-1}\delta_k$ & $(Z_{A,k}\eps_k)^{-1}\delta_k$ \\ 
$\tilde\lamb_k$ & $k^d V_{A,k}^{-1/2} (Z_{A,k}\eps_k)^{-3/2}\lamb_k$ & $k^d (Z_{C,k}Z_{A,k}\eps_k)^{-1}\lamb_k$ \\ 
$\tilde Z_{C,k}(\tilde n)$ & $(V_{A,k}Z_{A,k}\eps_k)^{-1/2} Z_{C,k}(n)$ & \\
$\tilde V_{A,k}$ & & $Z_{A,k}\eps_k Z_{C,k}^{-2} V_{A,k}$ \\ 
\hline\hline
\end{tabular}
\end{center}
\caption{Dimensionless variables ($Z_{C,k}\equiv Z_{C,k}(n_{0,k})$).}
\label{table_dimless}
\end{table}

To make the fixed point manifest when the system is critical, we use the dimensionless variables defined in table~\ref{table_dimless}.  The anomalous dimensions are defined by 
\begin{equation}
\begin{split}
\eta_{A,k} &= - \dl \ln Z_{A,k} , \\ 
\eta_{V,k} &= - \dl \ln V_{A,k} . 
\end{split}
\end{equation}
The dimensionless frequency variable $\tilde \w$ (Table~\ref{table_dimless}) allows us to define a (running) dynamical critical exponent $z_k=[\w]$ from $[Z_{A,k}]=-\eta_{A,k}$, $[V_{A,k}]=-\eta_{V,k}$, and $[\tilde\w]=0$, which gives 
\begin{equation}
z_k = 1 - \frac{\eta_{A,k}-\eta_{V,k}}{2} .
\end{equation}
Here $[X]$ denotes the scaling dimension of the variable $X$ (momenta having as usual scaling dimension 1). At the multicritical point, we expect $\eta_{A}=\eta_{V}$ and $z=1$. It is however possible that the regulator $R_k(\q)$, which does not satisfy the Lorentz invariance of the effective action at the multicritical point, modifies the expected critical behavior. Setting $Z_{C,k}(n)=0$ in the flow equations, we find
\begin{equation}
\eta_{V,k} = \eta_{A,k} - \frac{\eta_{A,k}^2}{d+2} .
\label{etadiff}
\end{equation}
Given the small value of the anomalous dimension in the $(d+1)$-dimensional XY  model ($d=2,3$), the results $\eta_A=\eta_V$ and $z=1$ are nevertheless satisfied to a very good accuracy (see below).

\subsubsection{2D multicritical point} 

\begin{figure}
\centerline{\includegraphics[width=6.cm,clip]{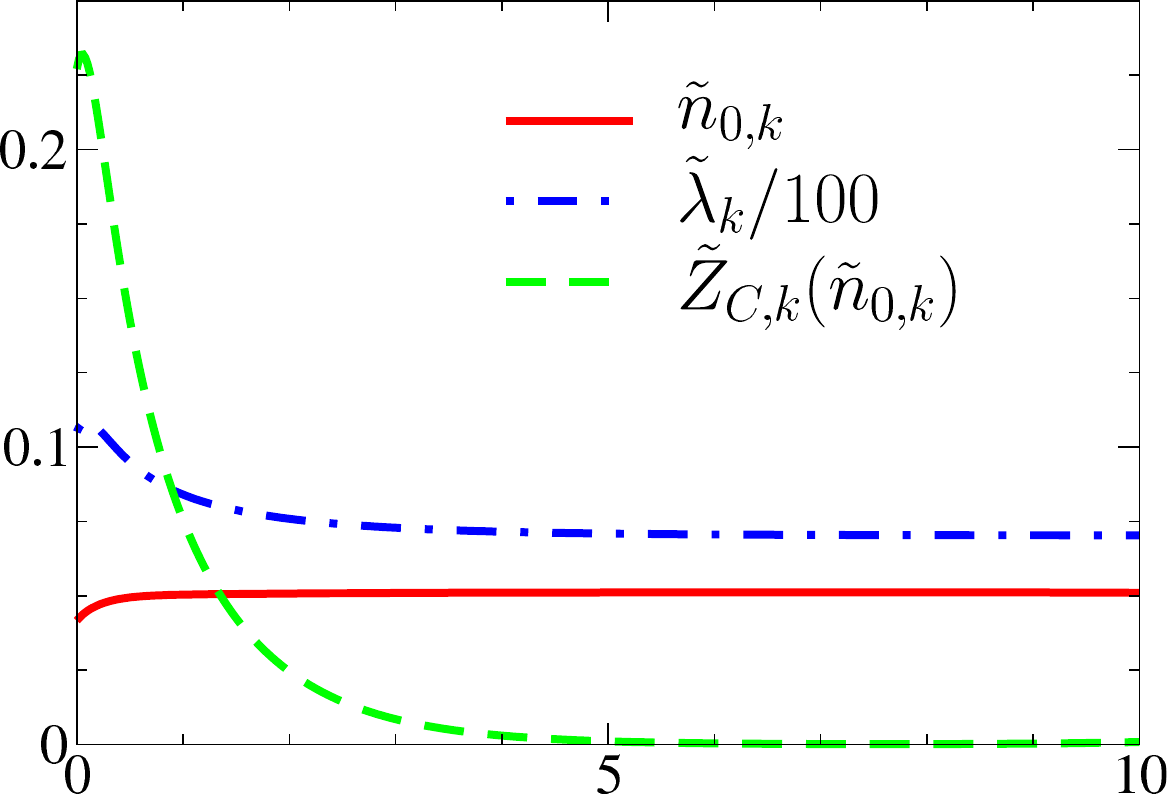}}
\vspace{0.15cm}
\centerline{\includegraphics[width=6.cm,clip]{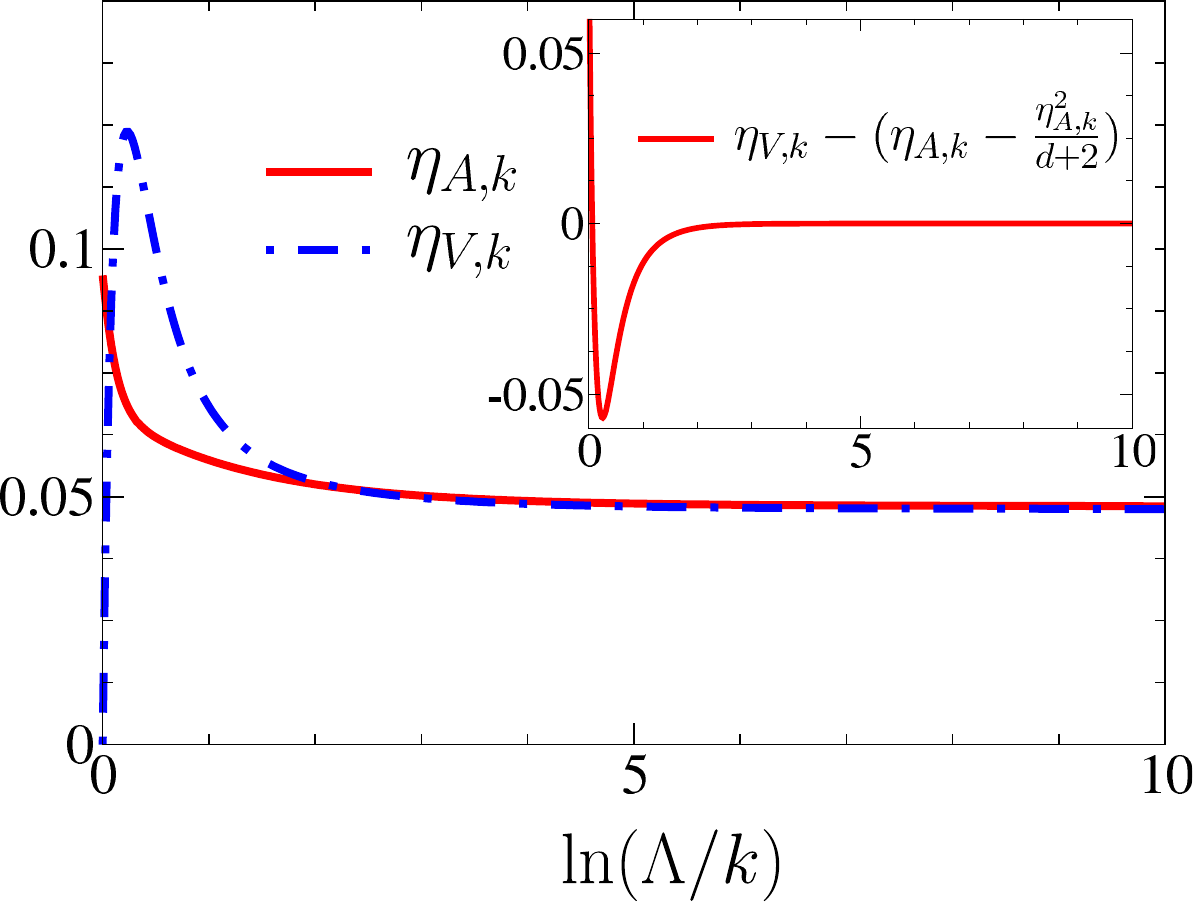}}
\caption{(Color online) (Top) Dimensionless condensate density $\tilde{n}_{0,k}$, coupling constant $\tilde{\lambda}_k$ and $\tilde Z_{C,k}(\tilde n_{0,k})$ vs $\ln(\Lambda/k)$ at the multicritical point $\bar n=1$ for $d=2$. (Bottom) Anomalous dimensions $\eta_{A,k}$ and $\eta_{V,k}$ vs $\ln(\Lambda/k)$. The inset shows that Eq.~(\ref{etadiff}) is satisfied when $k\to 0$.}
\label{fig_XY2D}
\vspace{0.25cm}
\centerline{\includegraphics[width=3.8cm,clip]{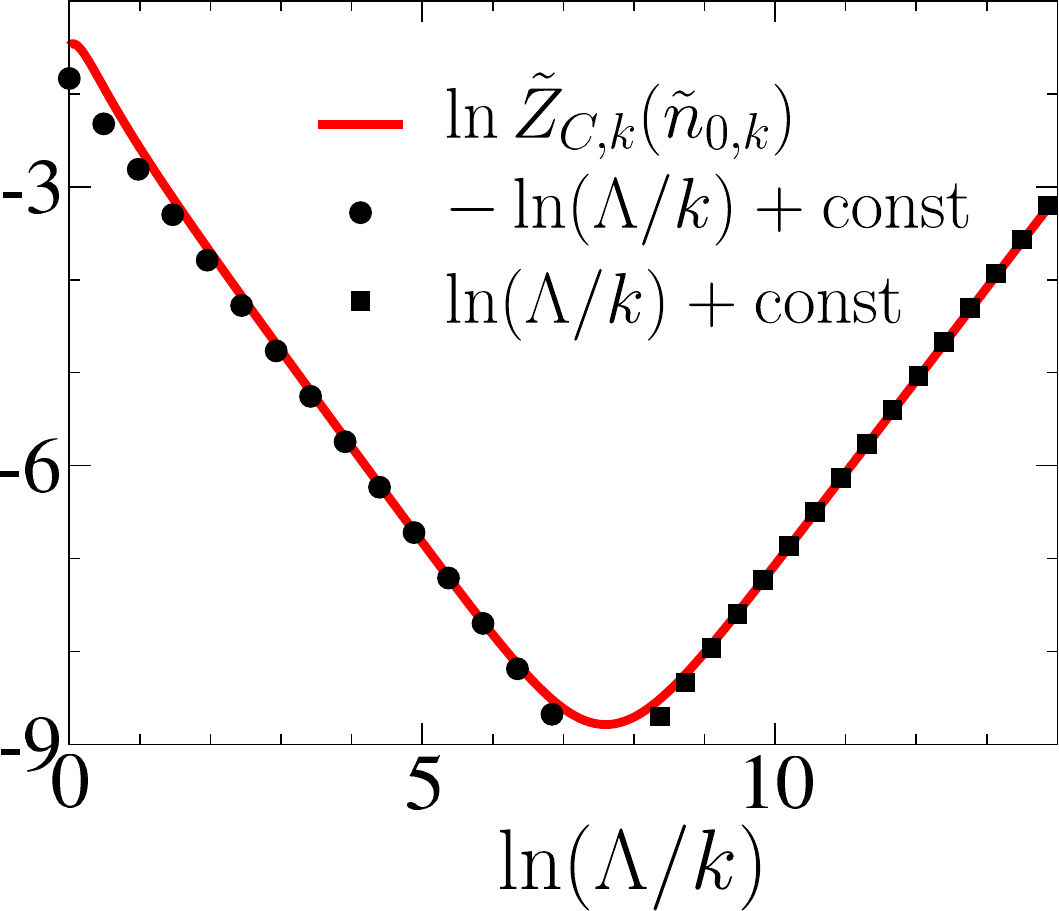}
\includegraphics[width=4.3cm,clip]{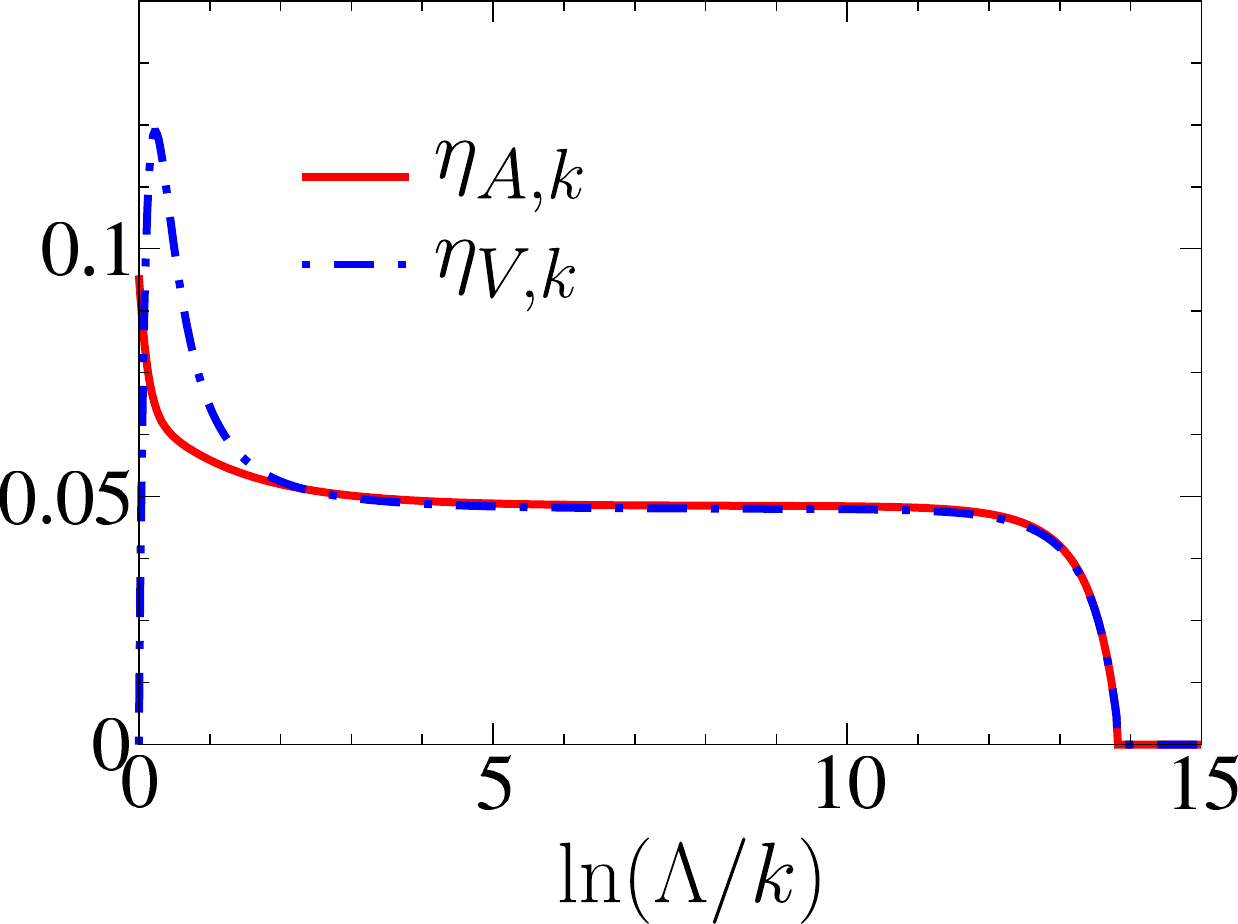}}
\caption{(Color online) (Left) $\ln \tilde Z_{C,k}(\tilde n_{0,k})$ vs $\ln(\Lambda/k)$ near the multicritical critical point. (Right) Anomalous dimensions $\eta_{A,k}$ and $\eta_{V,k}$. The end of the plateau determines the Josephson length $\xi_J=k_J^{-1}$.} 
\label{fig_crossexp}
\vspace{0.25cm}
\centerline{\includegraphics[width=6cm,clip]{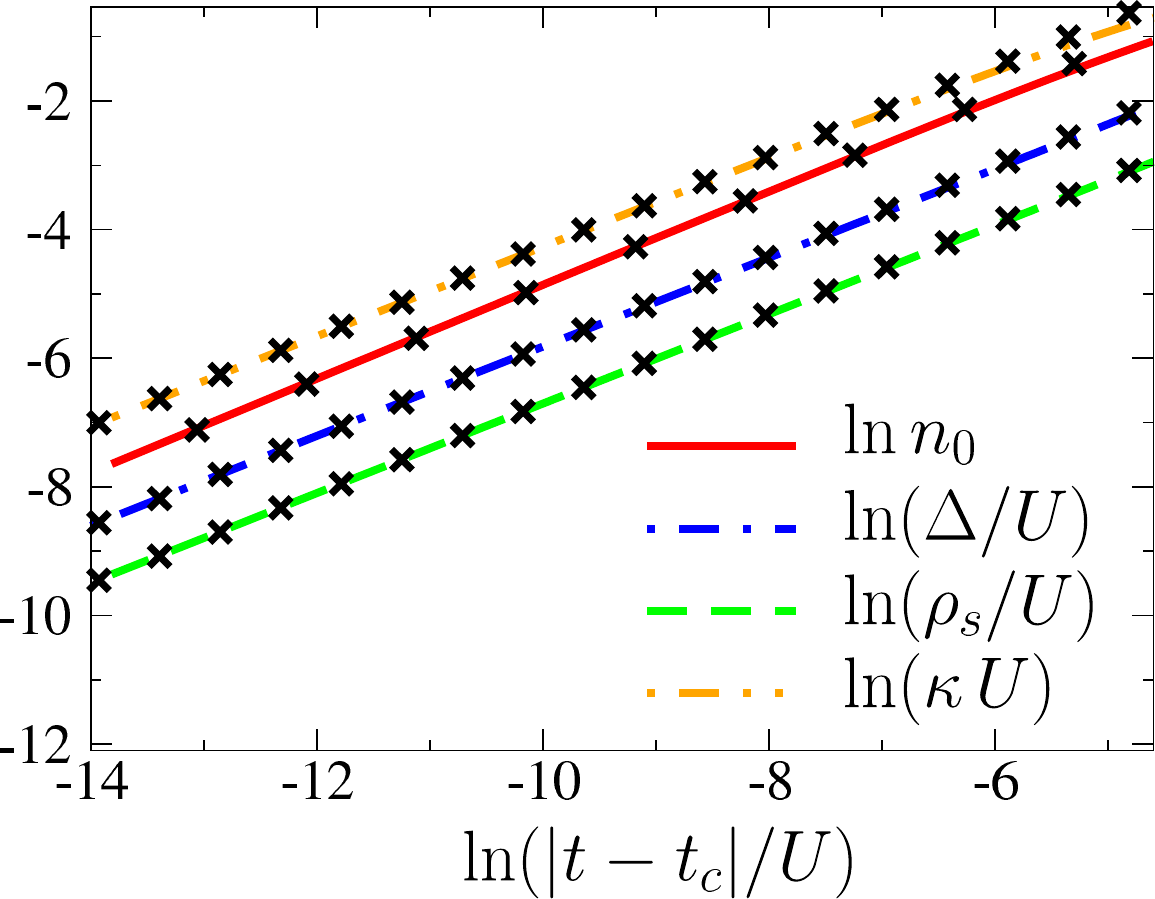}}
\caption{(Color online) Condensate density $n_0$, superfluid stiffness $\rho_s$, compressibility $\kappa$ and Mott gap $\Delta$ vs $|t-t_c|$ near the multicritical point  $(t_c,\mu_c)$ [$d=2$]. The crosses show the critical behavior~(\ref{n0critXY}-\ref{deltacritXY}) with $\nu\simeq 0.699$ and $\eta\simeq 0.049$.}
\label{fig_rhokappa}
\end{figure}

Let us first discuss the two-dimensional case. We find that $Z_{C}(n)$ vanishes for $\mu=0.382$, slightly away from the lobe tip located at $\mu=0.387$ (we now focus on the Mott insulating phase $\bar n=1$). We ascribe this slight discrepancy to the fact that the local gauge invariance (see Sec.~\ref{subsec_gauge_inv}), which leads to the Ward identity~(\ref{wardid4}), is not strictly satisfied in our approach since it is violated by both the BMW approximation and the derivative expansion. On the other hand, the fact that the multicritical point lies very close to the tip of the Mott lobe indicates that the local gauge invariance remains nearly satisfied, and all consequences discussed in Secs.~\ref{subsec_gauge_inv} and \ref{subsec_de} apply. 

Figure~\ref{fig_XY2D} shows the RG flow at the multicritical point. The plateaus observed for the dimensionless condensate density $\tilde n_{0,k}$ and coupling constant $\tilde\lamb_k$, as well as for the (running) anomalous dimensions $\eta_{A,k}$ and $\eta_{V,k}$, are characteristic of critical behavior. We clearly see the emergence of the Lorentz invariance as $k$ decreases: $\tilde Z_{C,k}(\tilde n_{0,k})\sim k$ is suppressed while $\eta_{A,k}$ and $\eta_{V,k}$ become nearly equal (implying $z_k\simeq 1$). We find the critical exponents $\nu=0.699$, $\eta_A=0.049$, $\eta_V=\eta_A(1-\eta_A/4)=0.049$ and $z=1.000$, to be compared with the best known estimates $\nu=0.671$ and $\eta=0.038$ for the three-dimensional XY model.\cite{Campostrini01} The exponent $\nu$ is deduced from the runaway flow from the critical surface when the system is nearly critical (e.g. $\tilde n_{0,k}-\tilde n_0^*\propto e^{-l/\nu}$ with $\tilde n_0^*$ the critical value of $\tilde n_0$).  

Figure~\ref{fig_crossexp} shows $|\tilde Z_{C,k}(\tilde n_{0,k})|$ near the multicritical point. It first decreases towards zero as the multicritical point is approached. Then $|\tilde Z_{C,k}(\tilde n_{0,k})|\sim 1/k\sim e^{-l}$ increases as the flow runs away from the critical surface, so that the critical exponent associated with the scaling field $\tilde Z_{C,k}(\tilde n_{0,k})$ is equal to one. The anomalous dimensions $\eta_{A,k}$ and $\eta_{V,k}$ show the momentum range where the flow is controlled by the multicritical point, as indicated by the plateaus in $\eta_{A,k}$ and $\eta_{V,k}$ in Fig.~\ref{fig_crossexp}. The end of the plateaus determines the Josephson length\cite{Josephson66} $\xi_J\equiv k_J^{-1}=k^{-1}$. 

\begin{figure}
\centerline{\includegraphics[width=6cm,clip]{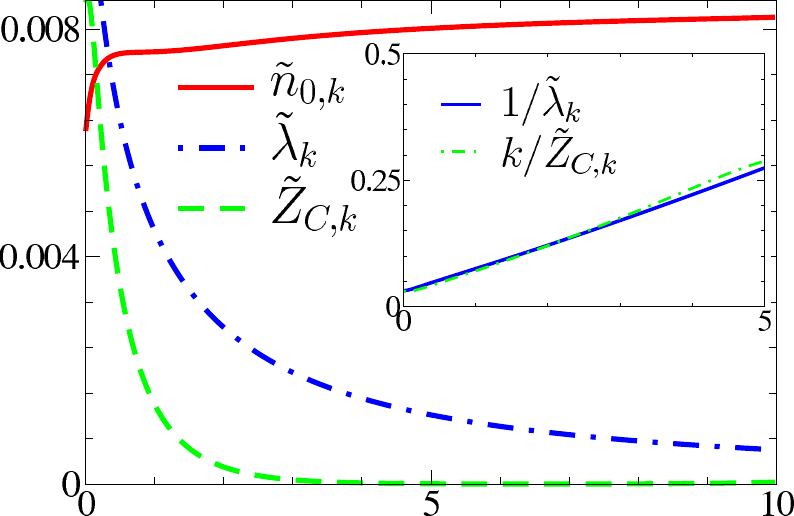}}
\centerline{\includegraphics[width=6cm,clip]{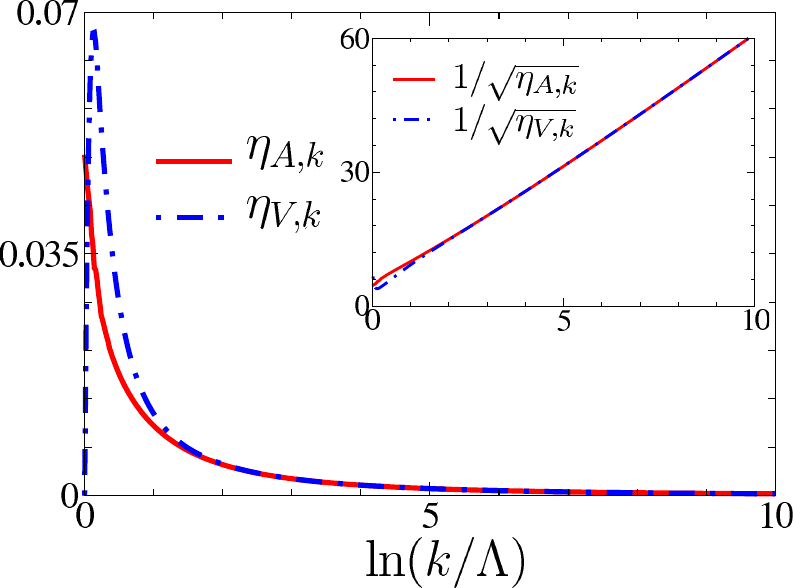}}
\caption{(Color online) Same as Fig.~\ref{fig_XY2D} but for a three-dimensional system.}
\label{fig_XY3D} 
\vspace{0.25cm}
\centerline{\includegraphics[width=6cm,clip]{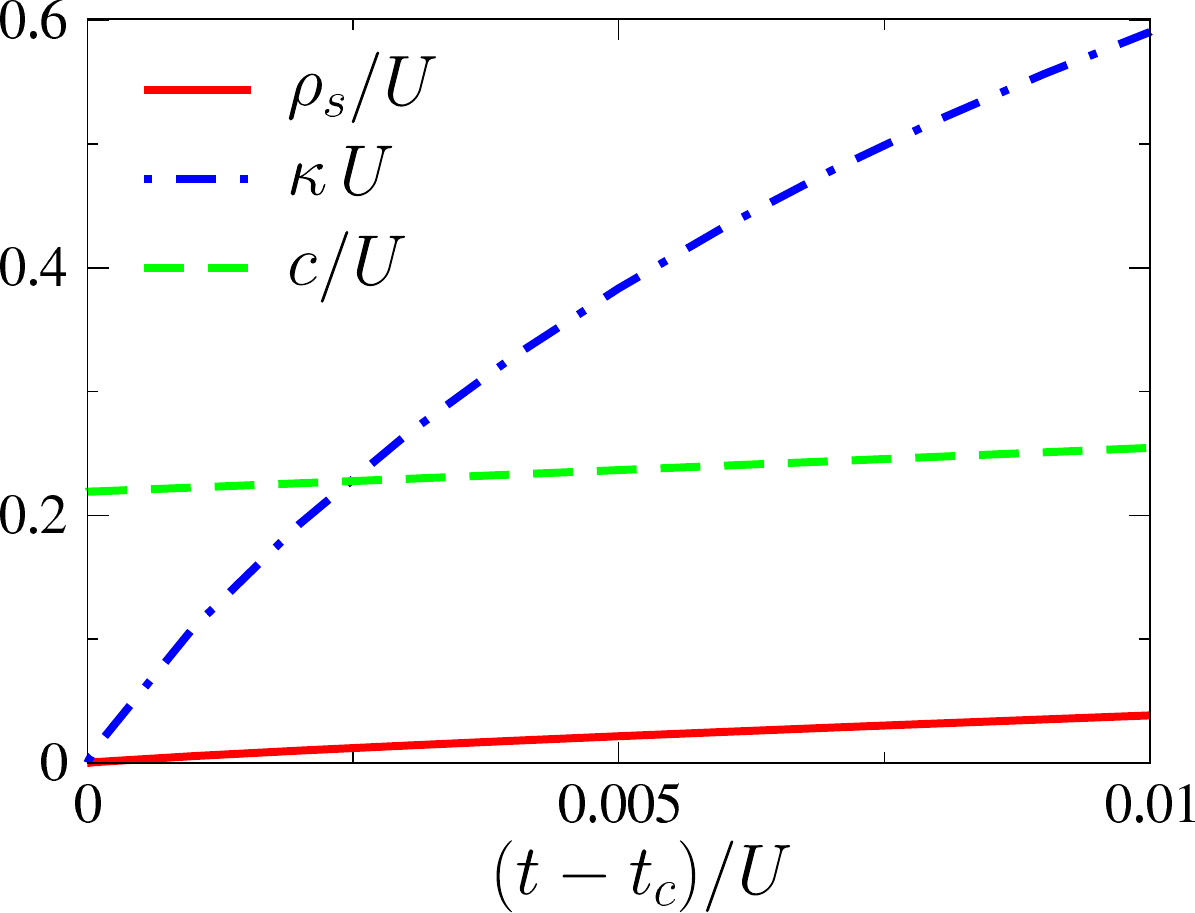}}
\caption{(Color online) Superfluid stiffness $\rho_s$, compressibility $\kappa$ and Goldstone mode velocity $c=\sqrt{\rho_s/\kappa}$ vs $t-t_c$ ($d=3$).}
\label{fig_XY3Da}
\end{figure}

We now discuss the behavior of the system for $\mu=\mu_c$ and $t\to t_c^+$. The condensate density must vanish with the critical exponent $2\beta=\nu(d+z-2+\eta)$, 
\begin{equation}
n_0 \sim (t-t_c)^{\nu(d+z-2+\eta)} .
\label{n0critXY} 
\end{equation}
From the scaling dimension $[\rho_s]=d+z-2$ of the superfluid stiffness and the fact that the Goldstone mode velocity $c=\sqrt{\rho_s/\kappa}$ remains finite due to the Lorentz invariance of the effective action $\Gamma_k$ in the limit $k\to 0$, we expect
\begin{equation}
\begin{split}
\rho_s &\sim (t-t_c)^{\nu(d+z-2)} , \\ 
\kappa &\sim (t-t_c)^{\nu(d+z-2)} .
\end{split}
\label{rhokappacritXY}
\end{equation}
Equations~(\ref{n0critXY},\ref{rhokappacritXY}) agree with the results obtained from the numerical solution of the RG equations (Fig.~\ref{fig_rhokappa}).

In the Mott phase, since the gap has scaling dimension $[\Delta]=z$, it must vanish as
\begin{equation}
\Delta \sim (t_c-t)^{\nu z} 
\label{deltacritXY}
\end{equation}
for $t\to t_c^-$, again in agreement with the results obtained from the RG equations (Fig.~\ref{fig_rhokappa}).\cite{note19}

\subsubsection{3D multicritical point}

In three dimensions, the system is at the upper critical dimension ($d+z=4$) and the transition is governed by the Gaussian fixed point with logarithmic corrections due to the marginally irrelevant coupling constant $\tilde\lamb_k$. The numerical solution of the flow equations show that $\tilde\lamb_k\sim 1/|\ln k|$, $\tilde Z_{C,k}(\tilde n_{0,k})\sim k/|\ln k|$ and $\eta_{A,k},\eta_{V,k}\sim 1/|\ln k|^2$, while $\tilde n_{0,k}$ converges to its fixed point value $\tilde n_0^*$ logarithmically (Fig.~\ref{fig_XY3D}). Figure~\ref{fig_XY3Da} shows $\rho_s$, $\kappa$ and $c$ as a function of $t-t_c$.

\subsection{Generic transition} 
\label{subsec_generic}

For all transition points away from the lobe tip, the $V_{A,k}\w^2$ term is irrelevant with respect to to the $Z_{C,k}\w$ term. The dynamical critical exponent is $z=2$ and a simple dimensional analysis shows that the upper critical dimension is $d_c^+=2$. The transition is therefore governed by the Gaussian fixed point (with logarithmic corrections for $d=2$) defined by $\tilde n_0^*=\tilde\lamb^*=\tilde V_A^*=0$ and $\eta_A^*=\eta_C^*=0$ where
\begin{equation}
\eta_{C,k} = -\dl \ln Z_{C,k}(n_{0,k}) . 
\end{equation}
The dimensionless variables used to study the generic transition are defined in Table~\ref{table_dimless}. Determining the dynamical critical exponent as in Sec.~\ref{subsec_multicrit}, we find  
\begin{equation}
z_k = 2-\eta_{A,k}+\eta_{C,k} . 
\end{equation}

\begin{figure}
\centerline{\includegraphics[width=6cm,clip]{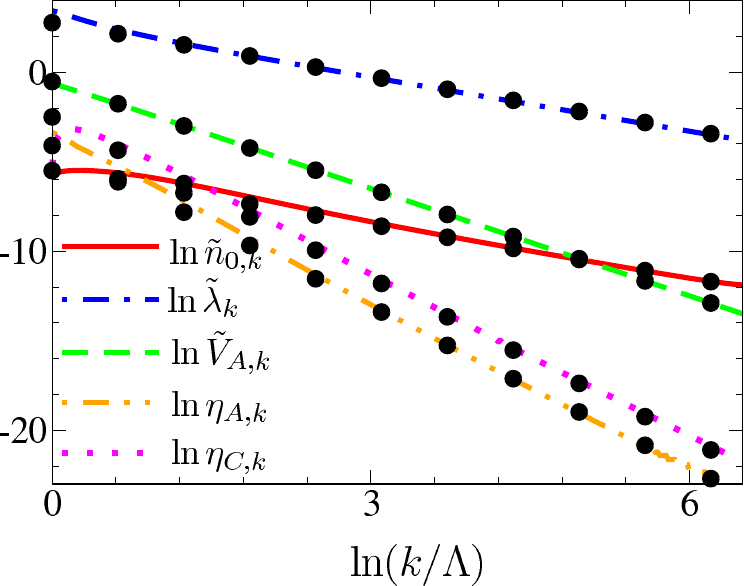}}
\caption{(Color online) RG flow at the three-dimensional generic transition. The dotted lines show fits to $\tilde n_{0,k}\sim k$, $\tilde \lamb_k\sim k$, $\tilde V_{A,k}\sim k^2$ and $\eta_{A,k},\eta_{C,k}\sim k^3$.}
\label{fig_gen3D} 
\vspace{0.25cm}
\centerline{\includegraphics[width=6cm,clip]{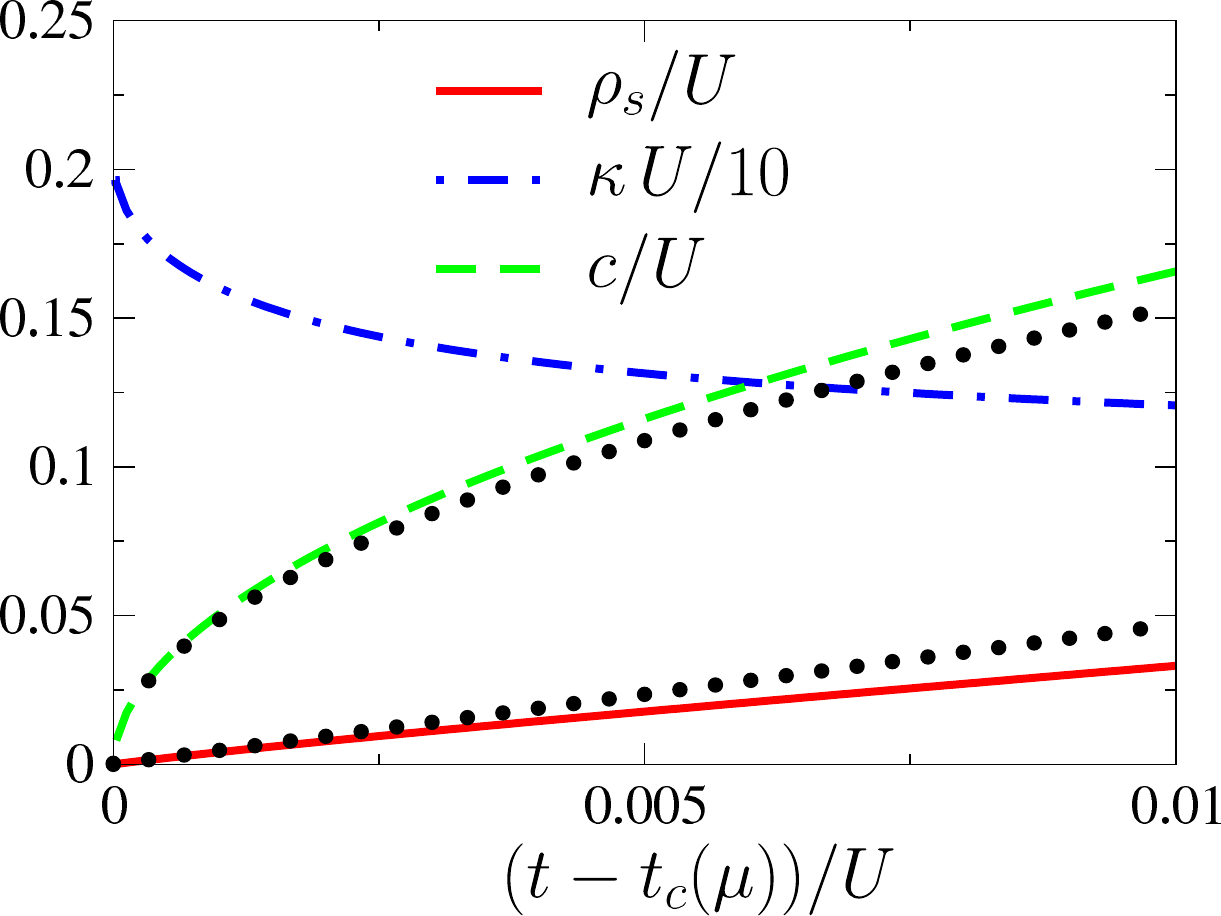}}
\caption{(Color online) Superfluid stiffness $\rho_s$, compressibility $\kappa$ and Goldstone mode velocity $c$ vs $t-t_c(\mu)$ near a generic transition point  $(t_c(\mu),\mu)$ [$\mu=0.7U$, $t_c(\mu) = 0.0219U$ and $d=3$]. The dotted lines show the mean-field critical behavior~(\ref{rhokappacritgen}).}
\label{fig_rhokappagen3D} 
\vspace{0.25cm}
\centerline{\includegraphics[width=6cm,clip]{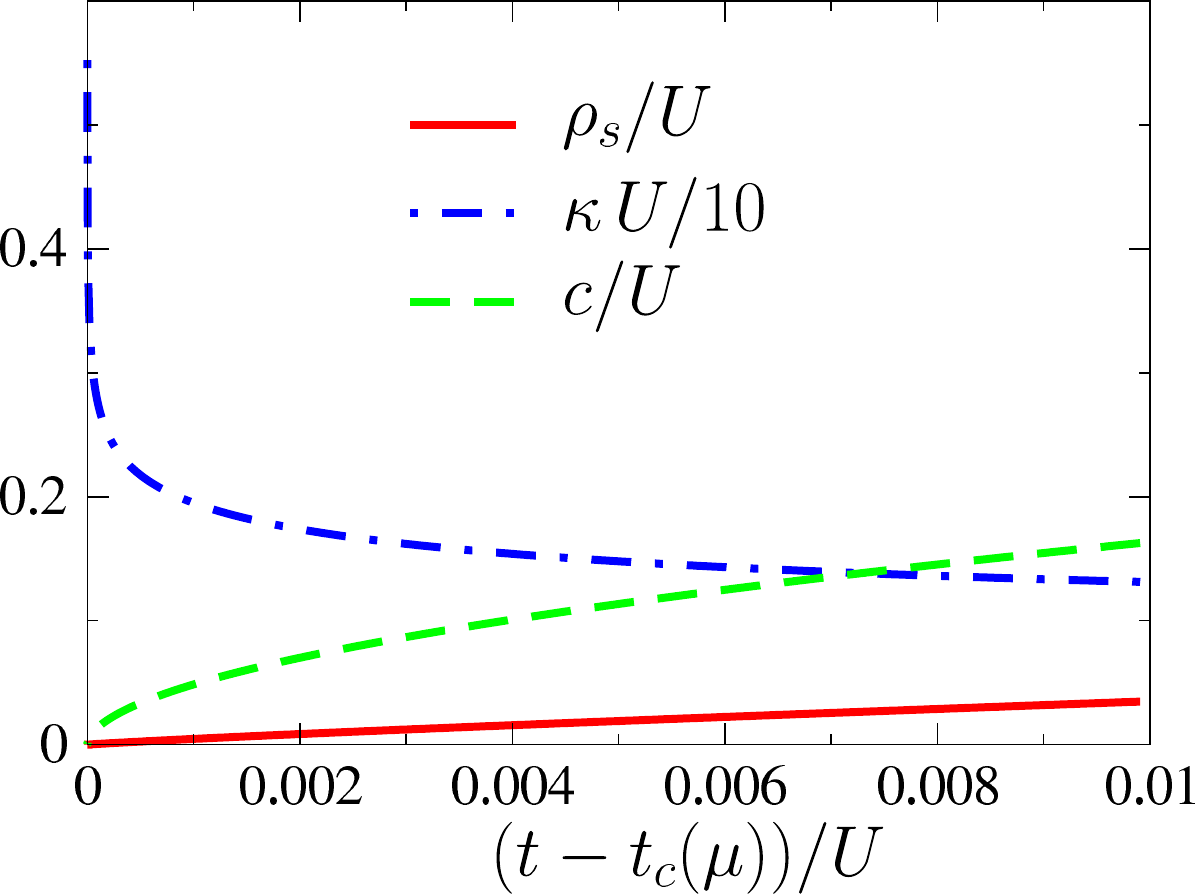}}
\caption{(Color online) Same as Fig.~\ref{fig_rhokappagen3D} but for the upper critical dimension $d_c^+=2$. The compressibility diverges when $t\to t_c(\mu)$.}
\label{fig_rhokappagen2D} 
\end{figure}

In three dimensions, linearization about the Gaussian fixed point gives 
\begin{equation}
\begin{gathered}
\dl \tilde n_{0,k} = -3 \tilde n_{0,k} + \frac{4}{3\pi^2} \tilde V_A \\ 
\dl \tilde \lamb_k = \tilde \lamb_k , \\ 
\dl \tilde V_{A,k} = 2 \tilde V_{A,k} ,
\end{gathered}
\end{equation}
and $\eta_{A,k}=\eta_{C,k}=0$. We deduce that $\tilde\lamb_k\sim k$, $\tilde V_{A,k}\sim k^2$ at the critical point, in agreement with the numerical solution of the flow equations (Fig.~\ref{fig_gen3D}). Figure~\ref{fig_gen3D} also shows that  $\eta_{A,k},\eta_{V,k}\sim k^3$ while the relevant variable $\tilde n_{0,k}$ vanishes linearly with $k$ at the critical point. 

When a generic transition point $(t_c(\mu),\mu)$ is approached on a path of constant chemical potential $\mu$ by varying $t-t_c(\mu)$, we observe the mean-field behavior 
\begin{equation}
\begin{split}
\rho_s &\sim t-t_c(\mu) , \\
\kappa &\sim \const ,
\end{split}
\label{rhokappacritgen} 
\end{equation}
for $t\to t_c(\mu)$ (Fig.~\ref{fig_rhokappagen3D}). The compressibility $\kappa$ remains finite at the transition and the velocity $c$ vanishes. 

At the upper critical dimension ($d=d_c^+=2$), the mean-field behavior is corrected by logarithmic terms. The marginally irrelevant variable $\tilde \lamb_k$ is suppressed as $|\ln k|^{-1}$, while the relevant variable $\tilde n_{0,k}$ vanish as $|\ln k|^{-1}$ at the critical point. We observe a divergence of the compressibility $\kappa$ as the phase transition is approached ($t\to t_c(\mu)$)  (Fig.~\ref{fig_rhokappagen2D}).

\section{Superfluid phase}
\label{sec_sf} 

In Sec.~\ref{subsec_dilute}, we show that our approach reproduces known results in the weakly-correlated (dilute) limit. In the following sections (Secs.~\ref{subsec_rgflow}-\ref{subsec_spectrum}), we discuss the properties of the two-dimensional superfluid phase. The three-dimensional superfluid phase is briefly discussed in Sec.~\ref{subsec_3dsf}. 

\subsection{The dilute limit}
\label{subsec_dilute}

At sufficiently low density, we expect the lattice to be irrelevant and the system to behave as a dilute superfluid gas of bosons with mass $m=1/2t$. The thermodynamics of a dilute Bose gas is controlled by the zero-temperature fixed point governing the quantum phase transition between the vacuum (\ie the Mott insulating phase with $\bar n=0$) and the superfluid phase.\cite{Sachdev_book} This quantum critical point is nothing but a particular case of the generic quantum critical point discussed in Sec.~\ref{subsec_generic}. 

Let us first consider the vacuum limit $\bar n_k=n_{0,k}=0$ and $\mu=-2dt$. The RG equations~(\ref{floweq}) give
\begin{equation}
\begin{gathered}
Z_{A,k}^{(\rm vac)}=Z_{C,k}^{(\rm vac)}=1, \\
V_{A,k}^{(\rm vac)}=0 , 
\label{dilute0}
\end{gathered}
\end{equation}
in agreement with the fact that the single-particle (normal) propagator is not renormalized: $G_{{\rm n},k}(q)=(i\w+\mu-t_\q)^{-1}$.\cite{Sachdev_book} The coupling constant $\lamb_k^{(\rm vac)}$ is given by 
\begin{equation}
\lamb_k^{(\rm vac)}= \llbrace 
\begin{array}{lc}
\dfrac{8\pi ta}{1-\frac{4}{3\pi} ka} & (d=3) , \\
- \dfrac{4\pi t}{\ln\left(\frac{ka}{2}\right) + C - \half}  & (d=2) , 
\end{array}
\right.
\label{dilute1}
\end{equation}
for $k\ll\Lambda$, where $C$ is the Euler constant and $a$ the ``s-wave'' scattering length for the Bose-Hubbard model,
\begin{equation}
a = \llbrace 
\begin{array}{lc}
\dfrac{1}{8\pi} \dfrac{1}{t/U+A} & (d=3) , \\ 
\dfrac{1}{2\sqrt{2}} e^{-4\pi t/U-C} & (d=2) 
\end{array}
\right. 
\label{dilute2}
\end{equation}
(recall we take the lattice spacing as the unit length) with $A\simeq 0.1264$. Equations~(\ref{dilute1},\ref{dilute2}) are derived in Appendix~\ref{app_vac}. The dimensionless coupling constant $\tilde\lamb^{(\rm vac)}_k=\lamb^{(\rm vac)}_k k^d/\eps_k$ (see table~\ref{table_dimless}) vanishes for $k\to 0$ when $d\geq 2$ and the quantum critical point governing the transition between the vacuum and the superfluid phase is Gaussian. The logarithmic vanishing of $\tilde\lamb^{(\rm vac)}_k$ in two dimensions agrees with $d=2$ being the upper critical dimension (Sec.~\ref{subsec_generic}).  

In the dilute limit, the finite condensate density $n_{0,k}$ can be ignored as long as $\eps_k\gg 2\lambda_k n_{0,k}$.\cite{note8} This defines the characteristic momentum scale
\begin{equation}
k_h \simeq \left(\frac{\lambda_{k_h}}{t} n_{0,k_h}\right)^{1/2} ,
\label{dilute2a}
\end{equation}
which is nothing but the inverse healing length of the superfluid: $k_h=\xi_h^{-1}$. The flow is governed by the Gaussian fixed point $\tilde n_0=\tilde\lamb=0$ for $k\gg k_h$, and is driven away from that fixed point when $k\ll k_h$ due to the finite boson density. For $k\gg k_h$, we can approximate the RG equations of $Z_{A,k}$, $Z_{C,k}$, $V_{A,k}$ and $\lamb_k$ by Eqs.~(\ref{dilute0},\ref{dilute1}) to leading order in $(k_h/k)^2=\lambda_{k_h}n_{0,k_h}/\eps_{k}$. Moreover, to this order, the variation with $k$ of the condensate density is determined by the equation
\begin{equation}
\dk (\lamb_k n_{0,k}) = 0 
\label{lambn0}
\end{equation}
so that 
\begin{equation}
n_{0,k_h} \simeq \frac{\lamb_\Lambda n_{0,\Lambda}}{\lamb_{k_h}} 
\label{dilute3}
\end{equation}
(see Appendix~\ref{app_vac}). This equation allows us to relate the chemical potential to the coupling constants at scale $k_h$. In the low-density limit,
\begin{equation}
V_\Lambda(n) = V_{\rm loc}(0) - (\mu+2dt) n + \dfrac{\lambda_\Lambda}{2} n^2 + \calO(n^3)  , 
\label{dilute4}
\end{equation}
where we have used $\lambda_\Lambda=V''_\Lambda(0)$ and Eq.~(\ref{Vexpand}) with $\bar n_{\rm loc}=0$. From Eqs.~(\ref{dilute3},\ref{dilute4}), we deduce
\begin{equation}
\mu+2dt = \lambda_\Lambda n_{0,\Lambda} \simeq  \lambda_{k_h} n_{0,k_h}
\end{equation}
for $\mu+2dt\geq 0$. 

Below the healing momentum scale $k_h$, the finite condensate density cannot be ignored. The Bogoliubov approximation amounts to ignoring any further renormalization as $k$ decreases from $k_h$ down to 0, that is approximating the $k=0$ effective action by its value at $k=k_h$. The initial value $\lamb_\Lambda$ and the first part of the RG flow ($k\gg k_h$) takes care of $T$-matrix renormalization of the coupling constant $\lamb_k$ (which is usually included in the Bogoliubov theory). This is what allows us to express the final results in terms of the scattering length $a$ rather than the bare interaction $U$. We expect the Bogoliubov approximation to be valid if the ratio between the mean interaction energy per particle and the typical kinetic energy, 
\begin{equation}
\gamma = \frac{\lambda_{k_h}\bar n}{t \bar n^{2/d}}  = \frac{\lambda_{k_h}}{t} \bar n^{1-2/d}, 
\end{equation}
is much smaller than unity.\cite{Petrov04} Note that in this equation we consider the coupling constant $\lamb_{k_h}$ at scale $k_h$. We can then define the dilute (or weak-coupling) limit by the conditions $\gamma\ll 1$ and $k_h\ll \Lambda$. The latter inequality ensures that the characteristic length scale associated with superfluid behavior is much larger than the lattice spacing (thus making the lattice irrelevant as far as the superfluid properties are concerned).\cite{note9}  

\begin{figure}
\centerline{\includegraphics[width=5cm,clip]{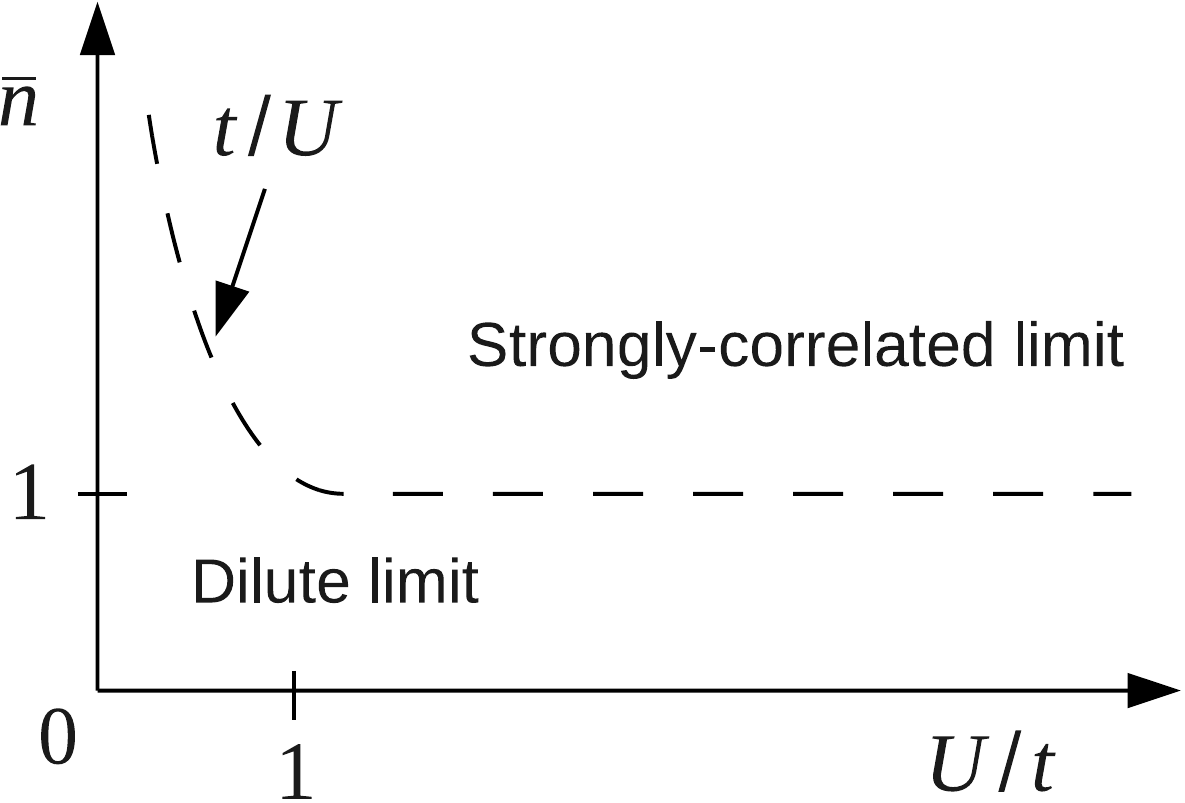}}
\caption{Crossover line between the dilute limit ($\gamma\ll 1$ and $k_h\ll\Lambda$) and the strongly-correlated limit.}
\label{fig_dilute_limit} 
\end{figure}

We are now in a position to reproduce the standard results in dilute Bose gases. In three dimensions, since $a\lesssim 0.31$ [Eq.~(\ref{dilute2})], the coupling constant $\lamb_k\simeq 8\pi ta$ is roughly constant for $k\ll\Lambda$. Given that $k_h\ll\Lambda$ in the dilute limit, we deduce 
\begin{equation}
\begin{gathered}
k_h \simeq \sqrt{8\pi a \bar n} , \\
\lamb_{k_h} \simeq 8\pi a t , \\
\mu+2dt \simeq 8\pi at \bar n , 
\end{gathered}
\end{equation} 
while the sound mode velocity takes the value
\begin{equation}
c \simeq \sqrt{2t\lamb_{k_h}n_{0,k_h}} \simeq 2t \sqrt{4\pi a \bar n} . 
\end{equation}
Since $Z_{A,k_h}\simeq 1$, the superfluid stiffness is given by
\begin{equation}
\rho_s \simeq 2t n_{0,k_h} \simeq 2t\bar n . 
\end{equation}
We have used $n_{0,k_h}=\bar n$ to leading order in the gas parameter $\gamma\sim a\bar n ^{1/3}$ of the three-dimensional dilute Bose gas. In the limit $U/t\ll 1$, where $a\sim U/8\pi t$ (the $T$-matrix renormalization of the coupling constant is negligible), one finds
\begin{equation}
\begin{gathered}
\lamb_{k_h} \simeq U , \\
k_h \simeq \sqrt{\frac{U}{t}\bar n} , \\
\mu+2dt \simeq U\bar n , \\
c \simeq \sqrt{2t U\bar n} . 
\end{gathered}
\label{Uweak}
\end{equation}
The domain of validity of the dilute limit is shown in Fig.~\ref{fig_dilute_limit}. 

In two dimensions, the logarithmic vanishing of $\lamb_k^{(\rm vac)}$ plays a crucial role. One finds 
\begin{equation}
\begin{gathered}
k_h \simeq \left( \frac{4\pi \bar n}{|\ln k_ha|}\right)^{1/2} \simeq \left( \frac{4\pi \bar n}{|\ln \sqrt{\bar n}a|}\right)^{1/2} , \\ 
\lamb_{k_h} \simeq \frac{4\pi t}{|\ln k_ha|} \simeq \frac{4\pi t}{|\ln \sqrt{\bar n}a|} ,
\end{gathered}
\label{khlambh}
\end{equation}
and 
\begin{equation}
\begin{gathered}
\mu+2dt \simeq \frac{4\pi t\bar n}{|\ln \sqrt{\bar n}a|} ,\\
c \simeq t \left( \frac{8\pi\bar n}{|\ln \sqrt{\bar n}a|} \right)^{1/2} ,
\end{gathered}
\end{equation}
in agreement with the results obtained by Schick for a dilute two-dimensional Bose gas.\cite{Schick71,Fisher88,note20} Note in particular that the small parameter $\gamma \sim 1/|\ln \sqrt{\bar n}a|$. When $U/t\ll 1$, the scattering length $a\sim e^{-4\pi t/U}$ is exponentially small [Eq.~(\ref{dilute2})] and one recovers Eqs.~(\ref{Uweak}).
The dilute limit is then simply defined by $k_h\sim \sqrt{(U/t)\bar n}\ll \Lambda$ (Fig.~\ref{fig_dilute_limit}). 

Even in the dilute limit $\gamma\ll 1$ and $k_h\ll \Lambda$, the Bogoliubov theory breaks down at the Ginzburg scale $k_G$ (see the discussion in the Introduction). The latter can be estimated from the one-loop correction to the Bogoliubov approximation. Using the results of Refs.~\onlinecite{Pistolesi04,Dupuis09b} with the bare interaction $U$ replaced by $\lamb_{k_h}$ to take into account fluctuations at momentum scales larger than $k_h$, we obtain 
\begin{equation}
k_G \sim \llbrace 
\begin{array}{lc}
k_h \exp\left(-\const/\sqrt{\bar na^3}\right) & (d=3) , \\ 
\dfrac{k_h}{|\ln \sqrt{\bar n}a|}  & (d=2) .
\end{array}
\right.  
\label{dilute4a}
\end{equation}
For $U\ll t$, Eqs.~(\ref{dilute4a}) become
\begin{equation}
k_G \sim \llbrace 
\begin{array}{lc}
k_h \exp\left[-\const/\sqrt{\bar n}(U/t)^{3/2}\right] & (d=3) , \\ 
\dfrac{U}{t} k_h  & (d=2) .
\end{array}
\right.  
\end{equation}
Thus, in the limit $\gamma\ll 1$, the Bogoliubov approximation remains valid in a large part of the momentum range $0\leq |\q|\ll k_h$ where the spectrum is linear, and breaks down only when $|\q|\ll k_G\ll k_h$. Thermodynamic quantities ($n_{0,k},\rho_{s,k},c_k$, etc.) are nevertheless insensitive to the Ginzburg scale and can be obtained from the Bogoliubov theory.\cite{Pistolesi04,Dupuis09b}

\subsection{RG flows}
\label{subsec_rgflow} 

\begin{figure}
\centerline{\includegraphics[width=6cm,clip]{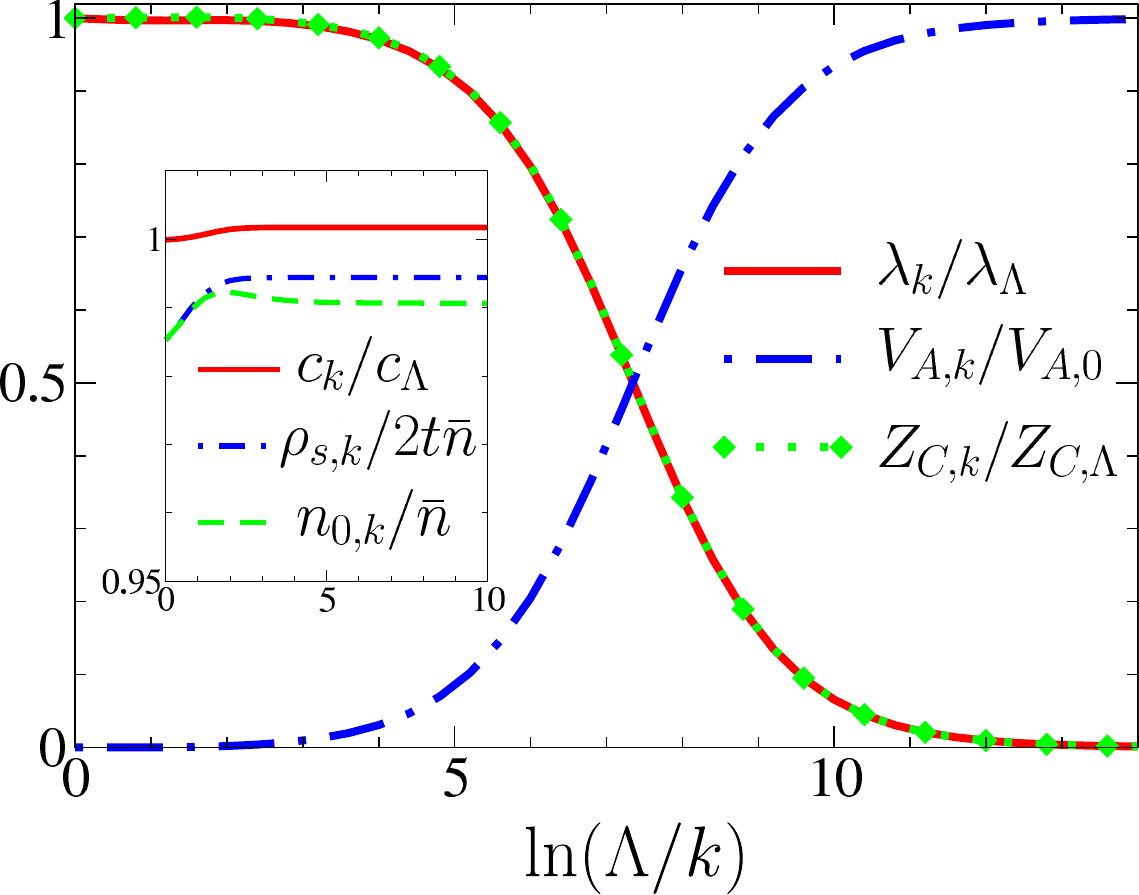}}
\caption{(Color online) RG flow in the superfluid phase, $t/U=10$ and $\bar n=1$ ($d=2$).} 
\label{fig_flowSF1} 
\vspace{0.25cm}
\centerline{\includegraphics[width=6cm,clip]{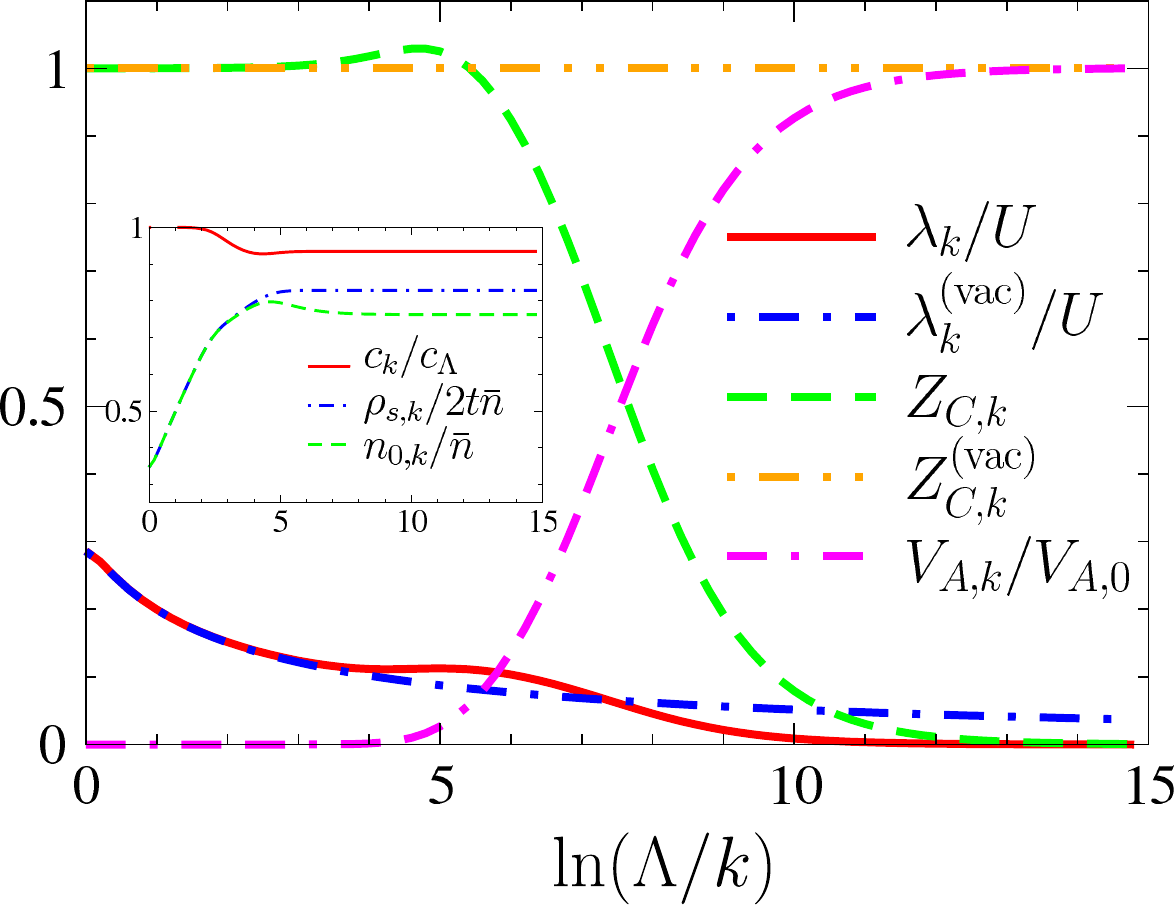}}
\caption{(Color online) Same as Fig.~\ref{fig_flowSF1}, but for $t/U=0.05$ and $\bar n\simeq 10^{-4}$. The RG flow of $\lamb_k^{(\rm vac)}$ and $Z_{C,k}^{(\rm vac)}$ in the vacuum ($\bar n=n_{0,k}=0$) is also shown.} 
\label{fig_flowSF2} 
\vspace{0.25cm}
\centerline{\includegraphics[width=6cm,clip]{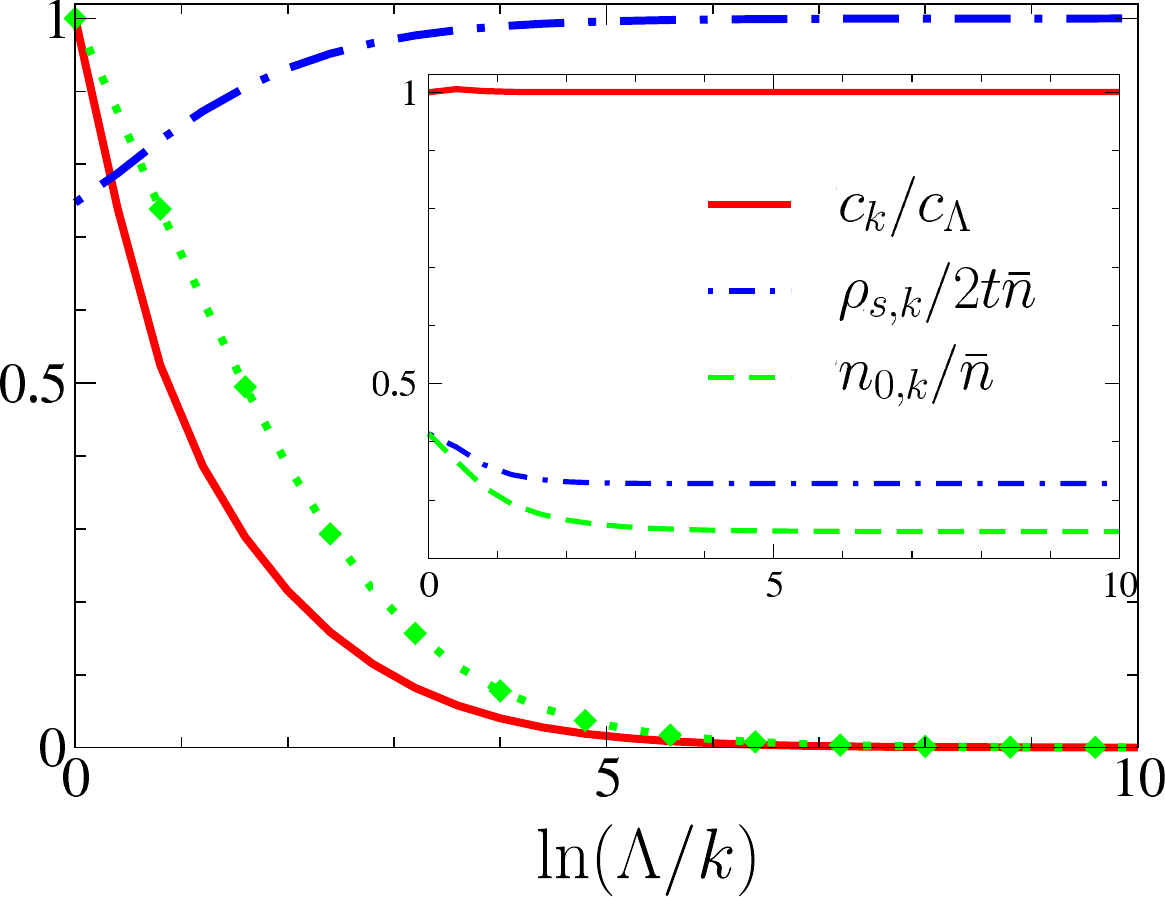}}
\caption{(Color online) Same as Fig.~\ref{fig_flowSF1}, but for $t/U\simeq 0.062$ and $\bar n=1$.} 
\label{fig_flowSF3} 
\end{figure} 

The RG flow is shown in Fig.~\ref{fig_flowSF1} for $t/U=10$ and $\bar n=1$ ($d=2$). For these values of $t/U$ and $\bar n$, the initial condition is well approximated by the Bogoliubov theory, 
\begin{equation}
\begin{gathered}
n_{0,\Lambda}\simeq \bar n,\quad \rho_{s,\Lambda}\simeq 2t\bar n, \quad c_{\Lambda}\simeq (2Ut\bar n)^{1/2}, \\ 
\lambda_{\Lambda}\simeq U,\quad Z_{C,\Lambda}\simeq 1, \quad V_{A,\Lambda}\simeq 0,
\end{gathered}
\label{init}
\end{equation}
and the $T$-matrix renormalization of $\lamb_k$ for $k\gg k_h$ is negligible. The healing scale is deduced from the numerical results and 
\begin{equation}
Z_{A,k_h}tk^2_h = \lamb_{k_h} n_{0,k_h} ,
\label{khdef1}
\end{equation}
which generalizes the definition~(\ref{dilute2a}) to cases where $Z_{A,k_h}$ may differ from unity. The thermodynamic quantities $n_{0,k}$, $\rho_{s,k}$ and $c_k$ vary weakly with $k$ and remain close to their Bogoliubov estimates~(\ref{init}). On the other hand the Ginzburg scale $k_G$ manifests itself by a strong variation with $k$  of $\lamb_k$, $Z_{C,k}$ and $V_{A,k}$. We determine $k_G$ from the inflection point in $V_{A,k}$,\cite{note10} 
\begin{equation}
\partial^2_l V_{A,k}\bigl|_{k=k_G} = 0 .
\label{kgdef}
\end{equation}
Both $k_h$ and $k_G$ [Eqs.~(\ref{khdef1},\ref{kgdef})] are in good agreement with $k_h\sim \sqrt{(U/t)\bar n}$ and $k_G\sim (U/t)k_h$.   

In the (perturbative) Bogoliubov regime $k_G\ll k\leq \Lambda$, $\lambda_k$, $Z_{C,k}$ and $V_{A,k}$ remain nearly equal to their initial values~(\ref{init}). We therefore expect the $k=0$ propagators to read
\begin{equation}
\begin{split}
G_{\rm ll}(q) &= - \frac{\eps_\q}{\w^2+E_\q^2} , \\ 
G_{\rm tt}(q) &= - \frac{\eps_\q+2U\bar n}{\w^2+E_\q^2} , \\ 
G_{\rm lt}(q) &= \frac{\w}{\w^2+E_\q^2} ,
\end{split} 
\end{equation}
for $|\q|\gg k_G$, which is the familiar Bogoliubov form with $E_\q=[\eps_\q(\eps_\q+2U\bar n)]^{1/2}$ the Bogoliubov excitation energy. The spectrum crosses over from a quadratic dispersion to a linear sound-like dispersion at the (healing) momentum scale $k_h$. For $k_G\ll |\q|\ll k_h$, $E_\q\simeq c_\Lambda|\q|$ with $c_\Lambda\simeq (2U\bar nt)^{1/2}$.

In the (nonperturbative) Goldstone regime $k\ll k_G$, $\lambda_k$, $Z_{C,k}\sim k$ vanish with $k\to 0$ while $V_{A,k}\simeq V_{A,k=0}$ takes a finite value. This regime is dominated by phase fluctuations, and characterized by the vanishing of the anomalous self-energy $\Sigma_{\text{an},k}(q=0)=\lambda_kn_{0,k}\sim k$ and the divergence of the longitudinal propagator (see Eq.~\eqref{green1} below).\cite{Nepomnyashchii75,Nepomnyashchii78,Dupuis11} The $k=0$ propagators~\eqref{green} are given by 
\begin{equation}
\begin{split}
G_{\rm ll}(q) &= -\frac{1}{2\lamb n_0} = - \frac{1}{2n_0C \sqrt{\w^2+c^2\q^2}} , \\ 
G_{\rm tt}(q) &= -\frac{1}{V_A(\w^2+c^2\q^2)} = - \frac{2n_0c^2}{\rho_s}\frac{1}{\w^2+c^2\q^2} ,  \\ 
G_{\rm lt}(q) &= \frac{Z_C\w}{2\lamb n_0V_A(\w^2+c^2\q^2)} = \frac{c^2}{\rho_s} \frac{dn_0}{d\mu} \frac{\w}{\w^2+c^2\q^2} ,
\end{split}
\label{green1}
\end{equation}
for $|\q|/|\w|/c\ll k_G$, where we have used $c=(Z_At/V_A)^{1/2}$, $\kappa=2n_0V_A$, $\rho_s=2tZ_A n_0$ and $\lim_{k\to 0}Z_{C,k}/\lamb_k=dn_0/d\mu$ (see Sec.~\ref{subsec_de}). The longitudinal propagator is obtained from $G_{{\rm ll},k}(q=0)=-1/(2\lamb_k n_{0,k})$ by replacing $\lamb_k\sim k$ with $C\sqrt{\w^2+c^2\q^2}$.\cite{Dupuis09b} In the Goldstone regime, the existence of a linear spectrum at low energy is due to the (relativistic) Lorentz invariance of the effective action ($Z_{C,k}\to 0$ while $V_{A,k}\to V_A>0$) and not to the finite value of the anomalous self-energy $\Sigma_{\rm an}(q=0)$ as in the Bogoliubov regime. Quite remarkably however, the value of the sound-mode velocity is insensitive to the Ginzburg scale $k_G$. These results agree with previous studies of interacting bosons in continuum models.\cite{Dupuis09a,Dupuis09b,Sinner09,Sinner10} 

In Fig.~\ref{fig_flowSF2}, we show the RG flow for $t/U=0.05$ and $\bar n\simeq 10^{-4}$. Although $t/U\ll 1$, the very small value of the density ensures that the system is in the dilute limit with $k_G\ll k_h\ll \Lambda$ ($l_h\simeq -3$ and $l_G\simeq -8$).\cite{note16} For $k\gg k_h$, the flow of the coupling constant $\lamb_k$ coincide with the flow in vacuum ($\lamb_k\simeq \lamb_k^{(\rm vac)}$). For $k_G\ll k\ll k_h$, the variation of $\lamb_k$ is weak: in the momentum range $k_G\ll |\q|\ll k_h$, the behavior of the system is well described by the Bogoliubov theory but with renormalized parameters ($\lamb_{k_h}$, $n_{0,k_h}$, etc.). In the Goldstone regime $k\ll k_G$, we recover the infrared behavior discussed above.

As $t/U$ decreases (at fixed density $\bar n$), the dimensionless coupling constant $\gamma$ increases and eventually becomes of order one. A typical flow in the strong-coupling regime $\gamma\gg 1$ is shown in Fig.~\ref{fig_flowSF3} for $t/U\simeq 0.062$ and $\bar n=1$. There is no Bogoliubov regime any more, $k_h\sim k_G\sim \Lambda$, and the condensate density $n_{0,k}$ and the superfluid stiffness $\rho_{s,k}$, are strongly suppressed. 

\begin{table}
\renewcommand{\arraystretch}{1.5}
\begin{center}
\begin{tabular}{lccccc}
\hline \hline
 & $Z_{A,k}$ & $V_{A,k}$ & $Z_{C,k}$ & $\lambda_k$ & $n_{0,k}$
\\ \hline
superfluid & $Z_A^*$ & $V_A^*$ & $k$ & $k$ & $n_0^*$ 
\\ 
multicritical point &  $k^{-\eta}$ & $k^{-\eta}$ & $k$ & $k^{1-2\eta}$ & $k^{1+\eta}$ 
\\ 
generic transition & $Z_A^*$ & $V_A^*$ & $Z_C^*$ & $|\ln k|^{-1}$ & $k^2|\ln k|^{-1}$ 
\\  
insulator & $Z_A^*$ & $V_A^*$ & $Z_C^*$ & $\lambda^*$ & 0 
\\ \hline \hline 
\end{tabular}
\end{center}
\caption{Infrared behavior of the two-dimensional Bose-Hubbard model. The stared quantities indicate nonzero fixed-point values and $\eta$ denotes the anomalous dimension at the three-dimensional $XY$ critical point. $Z_{C,k}$ stands for $Z_{C,k}(n_{0,k})$.}
\label{table_ir}
\end{table}

It is instructive to compare the infrared behavior in the superfluid phase or the Mott insulating phase with the critical behavior at the superfluid--Mott-insulator transition (Table~\ref{table_ir}). Both in the superfluid phase and at the multicritical points, the infrared behavior is characterized by a (relativistic) Lorentz invariance.

\subsection{Characteristic momentum scales: $k_h$ and $k_G$} 

\begin{figure}
\centerline{\includegraphics[width=5.5cm,clip]{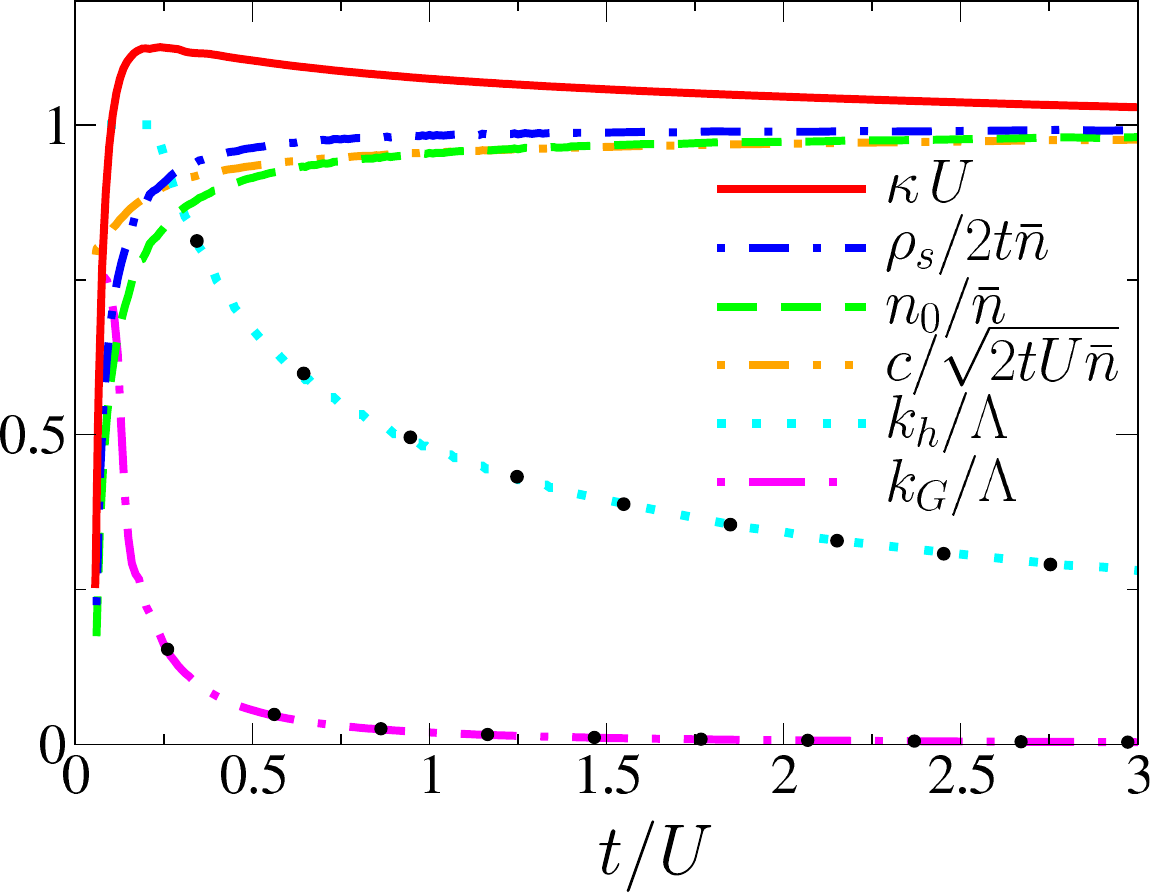}}
\caption{(Color online) Condensate density $n_0$, superfluid stiffness $\rho_s$, compressibility $\kappa$, velocity $c$, and characteristic scales $k_h$ and $k_G$ vs $t/U$ at fixed density $\bar n=1$ ($d=2$). Black dots show fits $k_h\propto \sqrt{\bar n U/t}$ and $k_G\propto\sqrt{\bar n(U/t)^3}$.} 
\label{fig_density} 
\vspace{0.25cm}
\centerline{\includegraphics[width=6cm,clip]{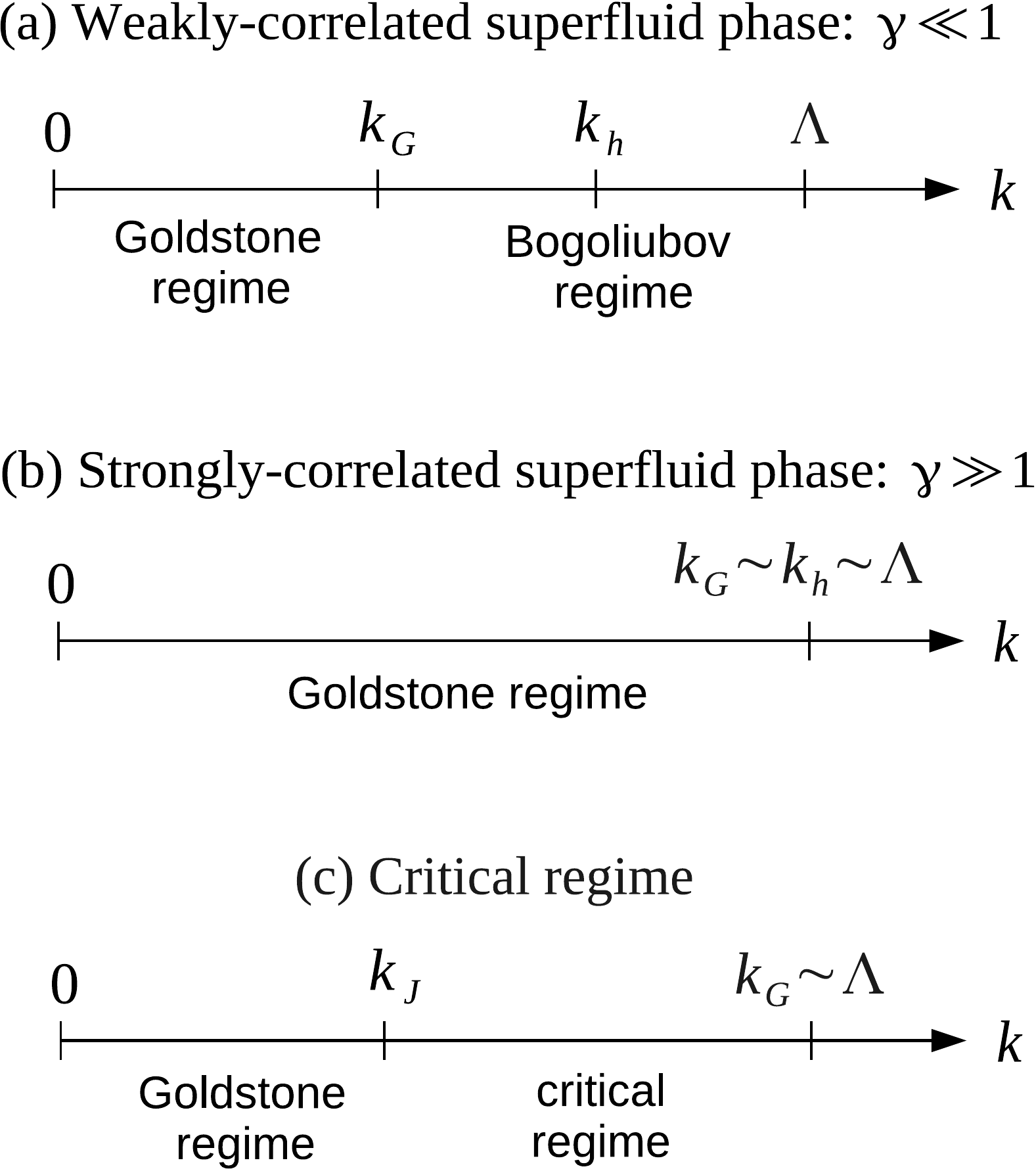}}
\caption{Behavior of the superfluid phase vs momentum scale $k$ at fixed commensurate density $\bar n$. $k_h$: healing scale, $k_G$: Ginzburg scale, $k_J$: Josephson scale.}
\label{fig_carac}
\end{figure}

Figure~\ref{fig_density} shows $k_G$, $k_h$, $n_0$, $\kappa$, $\rho_s$ and $c$ vs $t/U$ at fixed density $\bar n=1$. We see a sharp crossover between a weakly-correlated ($k_G\ll k_h$, $n_0\simeq \bar n$ and $\rho_s\simeq 2t\bar n$) and a strongly-correlated ($k_h\sim k_G\sim\Lambda$, $n_0\ll \bar n$ and $\rho_s\ll 2t\bar n$) superfluid phase as $t/U$ is decreased. Close to the multicritical point, there is a critical regime where the flow of $\tilde n_{0,k}$, $\tilde\lamb_k$, $\eta_{A,k}$ and $\eta_{V,k}$ shows plateaus characteristic of critical behavior (Sec.~\ref{sec_crit}). The critical behavior ends at the Josephson momentum scale $k_J$ (Sec.~\ref{subsec_multicrit}), and for $k\ll k_J$ we recover the Goldstone regime of the superfluid phase. The behavior in the superfluid phase at fixed commensurate density $\bar n$ ($\bar n$ integer) is summarized in Fig.~\ref{fig_carac}. Except for lattice effects (which force $k_h$, $k_G$ and $k_J$ to be at most of order $\Lambda$) we recover the behavior of the $(d+1)$-dimensional O($N$) model.\cite{Dupuis11} 

\begin{figure}
\centerline{\includegraphics[width=5.5cm,clip]{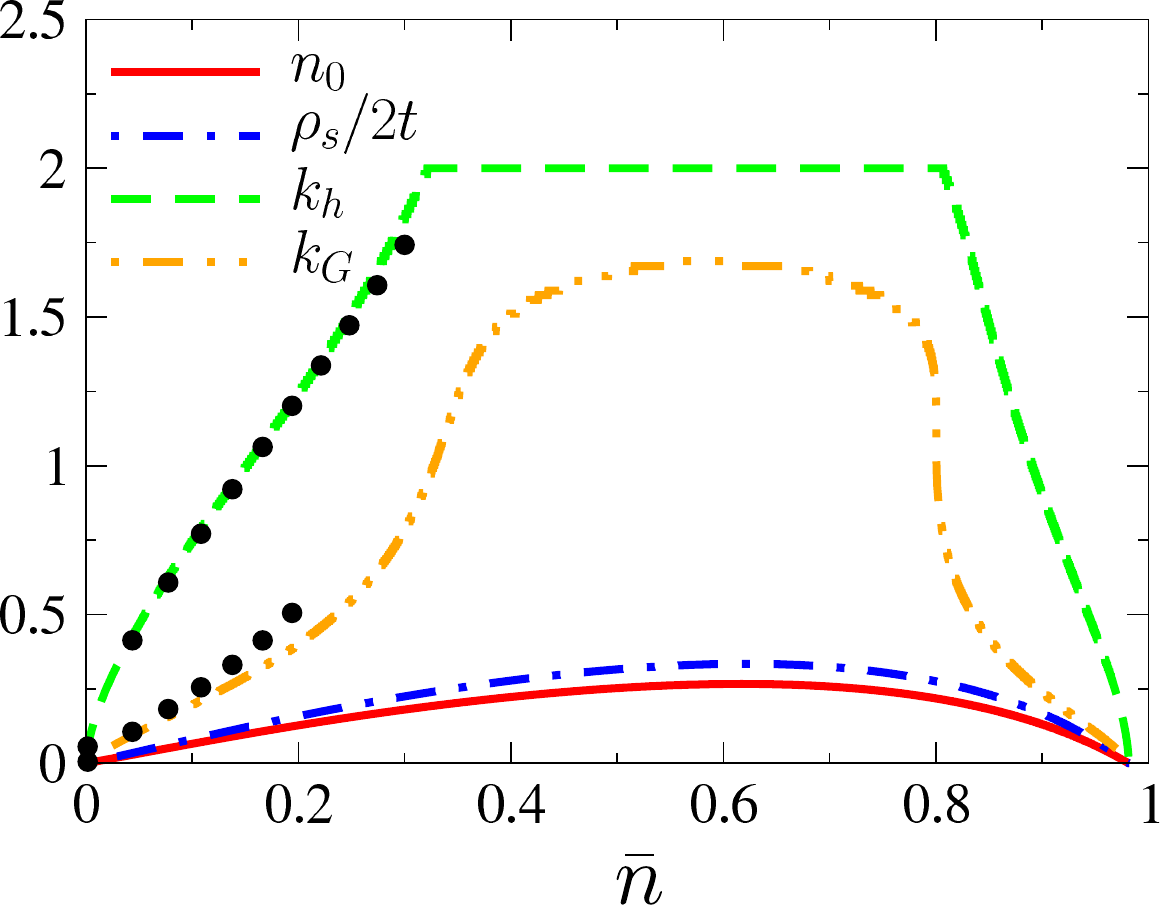}}
\caption{(Color online) Condensate density $n_0$, superfluid stiffness $\rho_s$ and characteristic scales $k_h$ and $k_G$ vs density $\bar n$ for $t/U=0.04$ ($d=2$). Black dots show fits $k_h\propto (\bar n/|\ln \sqrt{\bar na}|)^{1/2}$ and $k_G\propto k_h\lamb_{k_h}/t$ [Eq.~(\ref{dilute4a},\ref{khlambh})].} 
\label{fig_density1}
\end{figure}

Figure~\ref{fig_density1} shows $k_h$ and $k_G$ at fixed $t/U$ for $\bar n$ varying between 0 and 1. In the small density limit $\gamma\ll 1$, our numerical results for $k_h$ and $k_G$ agree with Eqs.~(\ref{dilute2a},\ref{khlambh}). For $\bar n \sim 0.5$, $k_G$ and $k_h$ become of the same order and $\gamma\gg 1$ as expected for a strongly-correlated superfluid phase. The behavior near the Mott insulating phase ($1-\bar n \ll 1$) is similar to the low-density limit and reflects the fact that the transitions from the superfluid phase to the vacuum or the Mott insulating phase $\bar n =1$ belong  to the same universality class. In particular, near the transition to the Mott insulating phase $\bar n=1$, the system is effectively in the weakly correlated limit ($\gamma=\lamb_{k_h}/t\ll 1$) in agreement with the fact that the quantum critical point is Gaussian for $d\geq 2$. The analogy between the limits $\bar n\ll 1$ and $1-\bar n\ll 1$ leads to interesting consequences which will be discussed elsewhere.\cite{note11}

\subsection{Low-energy spectrum}
\label{subsec_spectrum}

The knowledge of the infrared limit of the one-particle Green's function enables us to obtain the spectral function\cite{Dupuis09b}
\begin{align}
A(\q,\w) &= - \frac{1}{\pi} {\rm Im}\,G_{\rm n}(\q,\w+i0^+) \nonumber \\ 
&\simeq -\frac{1}{2\pi} {\rm Im}\left[ G_{\rm ll}(\q,\w+i0^+) +  G_{\rm tt}(\q,\w+i0^+) \right] 
\end{align}
in the low-energy limit. From Eqs.~(\ref{green1}) we deduce 
\begin{align}
A(\q,\w) ={}& \frac{n_0c}{2\rho_s|\q|} \left[\delta(\w-c|\q|)  - \delta(\w+c|\q|) \right] \nonumber \\
& + \frac{\sgn(\w)}{4\pi n_0C} \frac{\Theta(\w-c|\q|)}{\sqrt{\w^2-c^2\q^2}} 
\label{Aspec} 
\end{align}
for $|\q|,|\w|/c\ll k_G$. In addition to the delta peak due to the Goldstone mode, the spectral function exhibits a continuum of excitations which is a direct consequence of the singularity of the longitudinal propagator $G_{\rm ll}$. While the sound mode extends up to $|\q|\sim k_h$, the continuum is observed only at momenta and energies $|\q|,|\w|/c\lesssim k_G$. In the weak-coupling limit, where the lattice does not play an important role, these results are in complete agreement with Popov's hydrodynamic theory.\cite{Popov79,Dupuis11} The latter gives $C=(4t\bar n/cn_0)^{1/2}$ so that the ratio of spectral weights carried by the continuum and the sound mode is extremely small in the weak-coupling limit.\cite{Dupuis09b,[{Note that we have ignored the Beliaev damping\cite{Beliaev58a,Beliaev58b} which gives a life-time of order $|\q|^{-3}$ to the sound mode in two dimensions (see, e.g., }] [{). This damping makes the square-root singularity near $\w=c|\q|$ less visible but the continuum of excitations due to the singular longitudinal propagator remains visible away from $\w=c|\q|$, in particular in the strongly-correlated superfluid phase (see text).}] Kreisel08}

\begin{figure}
\centerline{\includegraphics[width=6cm,clip]{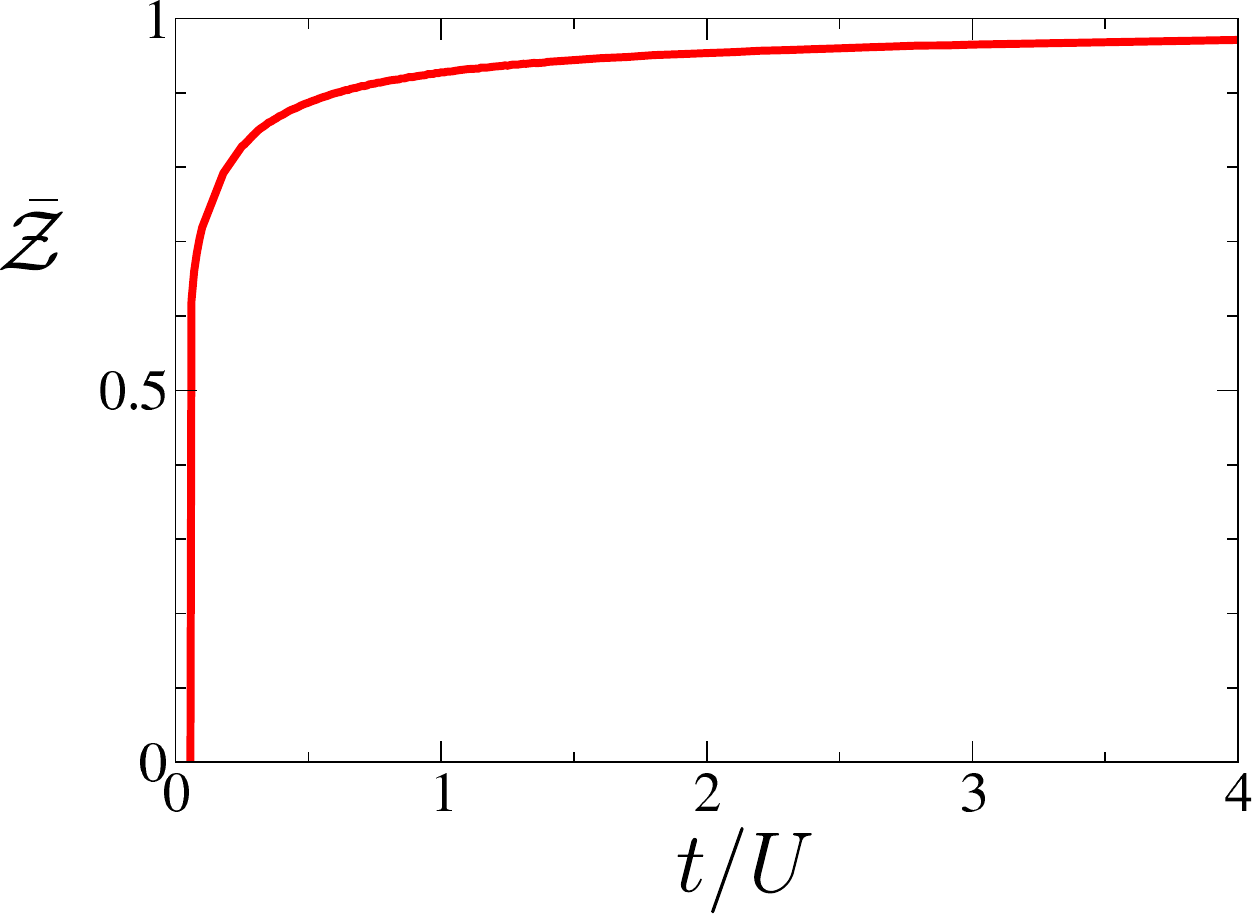}}
\caption{(Color online) Normalized spectral weight $\bar\calZ$ of the sound mode [Eq.~(\ref{weight})] vs $t/U$ ($\bar n=1$ and $d=2$).} 
\label{fig_spectral}
\end{figure}

It should be noted that there is no qualitative difference between the weakly- and strongly-correlated superfluid phases regarding the low-energy single-particle spectrum [Eq.~(\ref{Aspec})]. In the strong-coupling regime however, the continuum of excitations due to the singular longitudinal propagator is expected to extend up to momenta of order $k_G\sim\Lambda$ (i.e. over most part of the Brillouin zone) and carry a significant fraction of spectral weight. This expectation is confirmed by the suppression of spectral weight of the sound mode as the ratio $t/U$ is decreased at fixed density. Figure~\ref{fig_spectral} shows the spectral weight $\calZ_\q=n_0c/2\rho_s|\q|$ of the sound mode for $\bar n=1$, normalized by its value in the weakly-correlated limit $t\gg U$,  
\begin{equation}
\bar \calZ = 2|\q|\calZ_\q \sqrt{\frac{2t}{U\bar n}} = \frac{n_0c}{\rho_s} \sqrt{\frac{2t}{U\bar n}} . 
\label{weight} 
\end{equation}
$\bar\calZ$ remains close to one in the weakly-correlated superfluid phase but is strongly suppressed in the strongly-correlated regime. It vanishes at the transition to the Mott insulating phase ($t=t_c$) with a critical exponent $2\beta-\nu(d+z-2)=\nu\eta$,
\begin{equation}
\bar\calZ \sim (t-t_c)^{\nu\eta} 
\end{equation}
for $t\to t_c^+$. 

Being equivalent to the strong-coupling RPA, the initial effective action $\Gamma_\Lambda$ predicts the existence of a gapped mode in addition to  the sound mode.\cite{note12,Sengupta05,Ohashi06,Menotti08,Huber07} Whether this gapped mode is a true characteristic of the spectrum (which would then show up in the propagator $G_{k=0}(q;n_0)$) is an interesting question which however requires a more refined NPRG analysis~\cite{Dupuis09a,Dupuis09b,Sinner09,Sinner10} beyond the derivative expansion [Eq.~(\ref{gamde})].\cite{note17}

\subsection{The 3D superfluid phase} 
\label{subsec_3dsf}

\begin{figure}
\centerline{\includegraphics[width=5.5cm,clip]{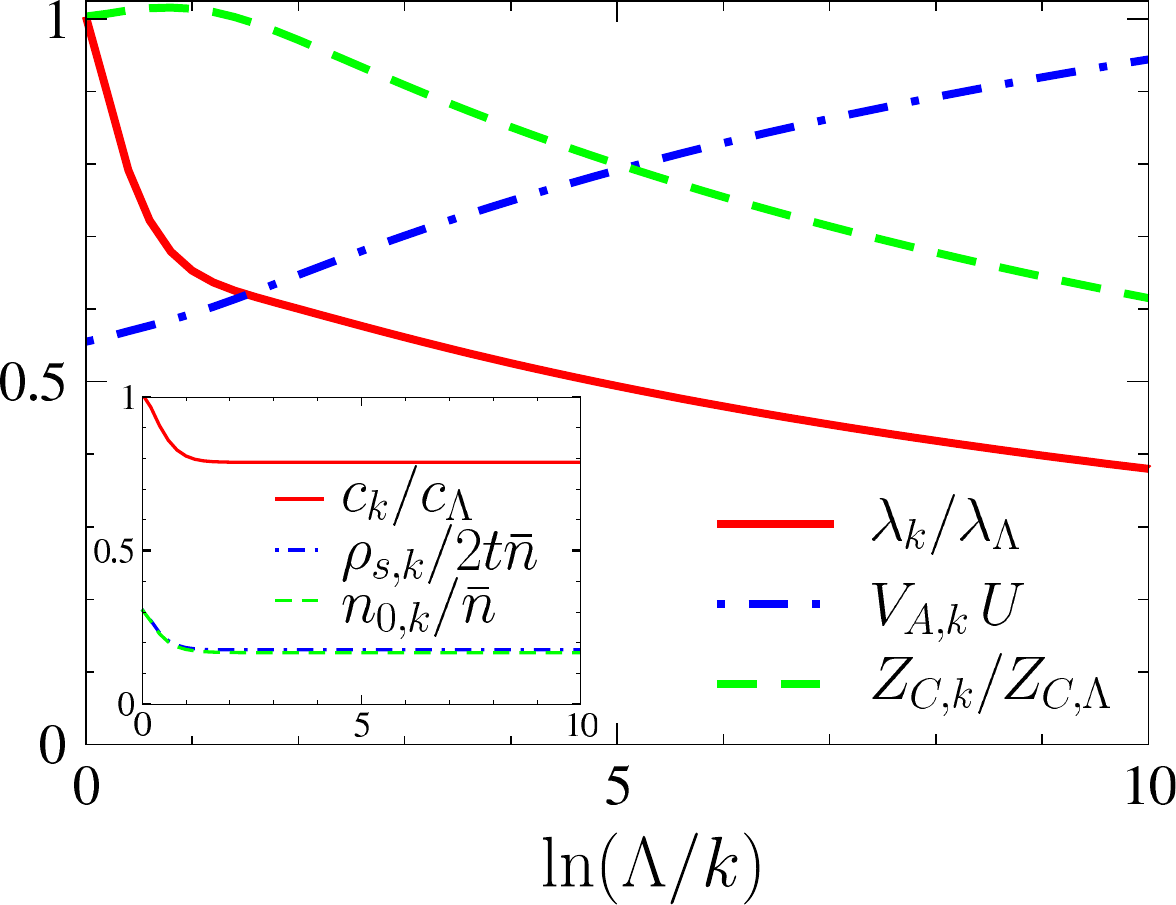}}
\caption{(Color online) RG flow in the three-dimensional superfluid phase for $t/U=0.034$ and $\bar n=1$.} 
\label{fig_sf3D}
\end{figure}

In the dilute limit $\gamma\ll 1$ (Sec.~\ref{subsec_dilute}), the initial conditions of the RG flow are given by 
\begin{equation}
\begin{gathered}
n_{0,\Lambda} \simeq \bar n, \quad \rho_{s,\Lambda} \simeq 2t\bar n , \quad c_\Lambda \simeq 2t \sqrt{4\pi a \bar n} , \\
\lamb_\Lambda \simeq 8\pi at, \quad Z_{C,\Lambda}(n) \simeq 1, \quad V_{A,\Lambda}(n) \simeq 0 ,
\end{gathered}
\end{equation}
and reproduce the Bogoliubov approximation. The flow for $k\leq \Lambda$ is logarithmic and therefore very slow. This explains why the Ginzburg scale is exponentially small in the dilute limit [Eq.~(\ref{dilute4a})] and irrelevant for most purposes. In the infrared limit, $G_{\rm tt}$ and $G_{lt}$ are given by~(\ref{green1}), while the longitudinal propagator  
\begin{equation}
G_{\rm ll}(q) \sim \ln \left(\frac{ck_h}{\sqrt{\w^2+c^2\q^2}}\right) 
\end{equation}
diverges logarithmically.

A typical RG flow in the strong-coupling limit, where $k_h,k_G\sim \Lambda$, is shown in Fig.~\ref{fig_sf3D} for $\bar n=1$. Thermodynamic quantities very rapidly converge to their $k=0$ values. On the other hand, the flow of $\lamb_k,Z_{C,k}(n_{0,k})\sim |\ln k|^{-1}$ and $V_{A,k}$ is logarithmic.

\section{Summary and conclusion}
\label{sec_conclu}

We have presented a detailed NPRG study of the Bose-Hubbard model. Although we have only considered the zero-temperature limit, it is straightforward to extend the analysis to finite temperatures. The lattice NPRG seems to be the only available technique which treats fluctuations at all length scales on equal footing: 

\begin{itemize}

\item the lattice NPRG takes into account on-site correlations which are responsible for the very existence of the superfluid--Mott-insulator transition. It is exact in the local limit (vanishing hopping amplitude), the latter corresponding to the initial condition of the NPRG at the microscopic scale $\Lambda=\sqrt{2d}$. 

\item the lattice NPRG also takes into account critical fluctuations at the superfluid--Mott-insulator transition. In this respect it is very similar to the standard implementation of the NPRG in continuum models, the cutoff function $R_k(\q)$ playing the role of an infrared regulator. 

\item as already known from previous studies in continuum models,\cite{Wetterich08,Dupuis07,Dupuis09a,Dupuis09b,Sinner09,Sinner10} the NPRG is a method of choice to study the superfluid phase. It is free of infrared divergences, satisfies the Hugenholtz-Pines theorem, and is able to describe both the (perturbative) Bogoliubov regime $k_G\ll k\leq\Lambda$ and the (nonperturbative) Goldstone regime $k\ll k_G$. The latter is characterized by a vanishing anomalous self-energy and a diverging longitudinal propagator. In the strong-coupling limit where $k_G\sim k_h\sim \Lambda$, there is no Bogoliubov regime and the whole RG flow becomes nonperturbative. 

\end{itemize}

Our results agree with known results on the Bose-Hubbard model. In particular we reproduce the phase diagram obtained from QMC calculations with a typical accuracy of 1-3~\% and at a very modest numerical cost.\cite{note18} Moreover, we recover the two universality classes of the superfluid--Mott-insulator transition. The lattice NPRG enables a detailed study of the critical behavior near multicritical or generic transition points, which confirms the original predictions of Fisher {\it et al.}\cite{Fisher89} based on scaling arguments.

\begin{acknowledgments}
We would like to thank B. Delamotte for useful discussions, B. Capogrosso-Sansone for providing
us with the QMC data shown in Figs.~\ref{fig_dia3D} and \ref{fig_dia2D}, and P. Anders for the DMFT data shown in Fig.~\ref{fig_dia3D}. 
\end{acknowledgments} 

\appendix

\section{Effective potential in the local limit}
\label{app_local} 

\begin{figure}
\centerline{\includegraphics[width=5cm]{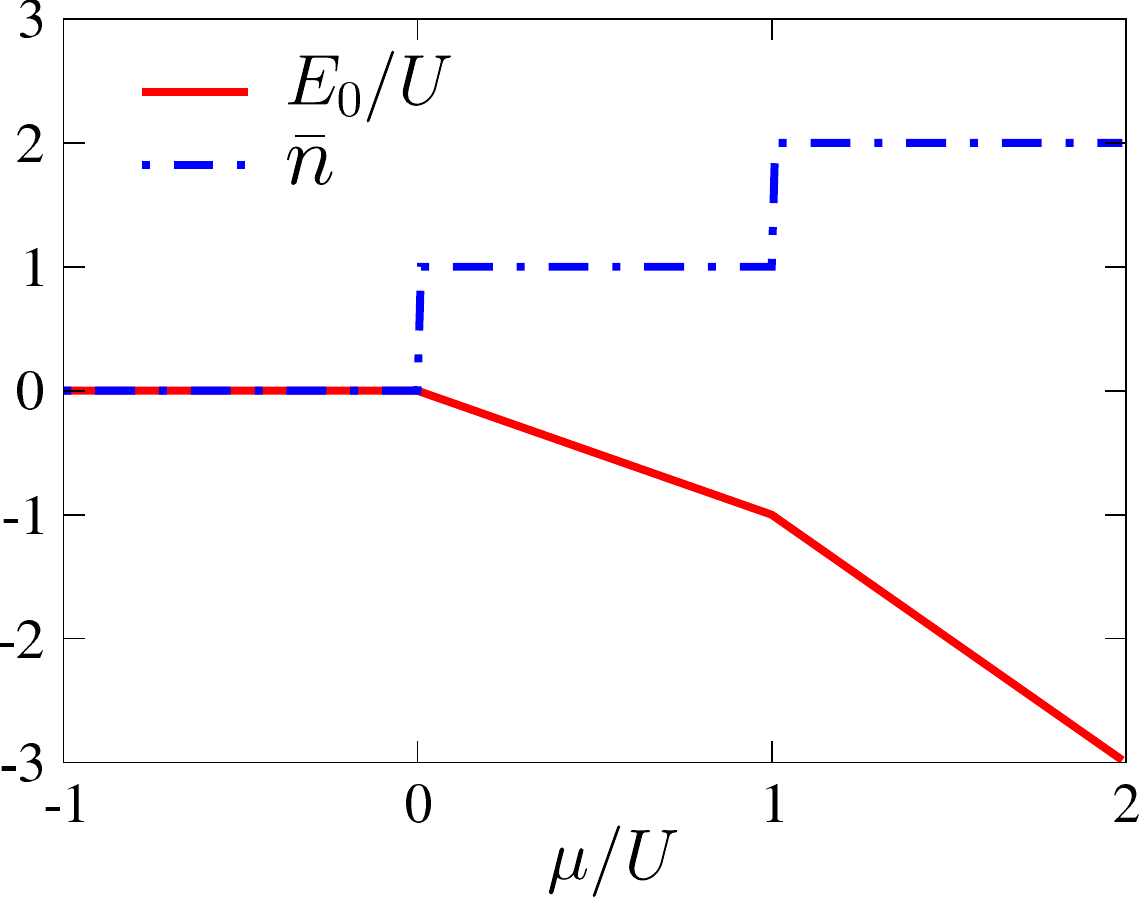}}
\caption{(Color online) Ground state energy $E_0$ and occupation number $\bar n$ vs $\mu/U$ in the local limit (with vanishing external source).}
\label{fig_local}
\end{figure}

In this appendix, we discuss the solution of the local Hamiltonian~(\ref{Hloc}) when the ground state is degenerate for vanishing external sources ($\mu/U$ integer). This degeneracy has important consequences for the effective potential $\Vloc(n)$. 

When $J^*=J=0$, the Hamiltonian is diagonal in the basis $\lbrace \ket{m}\rbrace$ (Sec.~\ref{subsec_init}). The ground state is the vacuum state $\ket{0}$ for $\mu<0$, and $\ket{m}$ for $m<\mu/U<m+1$. The ground state energy $E_0$ and occupation number $\bar n$ are shown in Fig.~\ref{fig_local} as a function of $\mu/U$. 
There is a quantum phase transition whenever $\mu/U$ is integer due to a level crossing. When $\mu/U=m$ ($m$ integer), the states $\ket{m}$ and $\ket{m+1}$ are degenerate. 

For an infinitesimal external source and $\mu/U=m$ ($m$ integer), it is sufficient to consider the degenerate states $\ket{m}$ and $\ket{m+1}$ to determine the ground state of the Hamiltonian~(\ref{Hloc}). In this subspace, 
\begin{equation}
\hat H \equiv \left( 
\begin{array}{lr}
\eps_m & -J^*\sqrt{m+1} \\ 
-J \sqrt{m+1} & \eps_{m+1} 
\end{array}
\right) , 
\label{app6} 
\end{equation}
with $\eps_m=\eps_{m+1}=-\frac{U}{2}m(m+1)$. Diagonalizing~(\ref{app6}), we find the two states 
\begin{equation}
\begin{split}
\ket{-} &= \frac{1}{\sqrt{2}} \left( \ket{m} + e^{i\theta} \ket{m+1} \right) , \\ 
\ket{+} &= \frac{1}{\sqrt{2}} \left( \ket{m} - e^{i\theta} \ket{m+1} \right) ,
\end{split}
\end{equation}
with eigenvalues 
\begin{equation}
\begin{split}
E_- &= \eps_m -|J|\sqrt{m+1} , \\ 
E_+ &= \eps_m +|J|\sqrt{m+1} , 
\end{split}
\end{equation}
where $\theta$ denotes the phase of the complex source $J=|J|e^{i\theta}$. The occupation number in the ground state $\ket{-}$ is 
\begin{equation}
\bra{-}\hat n\ket{-} = m + \half, 
\end{equation}
while the superfluid order parameter 
\begin{equation}
\bra{-}\hat b\ket{-} = \frac{e^{i\theta}}{2} \sqrt{m+1} 
\label{app7}
\end{equation}
is finite. We conclude that the U(1) symmetry is spontaneously broken whenever $\mu/U$ is integer, 
\begin{equation}
\lim_{|J|\to 0^+} \mean{\hat b} \neq 0  
\end{equation} 
although $\mean{\hat b}=0$ for $|J|=0$. 

\begin{figure}
\centerline{\includegraphics[width=6cm]{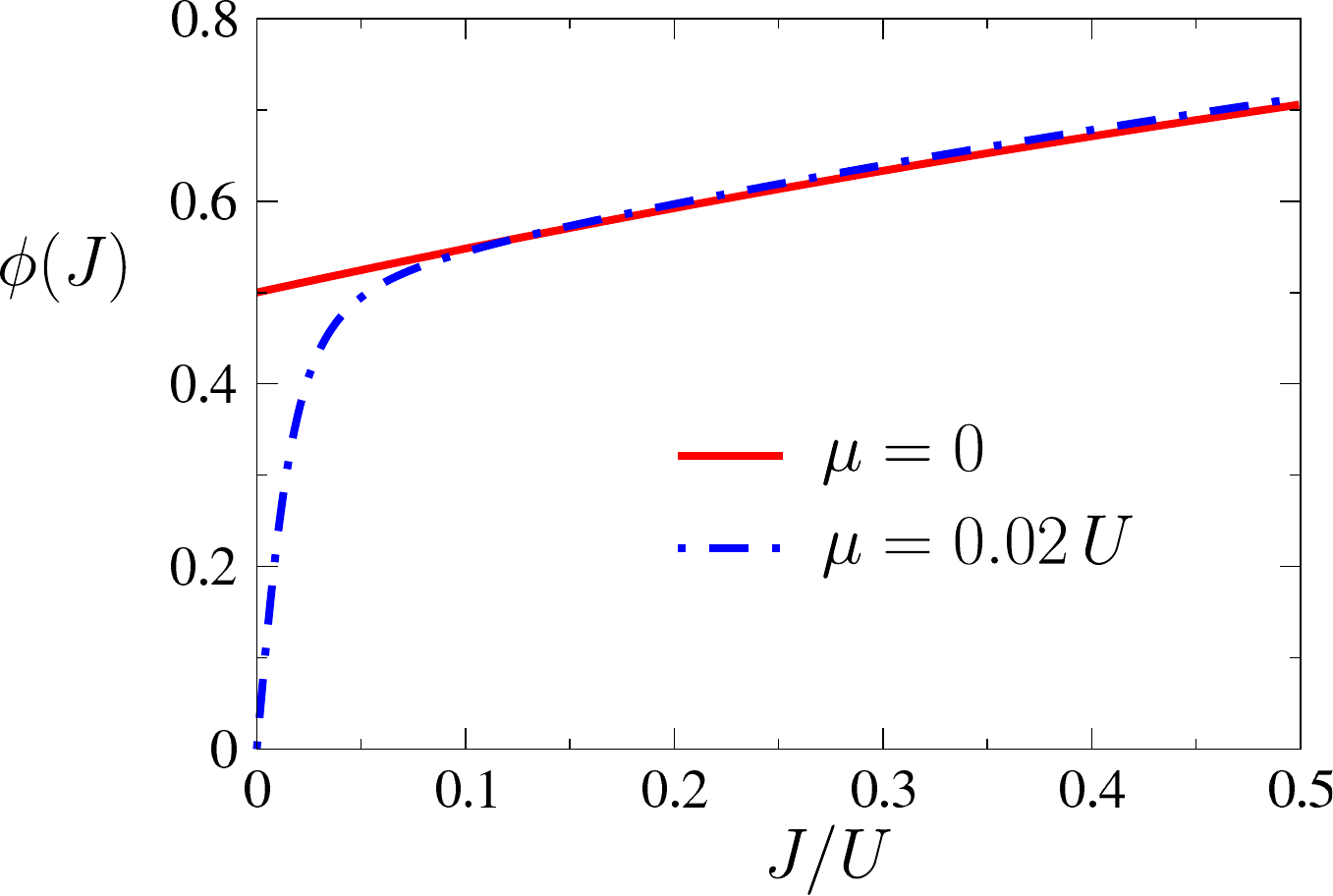}}
\centerline{\includegraphics[width=6cm]{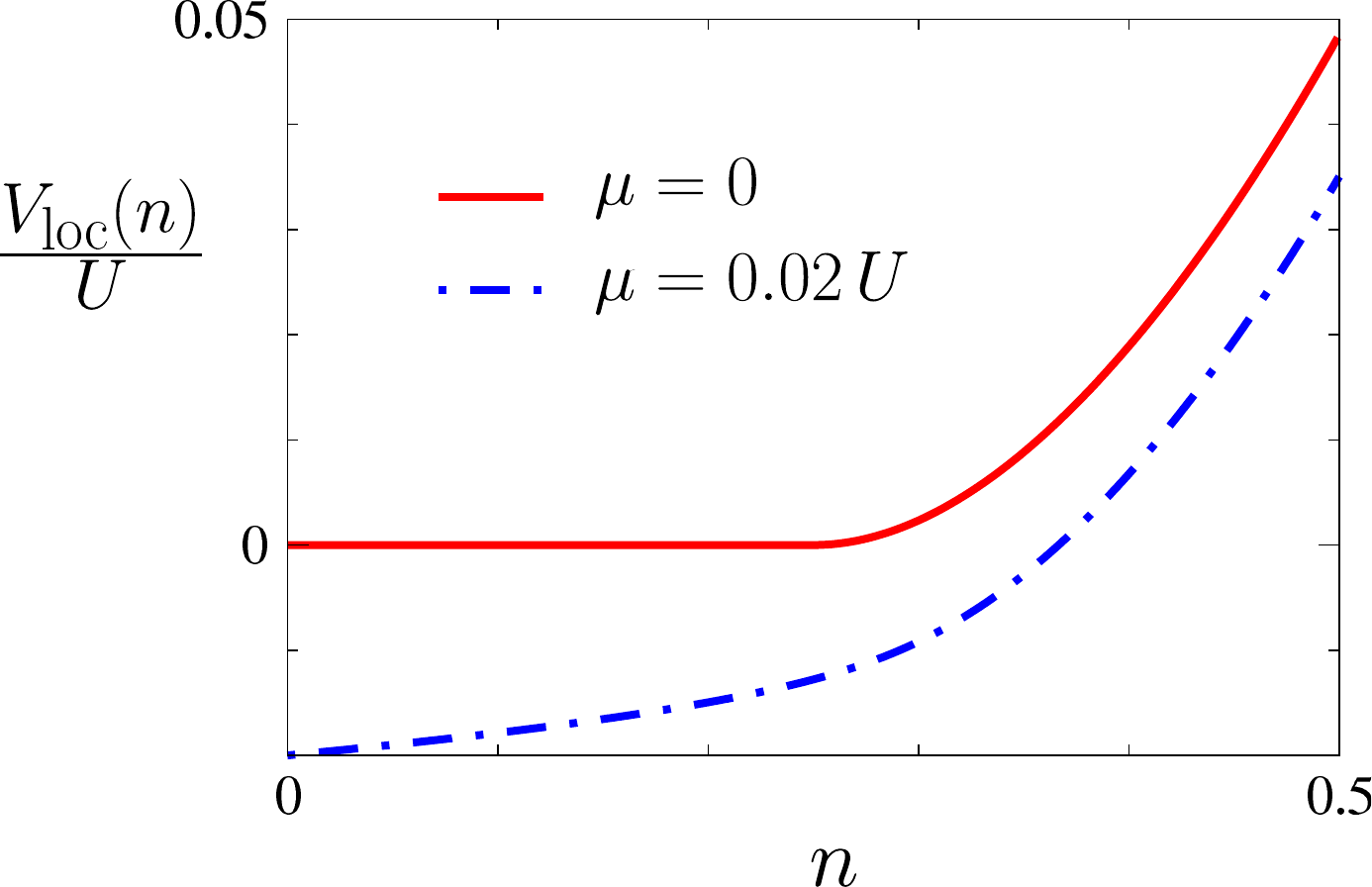}}
\caption{(Color online) Superfluid order parameter $\phi(J)=\mean{\hat b}$ (top) and effective potential $\Vloc(n)$ (bottom) in the local limit. The source $J$ is taken real.} 
\label{fig_Vloc_degenerate}
\end{figure}

Figure~\ref{fig_Vloc_degenerate} shows the superfluid order parameter $\phi(J)$ obtained from the numerical diagonalization of the full Hamiltonian~(\ref{Hloc}). For $\mu=0$, we find $|\phi(J=0^+)|=1/2$ in agreement with Eq.~(\ref{app7}). The effective potential $\Vloc(n)$ takes the usual form in a system with a spontaneous broken symmetry, with a flat part ($\Vloc'(n)=0$) for $n\leq 1/4$. By contrast, for $\mu/U=0.02$, the superfluid order parameter $\phi(J=0^+)$ vanishes and $\Vloc'(n)>0$ for all values of $n$.

\section{Large-field limit of the local effective potential $\Vloc(n)$}
\label{app_largefield} 

\begin{figure}
\centerline{\includegraphics[width=6cm,clip]{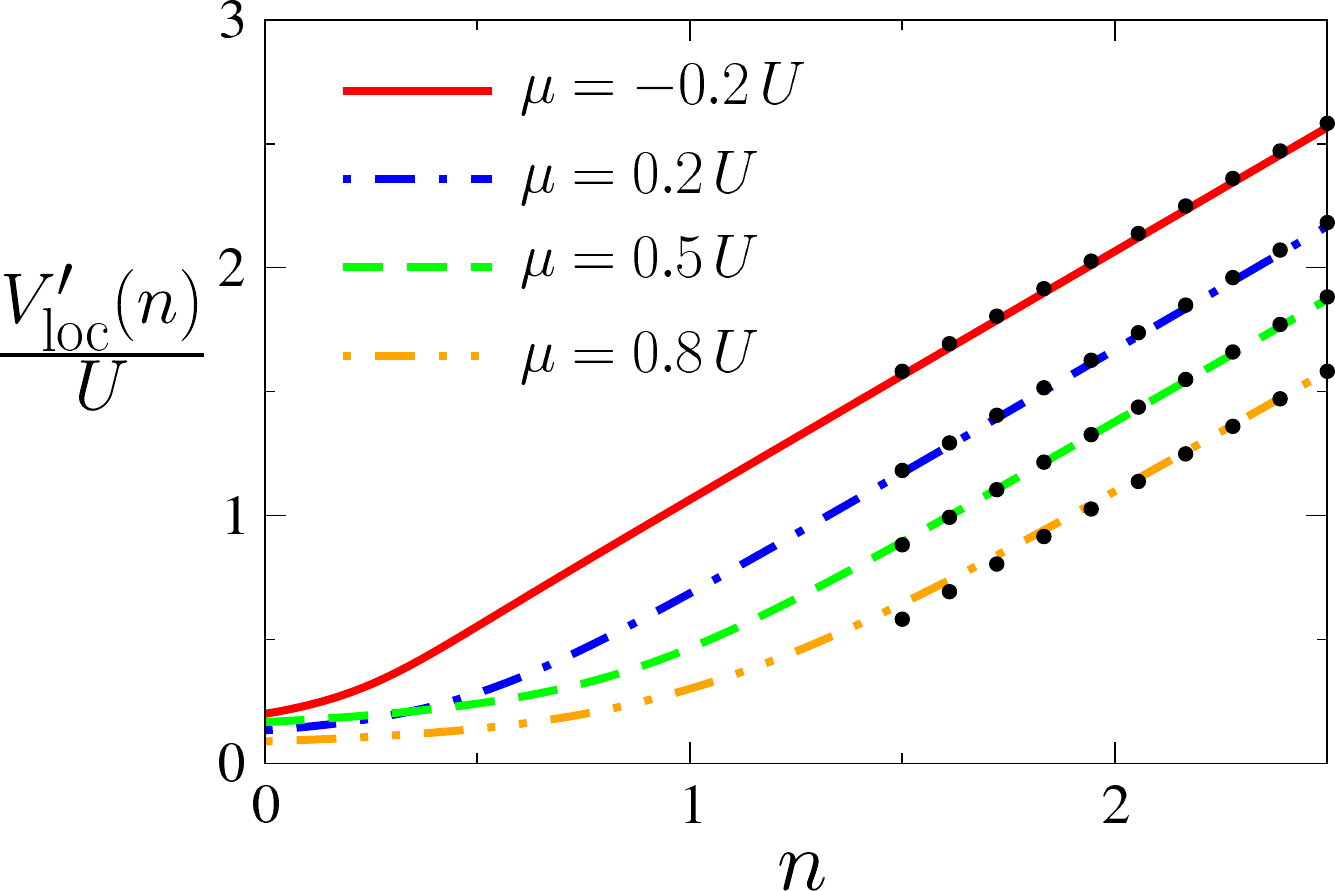}}
\caption{(Color online) Derivative $\Vloc'(n)$ of the local effective potential for various values of $\mu$. The dotted lines show the large-field limit~(\ref{app21}).}
\label{fig_pot_der}
\end{figure}

The large-field limit of the effective action $\Gamma_{\rm loc}[\phi^*,\phi]$ is obtained by considering the partition function 
\begin{equation}
\Zloc[J^*,J] = \int\calD[\psi^*,\psi] e^{-\Sloc[\psi^*,\psi]+\inttau (J^*\psi+J\psi)} 
\end{equation}
for $|J|\to\infty$. In this limit, we expect the field to weakly fluctuate about its saddle-point value $\psi_c$ defined by 
\begin{equation}
\frac{\delta \Sloc}{\delta\psi(\tau)}\biggl|_{\psi_c} = J^*(\tau), \quad 
\frac{\delta \Sloc}{\delta\psi^*(\tau)}\biggl|_{\psi_c} = J(\tau) .
\end{equation}
Let us compute the effective action by including Gaussian fluctuations about the saddle-point solution $\psi_c$ (one-loop order). The calculation is standard and gives\cite{[{See, e.g., }] Zinn_book} 
\begin{equation}
\Gamma_{\rm loc}[\phi^*,\phi] = \Sloc[\phi^*,\phi] + \half \Tr\ln \calG^{-1}_c[\phi^*,\phi] , 
\label{app20}
\end{equation}
where
\begin{multline}
\calG_c^{-1}[\tau,\tau';\phi^*,\phi] = -\delta(\tau-\tau') \\ \times \left(
\begin{array}{cc} 
\partial_{\tau'}-\mu+2U|\phi(\tau)|^2 & U \phi(\tau)^2 \\
U \phi^*(\tau)^2 & -\partial_{\tau'}-\mu+2U|\phi(\tau)|^2
\end{array}
\right) 
\end{multline}
is the inverse classical (local) propagator. By performing the trace in~(\ref{app20}) for a time-independent field $\phi$,\cite{[{See, for instance, Appendix G in }] Diener08}  we easily obtain the effective potential
\begin{align}
\Vloc(n) ={}& -\mu n + \frac{U}{2} n^2 \nonumber \\ & + \half \Bigl\lbrace \bigl[(\mu-2Un)^2-U^2 n^2 \bigr]^{1/2}+\mu - 2Un \Bigr\rbrace \nonumber \\ 
={}& - \bar \mu n + \frac{U}{2} n^2 + \calO(n^0) , 
\label{app21}
\end{align}
where 
\begin{equation}
\bar\mu = \mu + U\left(1-\frac{\sqrt{3}}{2}\right) . 
\end{equation}
To one-loop order, the effective potential is given by the microscopic action $\Sloc$ with a shift of the chemical potential. It is straightforward to verify that higher-order contributions (e.g. those coming from two-loop diagrams) are at most of order $\calO(n^0)$ in the large-field limit. Equation~(\ref{app21}) is in very good agreement with the numerical calculation of $\Vloc(n)$ (Fig.~\ref{fig_pot_der}).

\section{Derivative expansion of the local vertex $\Gamma^{(2)}_{\rm loc}$}
\label{app_de}

\begin{figure}
\centerline{\includegraphics[width=4cm,clip]{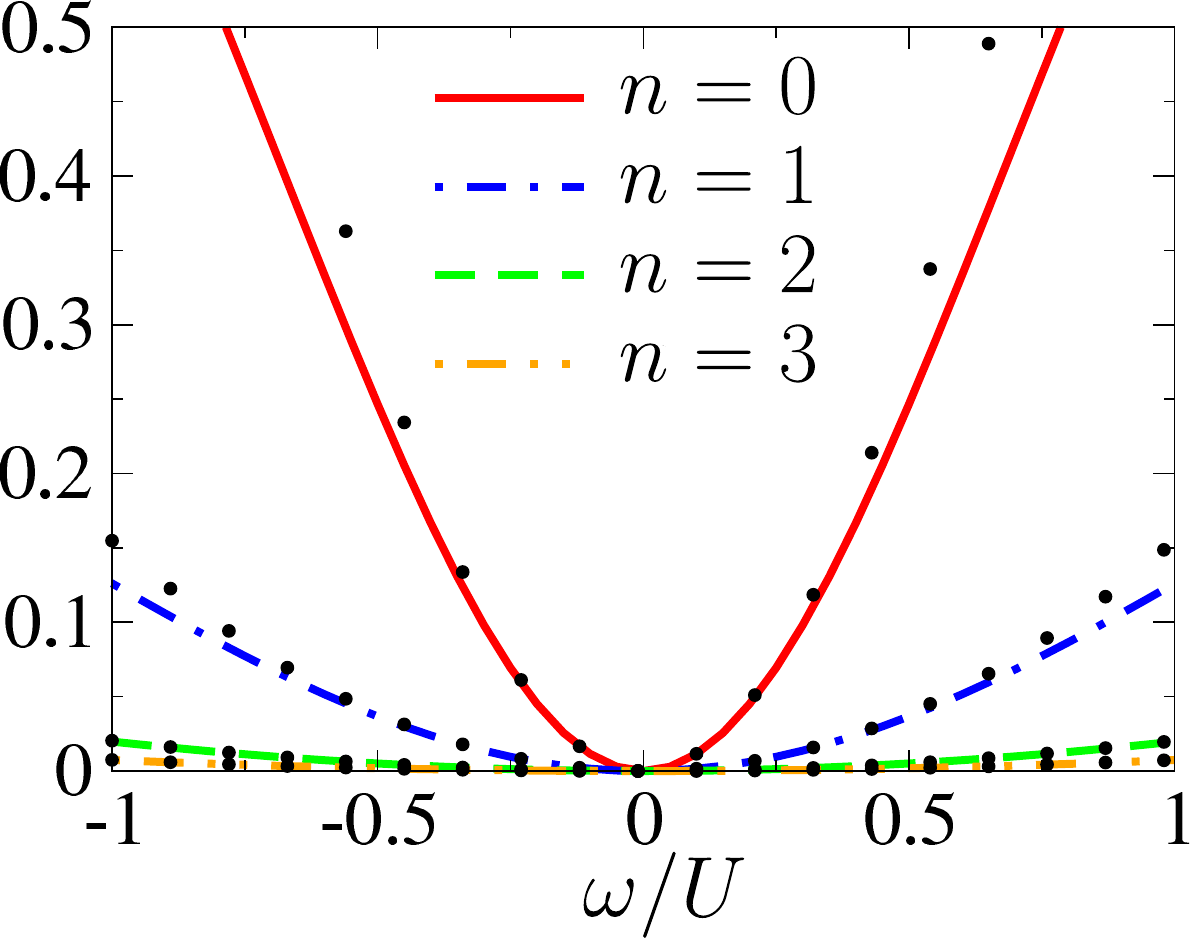}
\includegraphics[width=4cm,clip]{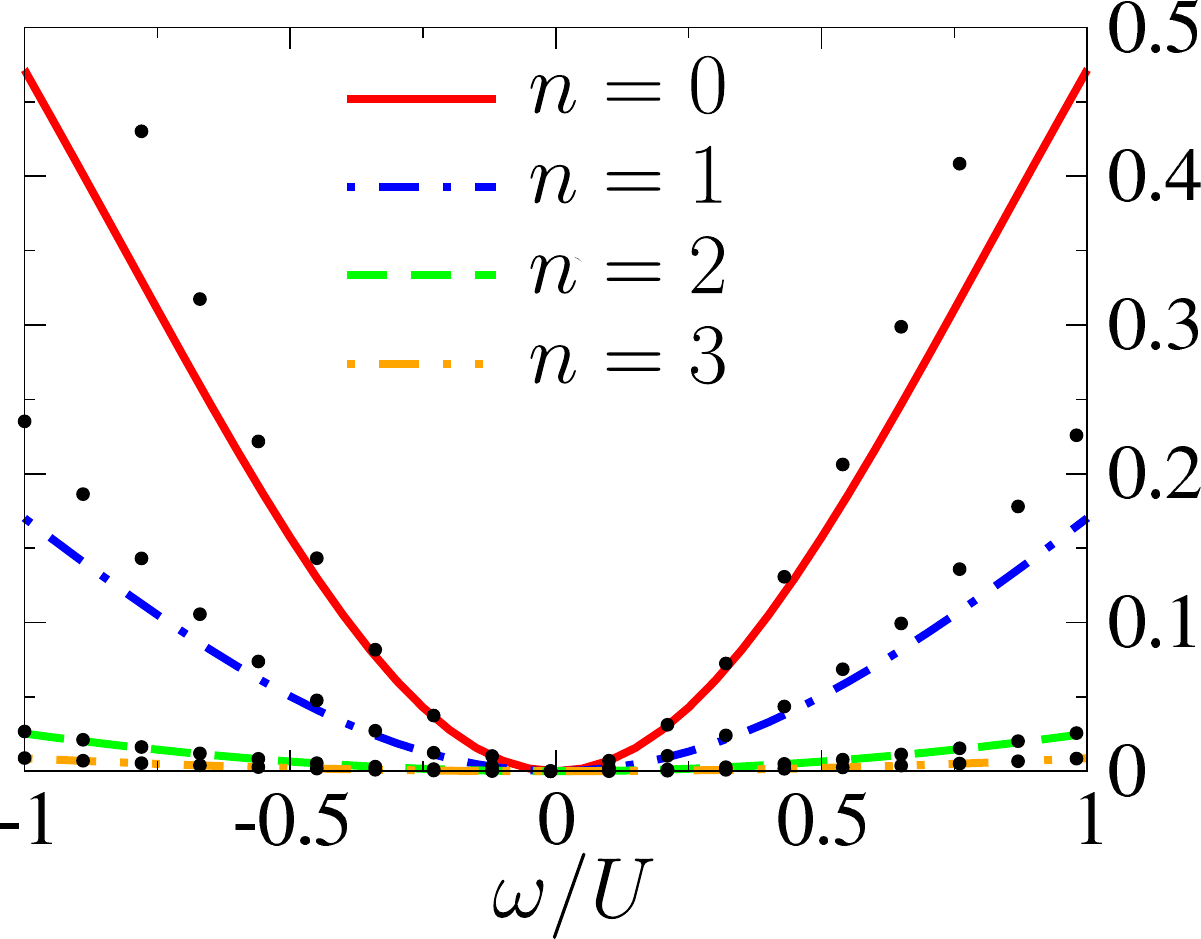}}
\caption{(Color online) $[\Gamma_{{\rm loc},A}(i\w;n)-V'_{\rm loc}(n)]/U$ vs $\w/U$ for various values of $n$. $\mu=0.2U$ (left) and $\mu=(\sqrt{2}-1)U$ (right). The dotted lines show the derivative expansion $V_A(n)\w^2$.}
\label{fig_deltaA_de}
\vspace{0.25cm}
\centerline{\includegraphics[width=4.cm,clip]{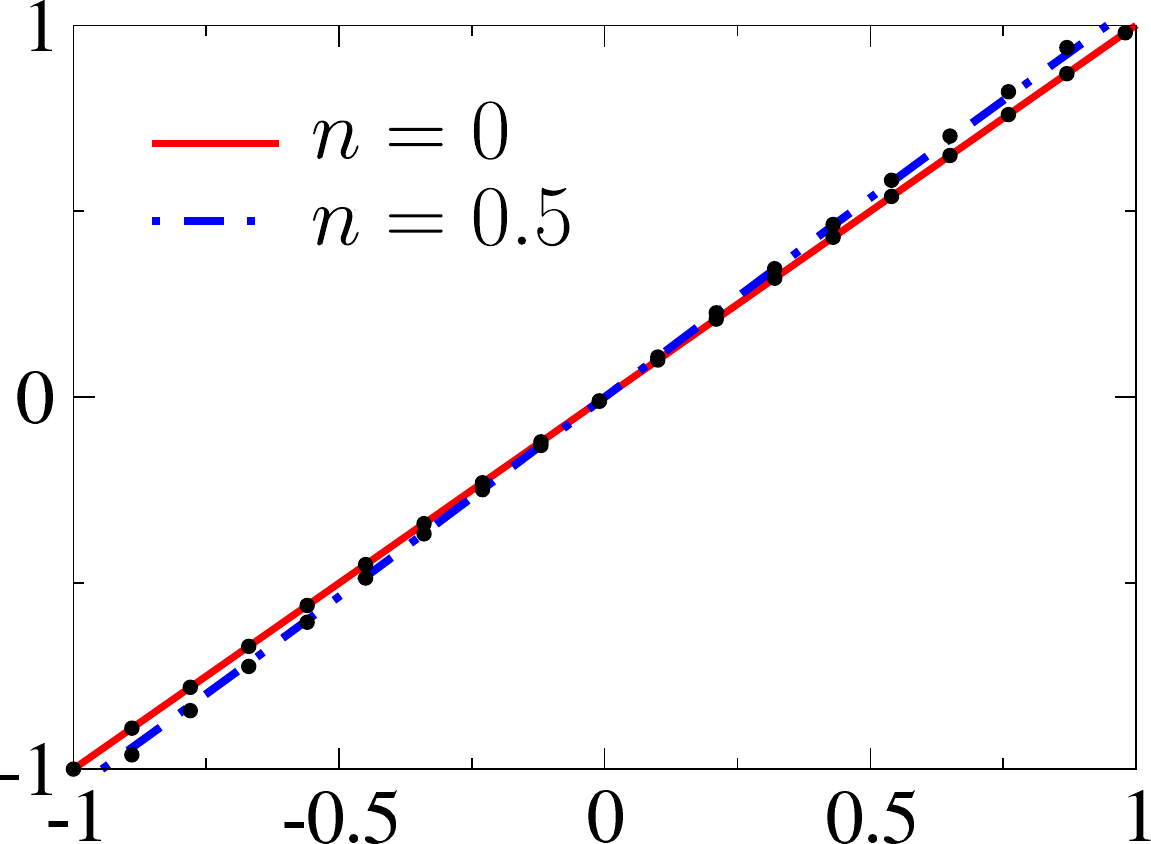}
\includegraphics[width=4.cm,clip]{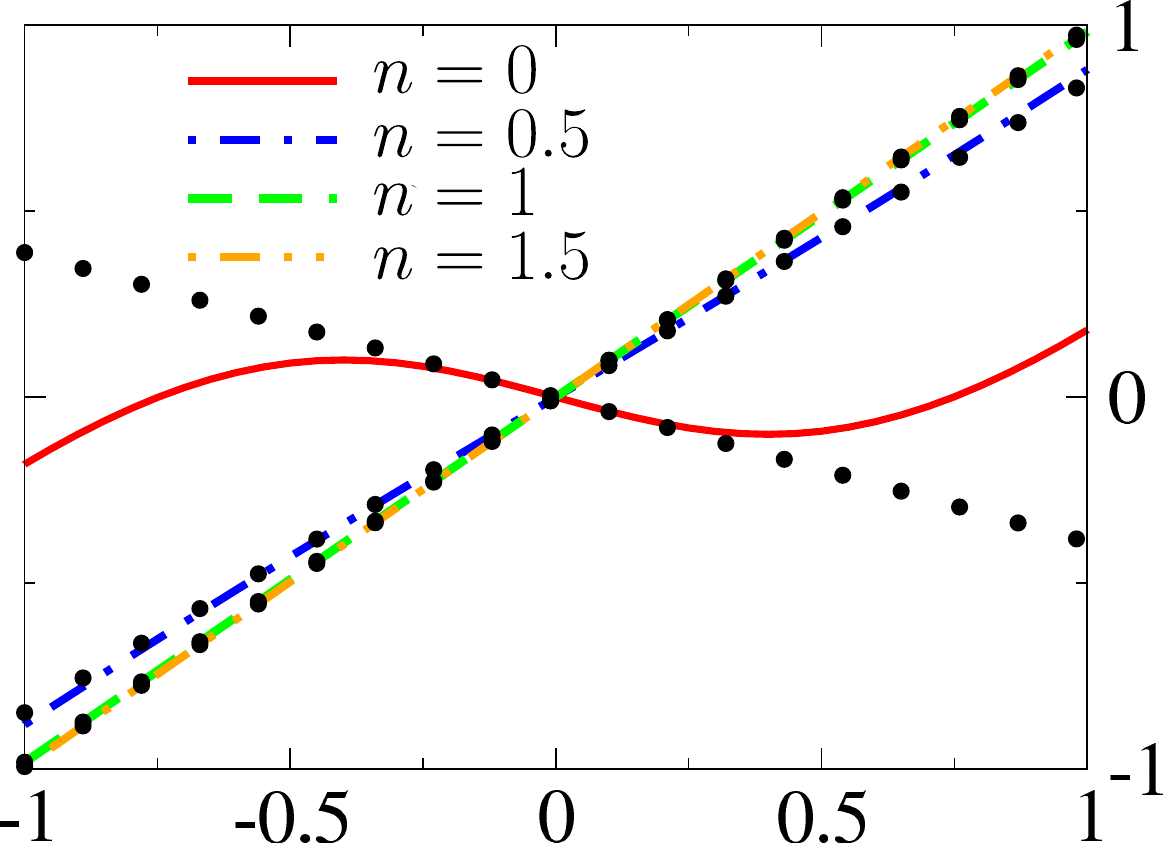}}
\centerline{\includegraphics[width=4.cm,clip]{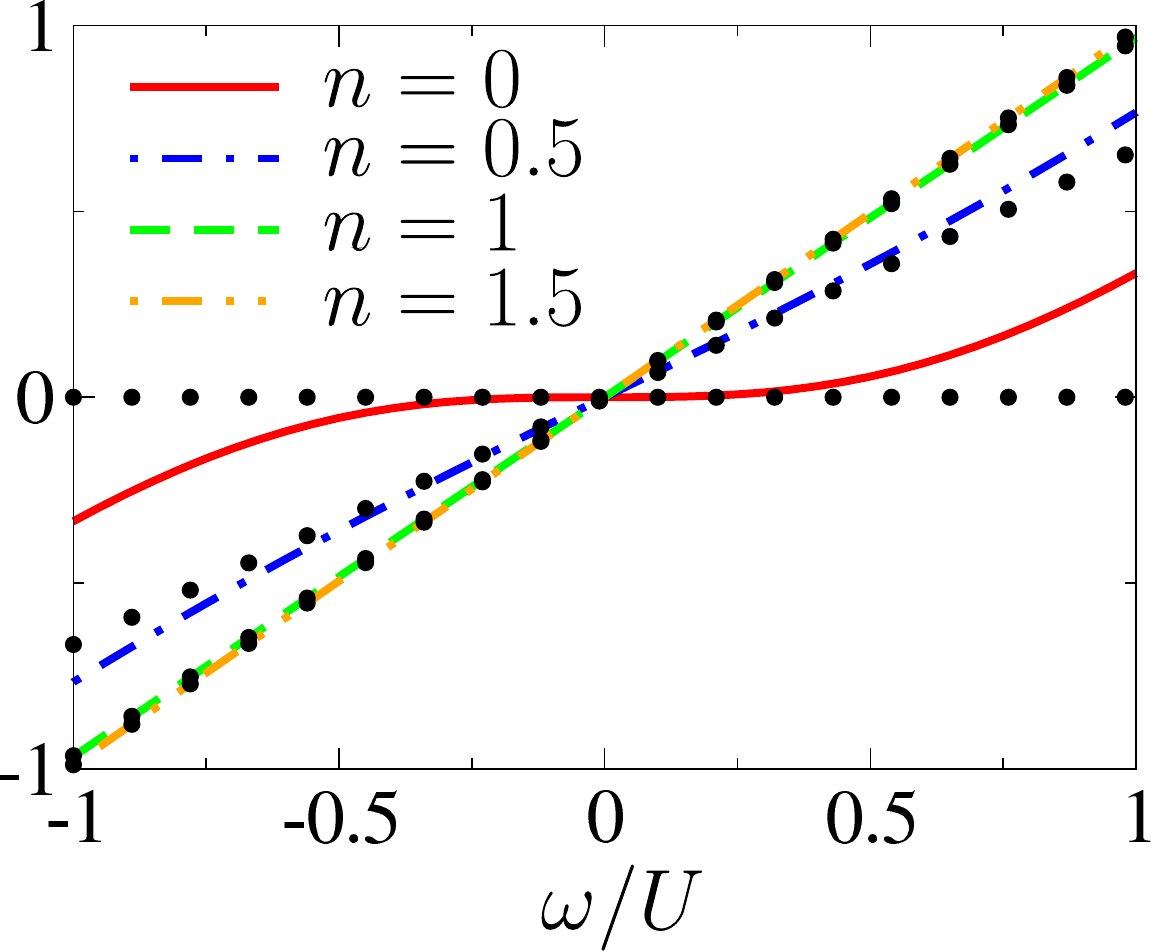}
\includegraphics[width=4.cm,clip]{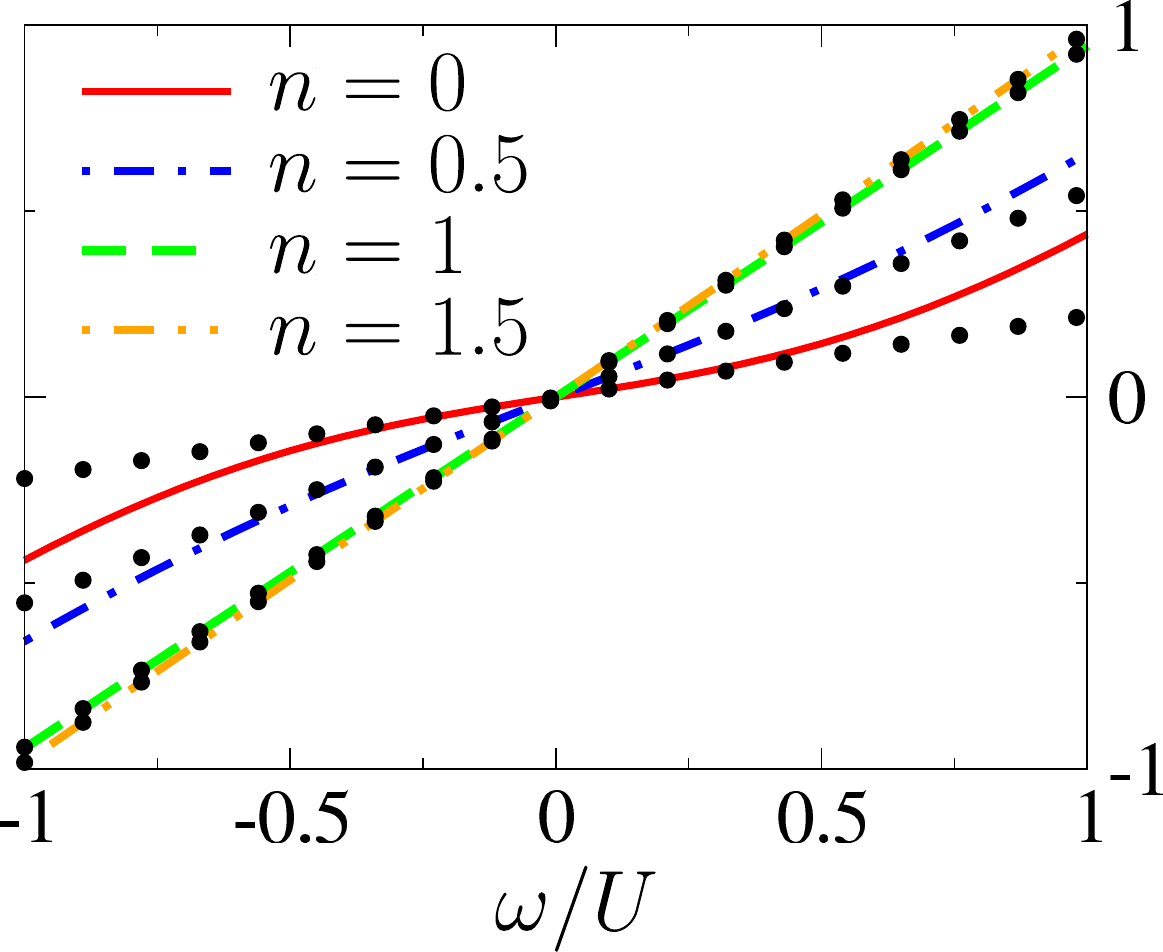}}
\caption{(Color online) $\Gamma_{{\rm loc},C}(i\w;n)/U$ vs $\w/U$ for various values of $n$. The dotted lines show the derivative expansion $Z_C(n)\w$. $\mu=-0.2U$, $0.2U$, $(\sqrt{2}-1)U$ and $0.6U$ (from top left to bottom right).}
\label{fig_deltaC_de}
\end{figure}

Figures~\ref{fig_deltaA_de} and \ref{fig_deltaC_de} show the local vertices $\Gamma_{{\rm loc},A}$ and $\Gamma_{{\rm loc},C}$ (also shown in Figs.~\ref{fig_deltaA} and \ref{fig_deltaC}) together with their derivative expansions
\begin{equation}
\begin{split}
\Gamma_{{\rm loc},A}(i\w;n) &= V_{A,\rm loc}(n)\w^2+V_{\rm loc}'(n) , \\ 
\Gamma_{{\rm loc},C}(i\w;n) &= Z_{C,\rm loc}(n) \w . 
\end{split}
\end{equation}
The derivative expansion is remarkably accurate whenever the chemical potential is negative or the condensate density large. In both limits, $\Gamma_{{\rm loc},A}(i\w;n)\simeq V_{\rm loc}'(n)$ and $\Gamma_{{\rm loc},C}(i\w;n) \simeq \w$. Since a negative $\mu$ or a large $n$ corresponds to a system deep in the superfluid phase, we conclude that the derivative expansion is fully justified in this limit. 

More generally, we see that the derivative expansion is always valid in the limit $|\w|\ll U$. As argued in Sec.~\ref{subsec_rgeq}, except deep in the Mott phase (where the strong-coupling RPA is a good approximation to the $k=0$ results), $U$ is a very large energy scale in the strong-coupling limit, and the knowledge of the vertices at energies $|\w|\ll U$ is sufficient to solve the flow equations. We therefore expect the derivative expansion to be justified also in the strong-coupling limit. 

\begin{figure}
\centerline{\includegraphics[width=6cm,clip]{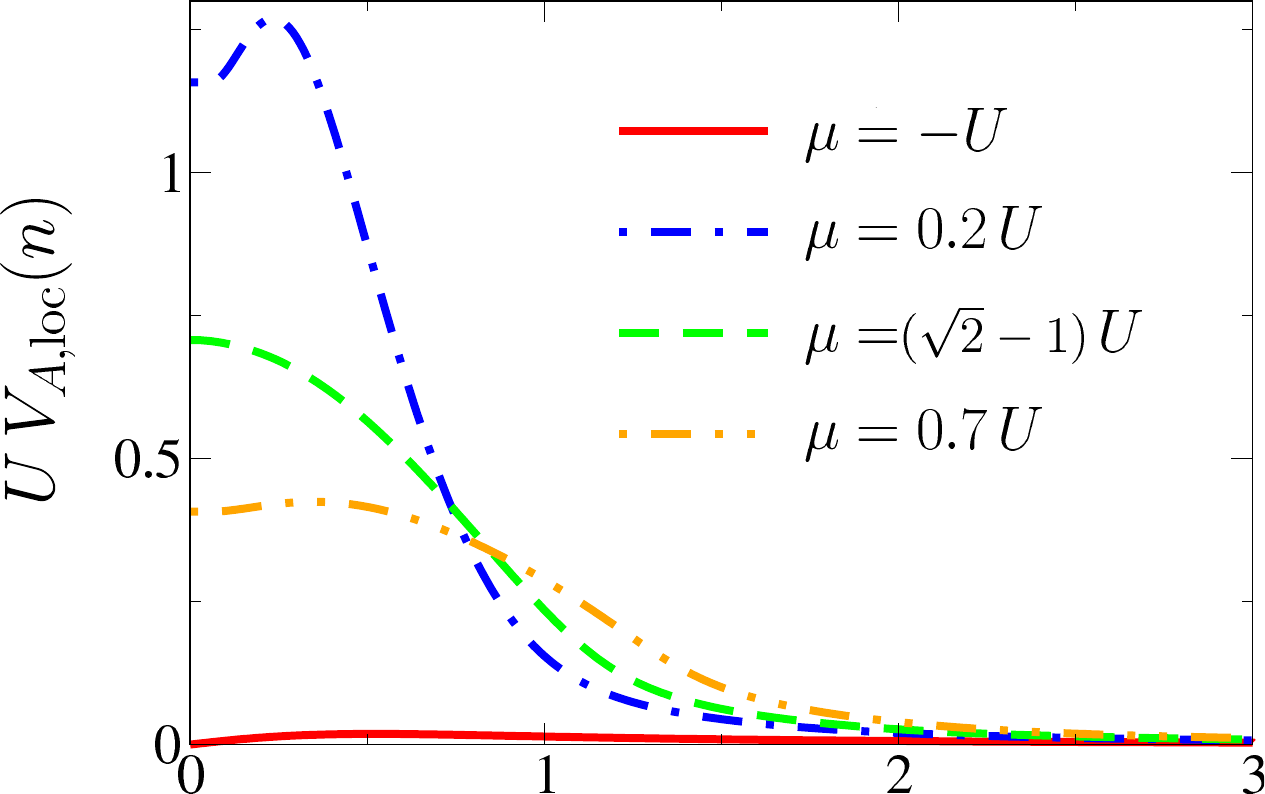}}
\centerline{\includegraphics[width=6cm,clip]{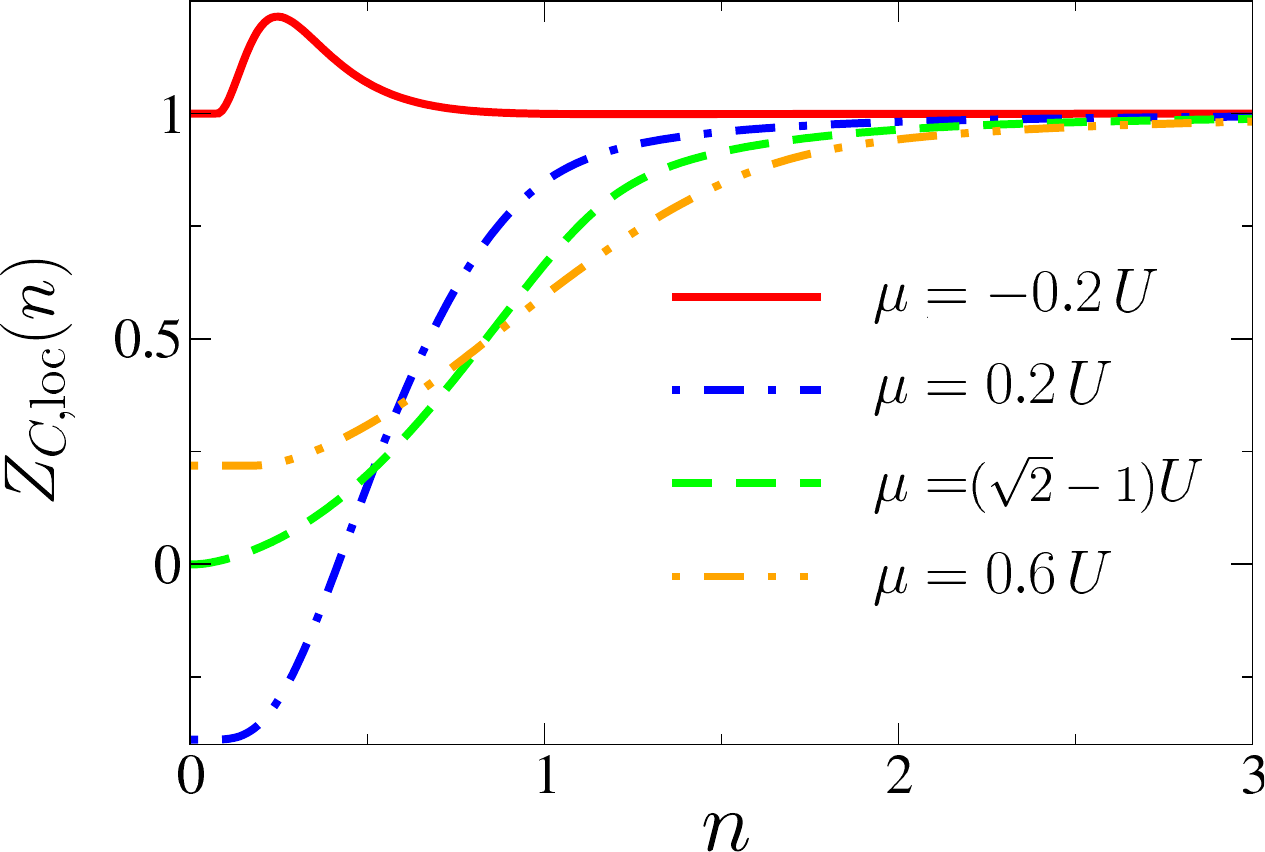}}
\caption{(Color online) $V_{A,{\rm loc}}(n)$ and $Z_{C,{\rm loc}}(n)$ vs $n$ for various values of the chemical potential $\mu$.}
\label{fig_VAZCloc}
\end{figure}

Figure~\ref{fig_VAZCloc} shows $V_{A,\rm loc}(n)$ and $Z_{C,\rm loc}(n)$ for various values of the chemical potential $\mu$.

\section{Flow equations} 
\label{app_floweq} 

The flow equations in the BMW scheme can be found in Appendix C.1 of Ref.~\onlinecite{Dupuis09b}. When $Z_{A,k}(n)$ and $V_{A,k}(n)$ are approximated by their values at the minimum $n_{0,k}$ of the effective potential, the equations simplify into 
\begin{widetext} 
\begin{align} 
\dl V_k(n) ={}& - \half \int_q \dl R_k(\q) \left[ G_{k,\rm ll}(q;n) + G_{k,\rm tt}(q;n) \right] , \\  
\dl Z_{C,k}(n) ={}& -2 n   V''_k(n)^2 \dw\left[ 3J_{\text{ll},\text{lt}}(q,n)-3 J_{\text{lt},\text{ll}}(q,n) + J_{\text{lt},\text{tt}}(q,n)- J_{\text{tt},\text{lt}}(q,n)\right]_{q=0} 
\nonumber\\ & 
-4 n^2 V^{(3)}_k(n) V''_k(n) \dw\left[ J_{\text{ll},\text{lt}}(q,n)- J_{\text{lt},\text{ll}}(q,n)\right]_{q=0}  -4 n^2   V^{(3)}_k(n) Z_{C,k}'(n) J_{\text{ll},\text{ll}}(q,n)
\nonumber\\ & 
-\half Z_{C,k}'(n) \left[I_{\text{tt}}(n)  +I_{\text{ll}}(n)\right]-I_{\text{ll}}(n) n Z_{C,k}''(n)  -2 n V''_k(n) Z_{C,k}'(n) J_{\text{ll},\text{tt}}(0,n) 
\nonumber \\ & 
-6 n V''_k(n) Z_{C,k}'(n)  J_{\text{ll},\text{ll}}(0,n)  \\
\dl V_{A,k} ={}&    -n_{0,k}\lamb_k^2 \dw^2\left[\Jbarlltt(q)    +2 \Jbarltlt(q) +\Jbarttll(q)\right]_{q=0} + 2 n_{0,k} \Jbarllll(0) Z_{C,k}'(n_{0,k})^2 
\nonumber \\ & 
+ 4  n_{0,k} \lamb_k Z_{C,k}'(n_{0,k}) \dw \left[\Jbarlllt(q)  -\Jbarltll(q)\right]_{q=0}, \\ 
\eta_{A,k} ={}& 2\lamb_k^2 n_{0,k} Z_{A,k} \eps_k k^2 \llbrace \frac{\delta_{d,2}}{2\pi} + \int_\q \theta(\eps_k-\eps_\q) \left[\frac{\partial^2_{q_x} \eps_\q}{\eps_\q} - \frac{(\partial_{q_x}\eps_\q)^2}{\eps_\q^2} \right] \rrbrace 
\int_\w \left( \frac{1}{D_-^2} + \frac{1}{D_+^2} \right) ,
\label{etaA} 
\end{align}
where  
\begin{equation}
\begin{split}
D_- &= (Z_{A,k}\eps_k + V_{A,k} \w^2) (Z_{A,k}\eps_k + V_{A,k} \w^2+2n_{0,k} \lamb_k)+Z_{C,k}(n_{0,k})^2\w^2 , \\
D_+ &= [Z_{A,k}(4dt-\eps_k) + V_{A,k} \w^2] [Z_{A,k}(4dt-\eps_k)+ V_{A,k} \w^2+2n_{0,k} \lamb_k]+Z_{C,k}(n_{0,k})^2\w^2 ,
\end{split}
\end{equation}
$Z_{A,k}\equiv Z_{A,k}(n_{0,k})$, $V_{A,k}\equiv V_{A,k}(n_{0,k})$, $\eta_{A,k}=-k\partial_k \ln Z_{A,k}$, and $l=\ln(k/\Lambda)$.

When $Z_{C,k}(n)$ is approximated by $Z_{C,k}(n_{0,k})\equiv Z_{C,k}$ and $V_k(n)$ truncated to quadratic order [Eq.~(\ref{trunc})], we obtain 
\begin{equation}
\begin{split}
\dl V_{0,k} &= - \half\int_q \dl R_k(\q) \left[ \gllb(q) +\gttb(q) \right] , \\ 
\dl n_{0,k} &= \frac{3}{2} \Ibarll + \half \Ibartt \quad \mbox{if}  \quad n_{0,k}>0,  \\ 
\dl \delta_k &= -2 \lamb_k \Ibarll  \quad \mbox{if}  \quad  n_{0,k}=0 , \\ 
\dl \lambda_k &= - \lambda_k^2 \bigl[9\Jbarllll(0) - 6\Jbarltlt(0) + \Jbartttt(0) \bigr] , \\
\dl Z_{C,k} &= 2 \lamb_k^2 n_{0,k} \frac{\partial}{\partial \w} \bigl[ \Jbarttlt(q) - \Jbarlttt(q) 
-3 \Jbarlllt(q) +3\Jbarltll(q \bigr]_{q=0} \\  
\dl V_{A,k} &= - 2 \lamb_k^2 n_{0,k} \frac{\partial}{\partial \w^2} \bigl[ \Jbarlltt(q) + \Jbarttll(q) 
+ 2\Jbarltlt(q) \bigr]_{q=0} ,
\end{split}
\label{floweq} 
\end{equation}
\end{widetext}
with $\eta_{A,k}$ given by~(\ref{etaA}). We have introduced the coefficients
\begin{equation}
\begin{split}
I_\alpha(n) &= \int_q  \tilde\dl G_{k,\alpha}(q;n) , \\ 
J_{\alpha\beta}(q;n) &= \int_{q'} [\tilde\dl G_{k,\alpha}(q';n)] G_{k,\beta}(q+q';n) ,  
\end{split}
\label{IJdef}
\end{equation}
where $\alpha,\beta={\rm ll,tt,lt}$. To alleviate the notations, we have omitted the subscript $k$ in $I_\alpha$ and $J_{\alpha\beta}$. The notation $\bar I_\alpha$, $\bar J_{\alpha\beta}$ and $\bar G$ means that these quantities are evaluated for $n=n_{0,k}$. The Green functions $G_{k,\alpha}$ in~(\ref{IJdef}) are defined as $-(\Gamma^{(2)}_k+R_k)^{-1}$, with $\Gamma^{(2)}_k$ approximated by its derivative expansion~(\ref{gamde},\ref{gamde1}). With $Z_{A,k}(n)$ and $V_{A,k}(n)$ approximated by their values at the minimum $n_{0,k}$ of the effective potential, this gives 
\begin{equation}
\begin{split}
\Gamma_{A,k}(q;n) &= Z_{A,k}\eps_\q + V_{A,k}\w^2 + V_k'(n)  , \\ 
\Gamma_{B,k}(q;n) &= V_k''(n) , \\
\Gamma_{C,k}(q;n) &= Z_{C,k}(n) \w 
\end{split}
\end{equation}
and 
\begin{equation}
\begin{split}
G_{k,\rm ll}(q;n) &= - \frac{\Gamma_{A,k}(q;n) + R_k(\q)}{D_k(q;n)} , \\ 
G_{k,\rm tt}(q;n) &= - \frac{\Gamma_{A,k}(q;n) + 2 n \Gamma_{B,k}(q;n)  + R_k(\q)}{D_k(q;n)} , \\ 
G_{k,\rm lt}(q;n) &= \frac{\Gamma_{C,k}(q;n)}{D_k(q;n)} ,
\end{split}
\end{equation}
where 
\begin{align}
D_k(q;n) ={}& [\Gamma_{A,k}(q;n) + R_k(\q) ]^2 + 2 n \Gamma_{B,k}(q;n) \nonumber \\ & \times [\Gamma_{A,k}(q;n) + R_k(\q) ] + \Gamma_{C,k}(q;n)^2 . 
\end{align}

\subsubsection*{Lattice regulator} 

The lattice cutoff function~(\ref{cutoff}) differs from cutoff functions used in the continuum, in particular due a symmetric treatment of the low- and high-energy parts of the spectrum (Fig.~\ref{fig_disp}). We can rewrite $R_k(\q)$ in the form 
\begin{equation}
R_k(\q) = - Z_{A,k}\eps_k \sgn(t_\q) y_\q r(y_\q), 
\end{equation}
where 
\begin{equation}
r(y) = \frac{1-y}{y} \Theta(1-y) 
\end{equation}
and 
\begin{equation}
y_\q = \llbrace 
\begin{array}{lcc}
\dfrac{\eps_\q}{\eps_k} & \mbox{if} & t_\q < 0 , \\
\dfrac{4dt-\eps_\q}{\eps_k} & \mbox{if} & t_\q > 0 .
\end{array}
\right.
\end{equation}
This gives 
\begin{align}
\dl R_k(\q) &= Z_{A,k}\eps_k \sgn(t_\q) y_\q [\eta_{A,k} r(y_\q) + 2y_\q r'(y_\q) ] \nonumber \\ 
&= Z_{A,k}\eps_k \sgn(t_\q) \Theta(1-y_\q) [\eta_{A,k}(1-y_\q)-2 ] . 
\label{app1}
\end{align} 
$R_k(\q)$ enters the flow equations always in the combination 
\begin{equation}
\Gamma_{A,k}(q;n)+R_k(\q)=Z_{A,k}\eps_\q+V_{A,k}\w^2+V_k'(n)+R_k(\q) ,
\end{equation}
where
\begin{equation}
Z_{A,k}\eps_\q+R_k(\q) = \llbrace 
\begin{array}{lcc} 
Z_{A,k}\eps_k & \mbox{if} & \eps_\q \leq \eps_k , \\ 
Z_{A,k}\eps_\q & \mbox{if} & \eps_\q \geq \eps_k 
\end{array}
\right.
\label{app2}
\end{equation}
for $t_\q<0$, and 
\begin{equation}
Z_{A,k}\eps_\q+R_k(\q) = \llbrace 
\begin{array}{lcc} 
Z_{A,k}(4dt-\eps_k) & \mbox{if} & \eps_\q \geq 4dt-\eps_k , \\ 
Z_{A,k}\eps_\q & \mbox{if} & \eps_\q \leq 4dt-\eps_k 
\end{array}
\right.
\label{app3}
\end{equation}
for $t_\q>0$. Equations~(\ref{app1},\ref{app2},\ref{app3}) lead to a significant simplification of the coefficients $I_\alpha(n)$ and $J_{\alpha\beta}(q;n)$. A typical contribution to $I_\alpha$ or $J_{\alpha\beta}$ reads  
\begin{equation}
F_k = \int_\q \int_\w \dl R_k(\q) f(Z_{A,k}\eps_\q+R_k(\q),\w) , 
\end{equation}
where $f$ is a product of propagators $G_k=-(\Gamma^{(2)}_k+R_k)^{-1}$. Since $\dl R_k(\q)$ restricts the momentum integral to the domain $y_\q\leq 1$ where $Z_{A,k}\eps_\q+R_k(\q)$ is independent of $\q$,
\begin{align}
F_k ={}& \int_\q \Theta(-t_\q) \dl R_k(\q) \int_\w f(Z_{A,k}\eps_k,\w) \nonumber \\ 
& + \int_\q \Theta(t_\q) \dl R_k(\q) \int_\w f(Z_{A,k}(4dt-\eps_k),\w) .
\label{app4}
\end{align} 
Introducing the lattice density of states
\begin{equation}
\calD(\eps) = \int_\q \delta(\eps-\eps_\q) \quad (0\leq \eps\leq 4dt) , 
\end{equation}
we obtain 
\begin{multline}
\int_\q \Theta(-t_\q) \dl R_k(\q) = \int_0^{2dt} d\eps \calD(\eps) \dl R_k(\q)  \\ 
= - Z_{A,k}\eps_k \int_0^{\eps_k} d\eps \calD(\eps) \left[\eta_{A,k}\left(1-\frac{\eps}{\eps_k}\right) -2\right] 
\label{app5}
\end{multline}
and 
\begin{multline}
\int_\q \Theta(t_\q) \dl R_k(\q) = \int_{2dt}^{4dt} d\eps \calD(\eps) \dl R_k(\q)  \\ 
= Z_{A,k}\eps_k \int_{4dt-\eps_k}^{4dt} d\eps \calD(\eps) \left[\eta_{A,k}\left(1-\frac{4dt-\eps}{\eps_k}\right) -2\right] .
\end{multline}
Since the hypercubic lattice density of states is symmetric, $\calD(\eps)=\calD(4dt-\eps)$, the last equation gives
\begin{equation}
\int_\q \Theta(t_\q) \dl R_k(\q) = -\int_\q \Theta(-t_\q) \dl R_k(\q) 
\end{equation}
which enables us to rewrite \eqref{app4} as 
\begin{multline}
F_k = \int_\q \Theta(-t_\q) \dl R_k(\q) \\
\times \int_\w \left[ f(Z_{A,k}\eps_k,\w) - f(Z_{A,k}(4dt-\eps_k),\w) \right] , 
\end{multline}
with the momentum integral given by~\eqref{app5}. 

In the limit $k\ll\Lambda$ (or $\eps_k\ll 2dt$), the function $f(Z_{A,k}(4dt-\eps_k),\w)$ involves propagators with a large gap and is therefore negligible with respect to $f(Z_{A,k}\eps_k,\w)$; the flow is then governed only by the low-energy modes. Moreover, for $\eps_\q\leq \eps_k\ll 2dt$, the lattice does not matter and we can approximate $\eps_\q\simeq t\q^2$, which leads to 
\begin{equation}
\calD(\eps) = 2 v_d \frac{\eps^{d/2-1}}{t^{d/2}} \qquad  (\eps\ll 2dt)  
\label{dosapp}
\end{equation}
and
\begin{equation}
\int_\q \Theta(-t_\q) \dl R_k(\q) = 8 \frac{v_d}{d} Z_{A,k}\eps_k k^d \left(1-\frac{\eta_{A,k}}{d+2}\right) , 
\end{equation}
where $v_d^{-1}=2^{d+1}\pi^{d/2}\Gamma(d/2)$. We then obtain 
\begin{equation}
F_k = 8 \frac{v_d}{d} Z_{A,k}\eps_k k^d \left(1-\frac{\eta_{A,k}}{d+2}\right)  \int_\w  f(Z_{A,k}\eps_k,\w) ,
\end{equation}
which is the usual form for models in the continuum limit with the theta cutoff function.\cite{Litim00} 

\section{The vacuum limit}
\label{app_vac}

\subsection{Scattering length on the lattice}

Let us first recall how the $s$-wave scattering length $a$ is computed from the low-energy behavior of the $T$ matrix in the continuum. For a contact interaction $U$, the retarded $T$ matrix is defined by 
\begin{equation}
\frac{1}{T^R(\w)} = \frac{1}{U} +  \Pi^R(\w) ,
\label{app10}
\end{equation}
where 
\begin{align}
\Pi^R(\w) &= \int_\q \frac{1}{2\eps_\q - \w - i0^+} \nonumber \\ 
&= \calP \int_\q \frac{1}{2\eps_\q - \w} + i\pi \int_\q \delta(\w-2\eps_\q) 
\label{app11}
\end{align}
($\calP$ denotes the principal part) and $\eps_\q=\q^2/2m$ is the dispersion of the free bosons. In three dimensions, 
\begin{equation}
T^R(\q^2/m) = \frac{4\pi a}{m} \frac{1}{1+i|\q|a} \quad (|\q|\to 0) ,
\label{app12}
\end{equation} 
while in two dimensions
\begin{equation}
T^R(\q^2/m) = - \frac{2\pi/m}{\ln\left(\frac{|\q|a}{2}\right)+C-i\frac{\pi}{2}} \quad (|\q|\to 0)  , 
\label{app13}
\end{equation}
where $C$ is Euler's constant. Equations~(\ref{app12}) and (\ref{app13}) define the $s$-wave scattering length $a$ in three and two dimensions, respectively. 

A scattering length can be defined similarly in the Bose-Hubbard model. At low energy ($\eps_\q\ll t$) the lattice does not matter and one can approximate the boson dispersion $\eps_\q=t_\q+2dt$ by $t\q^2$. The bosons then behaves as free particles with an effective mass $m=1/2t$. Thus, equations~(\ref{app10}-\ref{app13}) allow us to define a scattering length provided that we replace $m$ by $1/2t$. 

\subsubsection{$d=3$} 

In the low-energy limit $0\leq \w\ll t$,
\begin{align}
\Pi^R(\w) &\simeq \int_\q \frac{1}{2\eps_\q} + i\pi \int_\q  \delta(\w-2\eps_\q) \nonumber \\ 
&= \int_0^{8t} d\eps \frac{\calD(\eps)}{2\eps} + i\frac{\pi}{2} \calD\left(\frac{\w}{2}\right) ,
\label{app14}
\end{align}
where $\calD(\eps)=\int_\q \delta(\eps-\eps_\q)$ is the density of states of the cubic lattice. The last integral in (\ref{app14}) can be computed numerically while $\calD(\w)\simeq \sqrt{\w}/4\pi^2 t^{3/2}$ for small $\w$. This gives
\begin{equation}
\frac{1}{T^R(\w)} = \frac{1}{U} + \frac{A}{t} + \frac{i}{8\sqrt{2}\pi} \frac{\w^{1/2}}{t^{3/2}} 
\end{equation}
and 
\begin{equation} 
\frac{1}{T^R(2t|\q|^2)} = \frac{1+i|\q|a}{8\pi t a} 
\end{equation}
for $\q\to 0$ with 
\begin{equation}
a = \frac{1}{8\pi} \frac{1}{t/U+A} 
\label{app14c}
\end{equation}
and $A\simeq 0.1264$. 

\subsubsection{$d=2$} 

To compute $\Pi^R(\w)$ in two dimensions, we use
\begin{align}
\calP \int_\q \frac{1}{2\eps_\q - \w} ={}& \calP \int_\q \left(\frac{1}{2\eps_\q - \w} - \frac{1}{2t\q^2 - \w} \right) \nonumber \\ & + \calP \int_\q \frac{1}{2t\q^2 - \w} .
\label{app15}
\end{align}
Since the first integral in the rhs of Eq.~(\ref{app15}) is convergent, we can set $\w=0$, which gives
\begin{equation}
 \int_\q \left(\frac{1}{2\eps_\q} - \frac{1}{2t\q^2} \right) = \frac{1}{2t\pi^2} \left(G + \frac{\pi}{4} \ln \frac{8}{\pi^2} \right) , 
\label{app14a}
\end{equation}
where $G\simeq 0.916$ is the Catalan constant. As for the last integral in Eq.~(\ref{app15}) we obtain 
\begin{multline}
\calP  \int_\q \frac{1}{2t\q^2 - \w} = \calP  \int_\q \Theta(\pi-|\q|) \frac{1}{2t\q^2 - \w} \\ 
+ \calP  \int_\q \Theta(|\q|-\pi) \frac{1}{2t\q^2 - \w} \\ 
= \frac{1}{8\pi t} \ln \left(\frac{2t\pi^2}{\w}\right) + \frac{1}{4\pi t} \left(\ln 2 - \frac{2}{\pi} G \right) .
\end{multline} 
for $0\leq\w\ll t$. Since 
\begin{equation}
i\pi \int_\q \delta(\w-2\eps_\q) \simeq i\pi \int_\q \delta(\w-2t\q^2) = \frac{i}{8t} ,
\end{equation}
we finally obtain 
\begin{equation}
\frac{1}{T^R(2t|\q|^2)} = - \frac{1}{4\pi t} \left[ \ln\left(\frac{|\q|a}{2}\right)+C-i\frac{\pi}{2}\right] 
\end{equation}
for $\q\to 0$ with 
\begin{equation}
a = \frac{1}{2\sqrt{2}} e^{-4\pi t/U-C} . 
\label{app14b}
\end{equation}

\subsection{Coupling constant $\lamb_k$} 

The low-energy limit $k\ll \Lambda$ of the coupling constant $\lamb_k$ in vacuum ($\bar n_k=n_{0,k}=0$) can be expressed in terms of the scattering length $a$. $\lamb_k$ can be obtained from the RG equation~(\ref{floweq}) (see Sec.~\ref{subsec_rgeqvac}) or more simply from 
\begin{equation}
\frac{1}{\lamb_k} = \frac{1}{U} + \half \int_\q \frac{1}{\eps_\q+R_k(\q)} ,
\label{app16}
\end{equation}
where $\eps_\q+R_k(\q)$ is given by Eqs.~(\ref{app2},\ref{app3}) with $Z_{A,k}=1$. For $k\ll\Lambda$, we can ignore the effect of the cutoff function on the high-energy part of the spectrum ($t_\q>0$), 
\begin{equation}
\frac{1}{\lamb_k} = \frac{1}{U} + \half \int_\q \frac{\Theta(\eps_k-\eps_\q)}{\eps_k} + \half \int_\q \frac{\Theta(\eps_\q-\eps_k)}{\eps_\q}  .
\end{equation}

\subsubsection{$d=3$} 

Using 
\begin{equation}
\half \int_\q \frac{\Theta(\eps_k-\eps_\q)}{\eps_k} = \frac{k}{12\pi^2t} 
\end{equation}
and
\begin{align} 
\half \int_\q \frac{\Theta(\eps_\q-\eps_k)}{\eps_\q} &=  \half \int_\q \frac{1}{\eps_\q} - \half \int_\q \frac{\Theta(\eps_k-\eps_\q)}{\eps_\q} \nonumber \\ 
&\simeq \half \int_\q \frac{1}{\eps_\q} - \frac{k}{4\pi^2t} 
\end{align}
(we have approximated $\eps_\q\simeq t\q^2$ for $\eps_\q\leq \eps_k \ll\Lambda$), we finally obtain 
\begin{align}
\frac{1}{\lamb_k} &= \frac{1}{U} + \half \int_\q \frac{1}{\eps_\q} - \frac{k}{6\pi^2t} \nonumber \\ 
&= \frac{1}{8\pi ta} \left(1-\frac{4}{3\pi} ka \right) ,
\end{align}
where $a$ is the three-dimensional scattering length~(\ref{app14c}).

\subsubsection{$d=2$} 

In two dimensions, we rewrite Eq.~(\ref{app16}) as 
\begin{align}
\frac{1}{\lamb_k} ={}& \frac{1}{U} + \half \int_\q \frac{1}{t\q^2+\tilde R_k(\q)} \nonumber \\ &
+ \half \int_\q \left( \frac{1}{\eps_\q+R_k(\q)} - \frac{1}{t\q^2+\tilde R_k(\q)} \right) ,
\label{app17}
\end{align}
where $\tilde R_k(\q)$ is obtained from $R_k(\q)$ by replacing $\eps_\q$ with $t\q^2$. Since the last integral in (\ref{app17}) is convergent for $k\to 0$, we can set $k=0$ and use the result~(\ref{app14a}). The first integral in (\ref{app17}) can be expressed as 
\begin{multline}
\half \int_\q \frac{\Theta(k-|\q|)}{\eps_k} + \half \int_\q \frac{\Theta(\pi-|\q|)\Theta(|\q|-k)}{t\q^2} \\ + \half \int_\q \frac{\Theta(|\q|-\pi)}{t\q^2} \\ 
= \frac{k^2}{8\pi \eps_k} + \frac{1}{4\pi t} \ln\frac{\pi}{k} + \frac{1}{4\pi t} \left(\ln 2 - \frac{2}{\pi} G\right) . 
\end{multline}
We deduce 
\begin{align}
\frac{1}{\lamb_k} &= \frac{1}{U} + \frac{1}{8\pi t} + \frac{1}{4\pi t } \ln\left(\frac{4\sqrt{2}}{k}\right) \nonumber \\ 
&= - \frac{1}{4\pi t} \left[ \ln\left(\frac{ka}{2}\right) + C -\half \right] , 
\end{align} 
where $a$ is the two-dimensional scattering length~(\ref{app14b}).

\subsection{RG equation $\partial_k\lamb_k$} 
\label{subsec_rgeqvac}

In the vacuum, $Z_{A,k}=Z_{C,k}=1$, $V_{A,k}=0$ and $n_{0,k}=0$, the two-point vertex is defined by  
\begin{equation}
\Gamma_{A,k}(q) = \eps_\q , \quad \Gamma_{B,k}(q) = \lamb_k , \quad \Gamma_{C,k}(q) = \w .
\end{equation}
The RG equation satisfied by $\lamb_k$ takes the simple form
\begin{align}
\dk \lamb_k = - \lamb_k^2 \int_q & \frac{\dk R_k(\q)}{D^3} \bigl\lbrace 22\w^2[\eps_\q+R_k(\q)] \nonumber \\  &-10[\eps_\q+R_k(\q)]^3 \bigr\rbrace , 
\label{lambeq}
\end{align} 
where $D=[\eps_\q+R_k(\q)]^2+\w^2$.  We deduce 
\beq
\dk \lamb_k = \frac{\lamb_k^2}{2} \int_\q \frac{\dk R_k(\q)}{[\eps_\q+R_k(\q)]^2} 
\eeq 
and
\begin{align}
\frac{1}{\lamb_k} - \frac{1}{\lamb_\Lambda} &= \half \int_\q \left( \frac{1}{\eps_\q+R_k(\q)} - \frac{1}{\eps_\q+R_\Lambda(\q)} \right) \nonumber \\
&= \half \int_\q \frac{1}{\eps_\q+R_k(\q)} - \frac{1}{4dt} .
\label{app18}
\end{align}  
$\lamb_\Lambda$ can be computed from the local action $S_\Lambda=S_{\rm loc}$ in the vacuum ($\mu=-2dt$),
\begin{equation} 
\frac{1}{\lamb_\Lambda} = \frac{1}{U} + \int_\w G(i\w)G(-i\w) = \frac{1}{U} + \frac{1}{4dt} , 
\label{app19}
\end{equation}
where $G(i\w)=(i\w+\mu)^{-1}$ is the local (normal) propagator in vacuum. From Eqs.~(\ref{app18}) and (\ref{app19}), we recover Eq.~(\ref{app16}).

\subsection{RG equations in the dilute limit for $k\gg k_h$}

We now consider the RG equations at finite density but in the dilute limit ($k_h\ll\Lambda$) for $k\gg k_h$. To leading order in $\lamb_k n_{0,k}$, $\lamb_k$ satisfies the equation~(\ref{lambeq}). To obtain the equation satisfied by $n_{0,k}$, we must expand the propagator to first order in $\lamb_k n_{0,k}$,
\begin{equation}
\begin{split}
G_{\rm ll}(q) &= - \frac{\eps_\q+R_k(\q)}{D} + \frac{2\lamb_k n_{0,k}}{D^2} [\eps_\q+R_k(\q)]^2 , \\
G_{\rm tt}(q) &= - \frac{\eps_\q+R_k(\q)}{D} - \frac{2\lamb_k n_{0,k}}{D^2} \w^2 ,
\end{split}
\end{equation}
where $D$ is defined in Sec.~\ref{subsec_rgeqvac}. We will show below that the flow of $Z_{A,k}$, $Z_{C,k}$ and $V_{A,k}$ leads to higher-order corrections. We can therefore set $Z_{A,k}$, $Z_{C,k}$ and $V_{A,k}$ to their vacuum values. This gives
\begin{align}
\dk n_{0,k} = \int_q & \frac{\dk R_k(\q)}{D^3} 2 \lamb_k n_{0,k} \bigl\lbrace 5\w^2[\eps_\q+R_k(\q)] \nonumber \\ & -3 [\eps_\q+R_k(\q)]^3 \bigr\rbrace .
\label{n0eq}
\end{align}
Eq.~(\ref{lambn0}) follows from (\ref{lambeq}) and (\ref{n0eq}).

Let us now show that the flow of $Z_{A,k}$, $Z_{C,k}$ and $V_{A,k}$ give subleading contributions to Eq.~(\ref{n0eq}). To leading order in $\lamb_k n_{0,k}$, the RG equation of $Z_{C,k}$ reads
\begin{align}
\dl Z_{C,k} &= -\lamb_k^2 n_{0,k} \int_\q \frac{\dl R_k(\q)}{[\eps(\q)+R_k(\q)]^3} \nonumber \\ 
&= -\lamb_k^2 n_{0,k} \int_\q \Theta(-t_\q) \dl R_k(\q) \left[ \frac{1}{\eps_k^3} - \frac{1}{(4dt-\eps_k)^3} \right] .
\end{align}
To estimate the order of magnitude of $\dl Z_{C,k}$, we can ignore the term $(4dt-\eps_k)^{-3}$ (which is smaller than $\eps_k^{-3}$) and use the approximate density of states~(\ref{dosapp}). This gives 
\begin{align}
\dl Z_{C,k} &\sim - 8\frac{v_d}{d} \lamb_k^2 n_{0,k} k^d \eps_k^{-2} \nonumber \\ 
&\sim - 8\frac{v_d}{dt} \lamb_k k_h^2 k^{d-4} ,
\label{ZCeq}
\end{align}
where we have used $\lamb_kn_{0,k}=tk_h^2$ to leading order. Integrating this equation between $k=\Lambda$ and $k\ll\Lambda$, we obtain
\beq
Z_{C,k} -1 \sim \frac{\lamb_k k^2_h}{t k^{4-d}} 
\eeq
(ignoring the dependence of $\lamb_k$ on $k$). In the dilute limit and for $k\gtrsim k_h$, the rhs is always small. For example, in three dimensions, one finds $Z_{C,k_h} -1\sim k_ha\ll 1$. 

Similarly, we obtain
\begin{align}
\dl V_{A,k} &= - \frac{3}{4} \lamb_k^2 n_{0,k} \int_\q \frac{\dl R_k(\q)}{[\eps(\q)+R_k(\q)]^4} \nonumber \\
&\sim - 6\lamb_k \frac{v_d}{dt^2} k_h^2 k^{d-6} 
\label{VAeq}
\end{align}
and therefore
\beq
V_{A,k} \sim \frac{\lamb_k k^2_h}{t^2 k^{6-d}} 
\label{VAeq1}
\eeq
for $k\ll\Lambda$. Since the frequency integrals extend up to $\w\sim tk^2$ while $\eps_\q \sim tk^2$, the $V_{A,k}\w^2$ term in the propagators can be neglected if $V_{A,k}\ll 1/tk^2$. Equation~(\ref{VAeq1}) shows that this is indeed the case. For example, in three dimensions $tk_h^2 V_{A,k_h} \sim k_ha\ll 1$. 

Finally
\begin{align}
\eta_{A,k} \sim 4\lamb_k \frac{v_d}{dt} k_h^2 k^{d-4} , 
\label{etaAeq}
\end{align}
where we have calculated the integral over $\q$ in~(\ref{etaA}) using $\eps_\q=t\q^2$. As $Z_{C,k}$, $Z_{A,k}$ remains close to unity for $k\gtrsim k_h$. 

%


\begin{thebibliography}{85}
\bibitem [{not({\natexlab{b}})}]{note13}%
  \BibitemOpen
  \href@noop {} {} \bibinfo {note} {In general the
  Bogoliubov theory goes beyond the mere approximation
  $\Gamma[\phi^*,\phi]=S[\phi^*,\phi]$ as it includes $T$-matrix
  renormalization of the interaction $U$ (see Sec.~\ref{subsec_dilute}). This
  renormalization is however negligible in the limit $t/U\gg 1$.}\BibitemShut
  {Stop}%
\bibitem [{not({\natexlab{c}})}]{note4}%
  \BibitemOpen
  \href@noop {} {} \bibinfo {note} {Note that our definition
  of the compressibility differs from the usual one $\kappa=\bar n^{-2} d\bar
  n/d\mu$.}\BibitemShut {Stop}%
\bibitem [{not({\natexlab{d}})}]{note15}%
  \BibitemOpen
  \href@noop {} {} \bibinfo {note} {Note that in a gapless
  phase it is {\it a priori} not justified to deduce the finite $\q,\w$
  behavior of the $k=0$ propagator from the derivative expansion, since the
  latter requires $|\q|\lesssim k$ and $|\w|\lesssim \w_k^-$. For instance, if
  $Z_{C,k=0}=0$, then $Z_{C,k}$ will in general be nonzero for $k>0$. For
  $Z_{C,k}$ to drop out of the propagator $G_k$, one has to verify that it
  vanishes sufficiently rapidly with $k$ so that its contribution to
  $G_k(\q,i\w)$ with $|\q|\sim k$ and $|\w|\sim \w_k^-$ can be neglected. Using
  this procedure, one can show that all conclusions of Sec.~\ref{subsec_de} are
  correct (as confirmed by the numerical results of Secs.~\ref{sec_crit} and
  \ref{sec_sf}).}\BibitemShut {Stop}%
\bibitem [{not({\natexlab{e}})}]{note5}%
  \BibitemOpen
  \href@noop {} {} \bibinfo {note} {In the weak-coupling
  limit, $\Gamma^{(3)}_k$ and $\Gamma_k^{(4)}$ are nearly frequency independent
  and the derivative expansion of $\Gamma_k^{(2)}$ is essentially exact for
  $k_G\ll k\leq\Lambda$ ($k_G$ is the Ginzburg momentum scale).}\BibitemShut
  {Stop}%
\bibitem [{not({\natexlab{g}})}]{note7}%
  \BibitemOpen
  \href@noop {} {} \bibinfo {note} {Transcients effects
  associated with the approach to the multicritical points crucially depend on
  $Z_{C,k}(n)$ being a function (and not merely a number
  $Z_{C,k}(n_{0,k})$).}\BibitemShut {Stop}%
\bibitem [{not({\natexlab{h}})}]{note14}%
  \BibitemOpen
  \href@noop {} {} \bibinfo {note} {The fact that the
  transition at the Mott lobe tip occurs at constant density can be understood
  from the following argument.\cite{Fisher89} If the line $\bar n=p$ ($p$
  interger) in the superfluid phase joined the corresponding Mott insulating
  lobe at a point other than its tip, then the compressibility would be
  negative in the vicinity of the tip, which is physically not
  possible.}\BibitemShut {Stop}%
\bibitem{note19} From Fig.~\ref{fig_rhokappa}, we deduce $\nu\simeq 0.696$ and $2\beta=\nu(1+\eta)\simeq 0.7338$, which gives $\eta\simeq 0.053$.%
\bibitem [{not({\natexlab{i}})}]{note8}%
  \BibitemOpen
  \href@noop {} {} \bibinfo {note} {In the RG
  equations~(\ref{floweq}), the condensate density $n_{0,k}$ enters
  in the combination $\eps_k+2\lamb_kn_{0,k}$, so that the healing momentum
  scale is defined by $\eps_{k_h}\sim \lamb_{k_h}n_{0,k_h}$. $k_h$ can also be
  directly deduced from $\Gamma^{(2)}_k(q;n)$.}\BibitemShut {Stop}%
\bibitem [{not({\natexlab{j}})}]{note9}%
  \BibitemOpen
  \href@noop {} {} \bibinfo {note} {In practice, we find
  that the lattice can be ignored as soon as $k\lesssim \Lambda e^{-0.5}\simeq
  0.4\pi$.}\BibitemShut {Stop}%
\bibitem{note20} Schick's results\cite{Schick71} have also been obtained from an NPRG approach (in a continuum model) in Ref.~\onlinecite{Sinner10}.
\bibitem [{not({\natexlab{k}})}]{note10}%
  \BibitemOpen
  \href@noop {} {} \bibinfo {note} {Alternatively, one can
  determine $k_G$ from the inflection point in $\lamb_k$ or $Z_{C,k}$. The
  three criteria give similar results.}\BibitemShut {Stop}%
\bibitem [{not({\natexlab{n}})}]{note12}%
  \BibitemOpen
  \href@noop {} {} \bibinfo {note} {Note that the
  singularity of the longitudinal propagator and the continuum of excitations
  is overlooked in the strong-coupling RPA.}\BibitemShut {Stop}%
\bibitem [{not({\natexlab{o}})}]{note17}%
  \BibitemOpen
  \href@noop {} {} \bibinfo {note} {Recent calculations,
  based on the variational cluster approach, seem to indicate that a gapped
  mode is present in the superfluid phase.\cite{Knap11}}\BibitemShut {NoStop}%
\bibitem [{not({\natexlab{p}})}]{note18}%
  \BibitemOpen
  \href@noop {} {} \bibinfo {note} {It takes only a couple
  of seconds (depending on the approximation scheme) to solve
  numerically the NPRG equations (for $t$, $U$ and $\mu$ given) on a standard
  PC, so that the full determination of the phase diagram requires at most an
  hour.}\BibitemShut {Stop}%
\end{thebibliography}

\end{document}